\documentclass[12pt,a4paper]{article}

\usepackage{jheppub}
\graphicspath{{Img/}}

\usepackage{color}
\usepackage[dvipsnames, svgnames, x11names]{xcolor}
\usepackage[utf8]{inputenc}
\usepackage{float}
\usepackage{graphicx}
\usepackage{setspace}
\usepackage{subfigure}
\usepackage{bm}
\usepackage{bbm}
\usepackage{amsmath}
\usepackage{amssymb}
\usepackage{amsbsy}
\usepackage{amsthm}
\usepackage{amsfonts}
\usepackage{braket}
\usepackage{multirow}
\usepackage{ulem}
\usepackage{slashed}
\usepackage{dsfont}
\usepackage{csquotes}
\usepackage[compat=1.1.0]{tikz-feynman}

\usepackage{enumitem}
\newlist{senum}{enumerate}{1} 
\setlist[senum]{leftmargin=4.5cm ,align=left, label=\textbf{\Roman*}} 

\def\nn{\nonumber}
\def\Tr{\text{Tr}}
\def\sgn{\text{sgn}}
\def\J{\mathcal{J}}

\def\G{\mathcal{G}}
\def\O{\mathcal{O}}

\def\E{\mathcal{E}}

\def\avg#1{\left\langle#1\right\rangle}
\def\bra#1{\left\langle#1\right|}
\def\ket#1{\left|#1\right\rangle}

\def\abs#1{\left|#1\right|}
\def\kc#1{\left(#1\right)}
\def\kd#1{\left[#1\right]}
\def\ke#1{\left\{#1\right\}}

\def\Im{{\rm Im}}
\def\sgn{{\rm sgn}}
\def\mode{\text{ }{\rm mod}\text{ }}

\def\be{\begin{equation}}       \def\ee{\end{equation}}
\def\bea{\begin{eqnarray}}      \def\eea{\end{eqnarray}}
\def\ba{\begin{array}}
	\def\ea{\end{array}}
\def\bnum{\begin{enumerate} }
	\def\enum{\end{enumerate}}

\def\nn{\nonumber}

\def\=>{\Rightarrow}
\def\>{\rightarrow}

\def\eye2{Fathbb{I}}

\def\eff{\mathrm{eff}}

\def\Tr{\mathrm{Tr}}

\newcommand{\order}[1]{\mathcal{O}(#1)}

\newcommand{\twoptf}[1]{
	\feynmandiagram [inline=(a.base), layered layout, horizontal=a to b, small] {
		#1 ,
	};}

\title{Wormhole-induced effective coupling in SYK chains}

\author{Pablo Basteiro,$^{a}$}
\emailAdd{pablo.basteiro@uni-wuerzburg.de}

\author{Giuseppe Di Giulio,$^{a,b}$}
\emailAdd{giuseppe.di-giulio@fysik.su.se}

\author{Johanna Erdmenger,$^{a}$}
\emailAdd{johanna.erdmenger@uni-wuerzburg.de}

\author{Ren\'e Meyer,$^{a}$}
\emailAdd{rene.meyer@uni-wuerzburg.de}

\author{and Zhuo-Yu Xian$^{a,c,*}$ \note[*]{Corresponding author.}}
\emailAdd{zhuo-yu.xian@uni-wuerzburg.de}

\affiliation{$^a$Institute for Theoretical Physics and Astrophysics and W\"urzburg-Dresden Cluster of Excellence ct.qmat, Julius-Maximilians-Universit\"at W\"urzburg, Am Hubland, 97074 W\"urzburg, Germany
\\
$^b$ The Oscar Klein Centre and Department of Physics,
Stockholm University, AlbaNova,
\\
10691 Stockholm, Sweden
\\
$^c$ Department of Physics, Freie Universit{\"a}t Berlin, Arnimallee 14, 14195 Berlin, Germany}

\abstract{Inhomogeneous quantum chains have recently been considered in the context of developing novel discrete realizations of holographic dualities.
To advance this programme, we explore the ground states of infinite chains with large number $N$ of Majorana fermions on each site, which interact via on-site $q$-body Sachdev-Ye-Kitaev (SYK) couplings, as well as via additional inhomogeneous hopping terms between nearest-neighbour sites. 
The hopping parameters are either aperiodically or randomly distributed. Our approach unifies techniques to solve SYK-like models in the large $N$ limit with a real-space renormalization group method known as strong-disorder renormalization group (SDRG). 
We show that the SDRG decimation of SYK dots linked by a strong hopping induces an effective hopping interaction between their neighbouring sites. 
If two decimated sites are nearest neighbours, in the large $q$ limit their local ground states admit a holographic dual description in terms of eternal traversable wormholes.
At the end of the SDRG procedure, we obtain a factorised ground state of the infinite inhomogeneous SYK chains that we consider, which has a spacetime description involving a sequence of wormholes.
This amounts to a local near-boundary description of the bulk geometry in the context of discrete holography.
}

\begin{document}
\sloppy
\maketitle

\newpage

\section{Introduction}

The holographic principle \cite{tHooft:1993dmi,Susskind1995} has shaped our approach to high-energy physics. With the AdS/CFT correspondence \cite{Maldacena:1997re,Gubser:1998bc,Witten:1998qj,AHARONY2000183} as its best understood realization, holographic dualities have lead to a plethora of insights into the properties and phenomena arising in theories of quantum gravity. More recently, extensions of holography into discrete systems, motivated both from experimental and theoretical perspectives, have proven useful to investigate the regime of validity of the holographic principle. This \textit{discrete holography} programme \cite{Basteiro:2022zur,Basteiro:2022xvu,JahnCentralCharge,Jahn:2021kti,Erdmenger:2024jsb,Dey:2024jno,Basteiro:2022pyp,Asaduzzaman:2020hjl,Brower:2019kyh,Asaduzzaman:2021bcw,Brower:2022atv,Axenides:2013iwa,Axenides:2019lea,Axenides:2022cwy,Gubser:2016htz,Gubser:2016guj,Heydeman:2016ldy,Heydeman:2018qty,Hung:2019zsk,Chen:2021ipv,Chen:2021qah,Chen:2021rsy,Gesteau:2022hss} (see also \cite{Kollar:2019ngc,Yan:2018nco,Yan:2019quy,Petermann:2023zir,Chen:2023cad,FlickerBoyleDickens,Boyle:2024qzn,Lenggenhager:2024gmf,Chen:2023usl,Chen:2023nzk,Chen:2022uje,Attar:2022bwp,Stegmaier:2021chz,Lenggenhager:2021cps,Bienias:2021lem,Boettcher:2021njg,Boettcher:2019xwl,Sun:2024rxr,Tummuru:2023wuq,Shankar:2023tsw,Lenggenhager:2023kop,Bzdusek:2022vxe,Maciejko:2021czw,Maciejko:2020rad} for interdisciplinary studies on hyperbolic lattices relating to condensed-matter physics) aims to establish explicit dualities between theories on discretized  AdS spacetimes and lattice models at their boundary. In particular, disordered spin chains are a prime candidate for such boundary theories \cite{JahnCentralCharge,Jahn:2021kti,Basteiro:2022zur,Basteiro:2022xvu}, given that they naturally capture information about the discretization scheme of the bulk \cite{FlickerBoyleDickens,Boyle:2024qzn}. Analytical insights into the ground states of such disordered systems were obtained via a technique known as \textit{strong-disorder renormalization group} (SDRG) \cite{MDH79prl,MDH80prb} that involves iterative real-space decimations of high-energy degrees of freedom. This approach allows for the construction of tensor networks that capture the correlation structure, thus providing a first geometric description of the discrete boundary theory \cite{Basteiro:2022zur,Basteiro:2022xvu}. 
Nonetheless, the ground state in \cite{Basteiro:2022zur,Basteiro:2022xvu} is factorised into maximally entangled two-site states. Due to the flatness of entanglement spectrum, these two-site states do not enjoy a holographic description in a macroscopic wormhole geometry.
Thus, the quest for a spacetime dual within discrete holography is ongoing. 

To make progress in this direction, here we introduce Sachdev-Ye-Kitaev (SYK) \cite{SachdevYePRL1993,Kitaev:2015a1,Kitaev:2015a2} models into the framework of discrete holography: we consider
chains of SYK clusters with disordered nearest-neighbour hopping parameters. We study their ground-state correlations and find that the ground state is factorised in two-site states that are not maximally entangled. It was shown in \cite{Maldacena:2018lmt} that two coupled SYK clusters are holographically dual to wormhole. For our extended system of a chain of SYK clusters, this wormhole description arises only for two-site states involving nearest-neighbour sites. This leads to a sequence of wormholes along the chain. 

The SYK model is a quantum mechanics model describing the random $q$-body interactions of $N$ Majorana fermions \cite{SachdevYePRL1993,Kitaev:2015a1,Kitaev:2015a2}. In the large $N$ limit, the model is governed by exactly solvable Schwinger-Dyson (SD) equations. This system showcases a finite zero-temperature entropy due to the dense low-energy spectrum. Moreover, in the strong coupling regime, this system exhibits emergent conformal symmetry and a linear-in-temperature entropy described by the Schwarzian action \cite{Almheiri:2014cka,Maldacena:2016upp,Engelsoy:2016an,Kitaev:2017awl}. 
In parallel, the 2D dilaton theory denoted as Jackiw-Teitelboim (JT) gravity \cite{JACKIW1985343,TEITELBOIM198341} on nearly-AdS$_2$ spaces also exhibits these features \cite{Maldacena:2016upp} which holographically matches the low energy limit of the $(0+1)$-dimensional SYK model. Furthermore, the triple scaling limit of the SYK model corresponds to the quantum version of JT gravity \cite{Stanford:2017thb}.

In parallel, SYK lattices were shown to exhibit novel properties of a non-Fermi liquid, such as incoherent metallic behavior \cite{Gu:2016local,Gu:2017ohj}, linear resistivity \cite{Song:2017pfw,Chowdhury:2018sho,Patel:2017mjv,Patel:2019qce,Cha:2019zul}, many-body localization \cite{Jian:2017unn}, and low shear viscosity \cite{Ge:2018lzo}. Despite that some of these properties were explored in semi-holographic setting \cite{Jian:2017tzg,Doucot:2020fvy}, a precise dual gravitational description of generalized SYK models on extended lattices is still unclear. An exception to this is provided by the aforementioned two coupled SYK models that possess a clear holographic description at low energy as an eternal traversable wormhole (ETW) \cite{Maldacena:2018lmt}. In particular, the ETW ground state is nearly the thermofield-double (TFD) state constructed from two entangled SYK dots. From a dynamical point of view, a Majorana fermion can be transmitted from one site to the other due to hopping terms. The transmission is accelerated with respect to the free case by the local interactions with other fermions in each dot. The generalisation to four coupled SYK clusters was studied in \cite{Numasawa:2020sty}. It was found that the ground state where the four sites are pairwise connected by an ETW is thermodynamically dominant with respect to the four-throat wormhole solution. In this work, we show that such a factorisation into wormhole-like states arises for an infinite SYK chain with certain types of bond-disorder. 

Historically, the factorisation of ground states into two-site states is known to generally appear in disordered spin chains. In particular, these states are obtained from two strongly coupled sites which form a spin singlet. The aforementioned SDRG technique, based on perturbation theory in the Hamiltonian formalism, has proven extremely useful for obtaining insights into these factorised ground states, particularly regarding the distribution of singlet states. For models with randomly disordered bonds, the SDRG may be applied to a wide class of disorder distributions and has been successfully implemented to analyse the correlation functions and entanglement properties of random spin chains \cite{Fisher94SDRG,Igl_i_2018}. In addition, it was also used for the analysis of aperiodic spin chains \cite{Luck1993,HermissonGrimm97,vieira05,Vieira:2005PRB,Hermisson00,Juh_sz_2007,Iglói_2007}, that are deterministically disordered models constructed from substitution rules \cite{Baake2013,FlickerBoyleDickens}. A peculiarity of aperiodic disorder, due to its deterministic nature, is that it does not have a description in terms of a probability distribution, in contrast to random chains.

As is evident from the above-mentioned (non-exhaustive) literature, both large-$N$ methods for the SYK model and the SDRG have individually enjoyed great success. However, their mutual compatibility has not yet been investigated. In this work, we address this open question and provide a technical framework that extends the SDRG approach beyond the Hamiltonian formalism and unifies the notions of SDRG with the effective description of interacting systems in the large $N$ limit. We expect our results to have applications for studying further generalisations of the SYK model and their potential holographic duals.
We apply SDRG techniques to the SD equations of the bi-local fields in the large-$N$ model. In particular, we study the ground state of an SYK chain with spatially inhomogeneous hopping parameters and $q$-body on-site random interactions. We consider the general case where these on-site random interactions are spatially correlated within a distance given by a typical correlation length $\xi$. Based on the idea of SDRG, we first find the ground state of a block of sites with the strongest hopping interactions. According to perturbation theory with respect to the weaker hopping, this induces effective hopping interactions between the adjacent sites of this block of sites. 
To achieve our goal, we perturbatively analyse a four-site open chain with alternating hopping parameters in the sequence of weak-strong-weak. The solution is explicitly obtained in the limit $N\to\infty$ taken before $q\to\infty$. The perturbation theory in the weak hopping parameter naturally induces projections of the SD equations. We model the decimation of the two sites coupled by the strongest hopping parameter via their SD equations. We then derive an effective hopping between the remaining two sites and subsequently solve their projected SD equations.
Using these results, we reproduce the ground state on the central sites, which allows for a gravitational interpretation of it in terms of the ETW \cite{Maldacena:2018lmt}. 
Furthermore, after decimating the central sites, we discover an effective hopping between the left- and rightmost sites, induced by the fermionic correlations supported by the ETW between the central sites. We thus refer to this hopping as wormhole-induced effective coupling. Surprisingly, after this decimation, the effective hopping parameter is suppressed by the SYK interactions.
In contrast with what happens for the central sites, the induced correlations between the left- and rightmost sites do not admit such a wormhole description in the large $q$ limit. This is due to the fact that the corresponding SYK contributions to the self-energy vanish in the perturbative regime of validity of the SDRG.
A physical consequence of the presence of the ETW description is that the fermionic correlation between the central sites can be holographically interpreted as matter crossing through the wormhole. Thus, the propagation of fermions between the central sites is accelerated by the SYK interactions. Finally, the propagation of fermions between the left- and rightmost sites varies over time: at short times, the propagation is mediated by the correlation between the central sites and is therefore accelerated by the SYK interaction; at long times, the time it takes for a fermion to reach the opposite site scales inversely with the effective hopping parameter, resulting in a delay caused by the presence of SYK interactions.
We highlight that the factorization of the ground state in the large-$N$ limit is related to the emergence of different types of von Neumann operator algebras describing large-$N$ interacting Majorana fermions \cite{Basteiro:2024cuh}.

The decimation of the four-site inhomogeneous SYK chain mentioned above is the fundamental step that implements the SDRG  procedure. We show that the renormalization of the couplings is consistent with the original perturbative assumption, and, therefore, the decimation step can be iterated to realize the RG flow on infinite chains. To showcase the applicability of our techniques, we consider the aforementioned infinite SYK chain with random on-site interactions and either aperiodically or randomly distributed hopping parameters.
We show that the SDRG inherently generates the factorisation of the ground state into two-site states, a fraction of them being dual to ETWs. Considering these SYK chains as boundary theories in view of discrete holography, our results have a near-boundary spacetime description in terms of a sequence of wormholes, which is a new step towards a \textit{bona fide} duality.

As a by-product of our investigations, by considering the on-site random SYK interactions as a deformation of the free aperiodic chain,  
we find a line of fixed points for the SDRG flow of the system, which are parametrized by the specific values of the Hamiltonian's parameters. 
As for the chain with random bond-disorder,
we argue the presence of an attractive distribution of the hopping parameters, indicating that the SDRG method becomes asymptotically exact. 
This attractive distribution is the same as in the case without SYK interactions, signaling a similar low-energy physics between the free model and the chain in the presence of on-site random interactions.
Remarkably, the results for both aperiodic and random hopping distributions do not depend on the correlation length $\xi$ between random interactions at different sites.

The paper is structured as follows. In Sec.\,\ref{sec:setup}, we introduce the model, the relevant degrees of freedom in the large $N$ limit, and the SD equations.
In Sec.\,\ref{sec:foursite_SYK}, we perturbatively solve the four-site chain in the limit $q\to\infty$ taken after $N\to\infty$, and we interpret the solution as a real-space RG decimation. The decimation is then applied to infinite chains in Sec.\,\ref{sec:wormholenetwork}, with a specific focus on chains with aperiodically or randomly distributed hopping parameters. Finally, in Sec.\,\ref{sec:conclusions}, we discuss the results of this work, and we provide future related research avenues. Technical details and reviews of known results are reported in the appendices.

\section{Inhomogeneous SYK Chains}
\label{sec:setup}

\subsection{The Model}
\label{subsec:themodel}

\begin{figure}
    \centering
    \includegraphics[width=\linewidth]{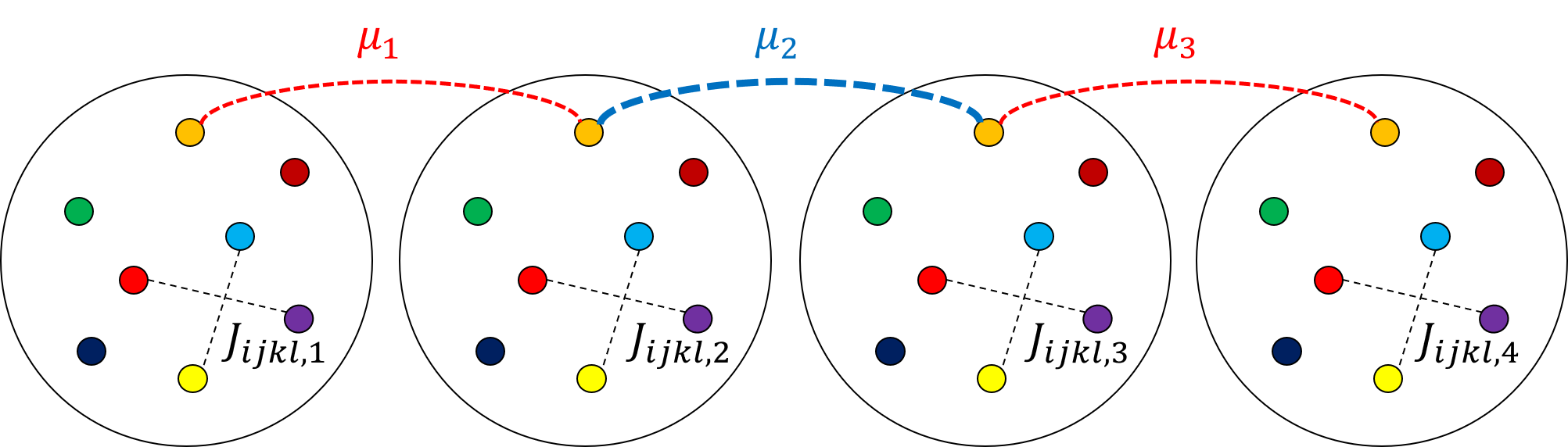}
    \caption{A SYK chain with $L=4,\,N=8,\,q=4$ and $\mu_1,\mu_3\ll\mu_2$. The Majorana fermions with different indices $j$ are represented by circles in different colours. The interactions are represented by dashed curves. In Sec.~\ref{sec:foursite_SYK}, we solve the SYK chain in this form with $N\gg q^2\gg 1$.}
    \label{fig:SYKChain}
\end{figure}

We begin by introducing the system under consideration for this work. The model consists of a chain of $L$ sites, where $N$ Majorana fermions $\psi^j_x$, $x=1,\dots,L$ with color labels $j=1,\dots,N$ are defined at each site. These Majoranas obey anti-commutation relations $\ke{ \psi^j_x, \psi^k_y}=\delta_{jk}\delta_{xy}$. With these fundamental degrees of freedom, we construct an SYK chain Hamiltonian with on-site $q$-body interaction with random couplings and nearest-neighbor $2$-body interactions (hopping) with inhomogeneous coupling, which reads
\begin{align}\begin{split}\label{eq:Hamiltonian}
    H=&\sum_{x=1}^L \Big( \mathrm{i}^{q/2}\sum_{1\leq j_1<j_2<\cdots<j_q\leq N} s_x J_{j_1j_2\cdots j_q,x}\psi_x^{j_1}\psi_x^{j_2}\cdots\psi_x^{j_q} +\mathrm{i}\sum_{1\leq j\leq N}  \mu_x \psi_x^{j}\psi_{x+1}^{j}\Big)\,.
\end{split}\end{align}
Here, $q$ is even and denotes the number of Majoranas in the all-to-all $q$-body SYK interaction term, and $s_x=(-1)^{(x-1)q/2}$. The model is pictorially represented in Fig.~\ref{fig:SYKChain}. This local interaction is controlled by the coupling parameter $J_{j_1j_2\cdots j_q,x}$ which is randomly drawn from a Gaussian distribution with zero mean value and correlation
\begin{align}
    \avg{J_{j_1j_2\cdots j_q,x}J_{j_1j_2\cdots j_q,y}}
    =\frac{(q-1)!}{N^{q-1}}J^2 e^{-\abs{x-y}^2/\xi^2}
    =\frac{(q-1)!2^{q-1}}{qN^{q-1}}\J^2e^{-\abs{x-y}^2/\xi^2},
    \label{eq:Average_SYK_coupling}
\end{align}
where $\xi$ is the dimensionless correlation length.
The hopping parameters $\{\mu_x\}$ are spatially inhomogeneous in general, and we will focus on two specific examples of such spatial distributions in the sections below. For finite $L$, we consider the chain to have open boundary conditions. However, we will ultimately be interested in the thermodynamic limit where $L\to\infty$.\\
A particular instance of the model \eqref{eq:Hamiltonian} is the case when $\J=0$, i.e. when the SYK interactions are turned off. Then, the hopping part $\mathrm{i}\sum_{1\leq j\leq N}  \mu_x \psi_x^{j}\psi_{x+1}^{j}$ of the Hamiltonian factorises into the individual colors and describes a collection of $N$ free Majorana fermions. Each of these can be mapped via a Jordan-Wigner transformation to an XX quantum spin chain. More precisely, the Hamiltonian \eqref{eq:Hamiltonian} for $\J=0$ is equivalent to $N/2$ decoupled XX chains
\begin{equation}
    H_{XX}=-\frac{1}{2}\sum_{k=1}^{N/2}\sum_{x} \mu_x\left(X^{(k)}_x X^{(k)}_{x+1}+Y^{(k)}_x Y^{(k)}_{x+1}\right) \,,
    \label{eq:genericNandp1_spinDOFform}
\end{equation}
where $X,Y$ denote the Pauli matrices and the index $k$ labels each of the $N/2$ copies of the chain. Note that the hopping parameters are still spatially distributed as they were on the original Hamiltonian, and thus \eqref{eq:genericNandp1_spinDOFform} describes $N/2$ copies of a disordered XX chain. Many exact results are known for this system, and we will often return to it throughout this manuscript as a consistency check for our calculations for $\J\neq0$.\\

In this work, we consider the model \eqref{eq:Hamiltonian} under a specific hierarchy of limits $N\gg (\beta \J,\beta \mu_x )\gg1 $. In other words, we consider the large $N$ limit first, and only afterwards, we take the low temperature limit.
A key quantity in our analysis will be the $L\times L$ matrix of two-point correlation functions (or Green's functions) in Euclidean time $\tau$ denoted by $G_{xy}(\tau)$. This matrix is defined as
\begin{equation}
    G_{xy}(\tau_{12})=\frac{1}{N}\sum_{j=1}^N\frac{1}{Z}\Tr[e^{-\beta H}T\psi^j_x(\tau_{1})\psi^j_y(\tau_2)]=\frac{1}{N}\sum_{j=1}^N\frac{1}{Z}\Tr[e^{-\beta H}T\psi^j_x(\tau_{12})\psi^j_y(0)]
    \label{eq:Definition_G}\,,
\end{equation}
where $Z=\Tr{e^{-\beta H}}$, $\psi_x^j(\tau)=e^{\tau H}\psi_x^je^{-\tau H}$, $\tau_{ij}=\tau_i-\tau_j$, and the operator $T$ denotes time-ordering $T\psi_x^j(\tau_1)\psi_y^k(\tau_2)=\Theta(\tau_{12})\psi_x^j(\tau_1)\psi_y^k(\tau_2)-\Theta(-\tau_{12})\psi_y^k(\tau_2)\psi_x^j(\tau_1)$.
Based on the properties of the time-ordering and the resulting anti-periodic boundary conditions, the Green's function \eqref{eq:Definition_G} enjoys the following symmetries
\begin{align}
\label{eq:symmetries_G}
    G_{xy}(\tau)
    =G_{yx}(\tau)^*=G_{xy}(\beta-\tau)^*=-G_{xy}(-\tau)^*\,,
\end{align}
from which we can derive further relations. If $G_{xy}(\tau)\in \mathds R$, we have $G_{xy}(\tau)=-G_{xy}(-\tau)$, while if $G_{xy}(\tau)\in \mathrm{i}\mathds{R}$, we have $G_{xy}(\tau)=G_{xy}(-\tau)$. In addition, $G_{xx}(\tau)\in \mathds R$. Due of the anti-commutation relation $\ke{\psi_x^j,\psi_y^j}=\delta_{xy}$, for $x=y$, $G_{xx}(\tau)$ is discontinuous at $\tau=0$, namely $G_{xx}(0^+)\neq G_{xx}(0^-)$, while, for $x\neq y$, $G_{xy}(\tau)$ is continuous at $\tau=0$, i.e. $G_{xy}(0^+)=G_{xy}(0^-)$. This implies $G_{xy}(0)\in \mathrm{i}\mathds R$.  Further relations rely on the form of hopping terms. For the chains with nearest-neighbor interactions considered in this paper, we will see that, for even $\abs{x-y}$, $G_{xy}(\tau)\in \mathds R$, while for odd $\abs{x-y}$, $G_{xy}(\tau)\in \mathrm{i}\mathds R$. When $\abs{x-y}=1$, $G_{xy}'(\tau)$ is discontinuous at $\tau=0$, while, when $\abs{x-y}\geq2$, $G_{xy}'(\tau)$ is continuous at $\tau=0$. Finally, for even $\abs{x-y}\geq2$, we have that $G_{xy}(0)=0$.

\subsection{Schwinger-Dyson equations}
\label{subsec:SDeqLargeN}

The quantum model introduced above is not analytically tractable for generic values of $N$, but it does admit an effective classical description in the large $N$ limit. More precisely, the partition function at finite $N$ can be written as the path integral
\begin{align}
    Z=\Tr[e^{-\beta H}]=\int D\psi e^{-S},\quad
    S=\int_0^\beta d\tau\kc{ \sum_{xj}\frac12\psi_x^j\partial_\tau \psi_x^j+H}\,,
    \label{eq:path_integral}
\end{align}
where $H$ is to be though of as a function of the real Grassmann variable $\psi_x^j$ as given by \eqref{eq:Hamiltonian} and $\psi_x^j(\tau)$ obeys anti-periodic boundary conditions, i.e. $\psi_x^j(\tau+\beta)=-\psi_x^j(\tau)$. The ground state of this model can be accessed by further considering the $\beta\to\infty$ limit. Following the treatment in \cite{SachdevYePRL1993,Maldacena:2016remarks,Sarosi:2017ykf}, the random disorder can be integrated out due to the Gaussian form of the coupling distribution for $J_{j_1\dots j_q}$ with variance given by \eqref{eq:Average_SYK_coupling}. Moreover, we can introduce an auxiliary field, the self-energy $\Sigma_{xy}(\tau)$, which acts as a Lagrange multiplier and allows us to write down an effective action for the model, which reads
\begin{align}
    -S/N=&\log \text{PF}(\partial_\tau \delta_{xy}-\Sigma_{xy})\nonumber\\
    &-\frac12\sum_{xy}\int_0^{\beta} d\tau_{1}d\tau_{2}\kd{G_{xy}(\tau_{12})\Sigma_{xy}(\tau_{12})-s_{xy}\frac{\J^2}{2q^2}\kc{2G_{xy}(\tau_{12})}^q}\nonumber\\
    &
    +\frac{1}{2}  \sum_x \mu_x\int_0^{\beta} d\tau (-2\mathrm{i}G_{x,x+1}(0))\,,
    \label{eq:eff_action2}
\end{align}
where we have defined $s_{xy}
=\mathrm{i}^{(x+y)q}e^{-\abs{x-y}^2/\xi^2}$. In the large $N$ limit, we assume that only replica diagonal diagrams dominate when taking the disorder average in the Green's function \cite{Gu:2016local,Maldacena:2018lmt}. The large $N$ behavior of the two-point function is governed by the saddle point of the path integral \eqref{eq:path_integral}, and the dynamics of the master fields $G$ and $\Sigma$ are thus given by the equations of motion.  From the effective action \eqref{eq:eff_action2}, we can derive these equations of motion by imposing that the variations $\frac{\delta S}{\delta G},\,\frac{\delta S}{\delta \Sigma}$ vanish. This yields the so-called \textit{Schwinger-Dyson} (SD) equations
\begin{subequations}\label{eq:SD_eqs_full_form}
\begin{align}
    &\partial_{\tau_1}G_{xy}(\tau_{12})-\sum_z\int_0^\beta d\tau_3 \Sigma_{xz}(\tau_{13})G_{zy}(\tau_{32})=\delta_{xy}\delta_\beta(\tau_{12})\,,
\label{eq:SD_eqs_full_form_G}\\
    &\Sigma_{xy}(\tau_{12})
    =s_{xy}\frac{\J^2}{q}(2G_{xy}(\tau_{12}))^{q-1}-\mathrm{i} \mu_{xy}\delta_\beta(\tau_{12})
    \equiv\Sigma_{\J\,xy}(\tau_{12})+\Sigma_{\mu\, xy}(\tau_{12})\,,
\label{eq:SD_eqs_full_form_Sigma}
\end{align}
\end{subequations}
where we have introduced the hopping matrix $\mu_{xy}=\mu_x\delta_{x+1,y}-\mu_y\delta_{x,y+1}$ and the anti-periodic delta function $\delta_\beta(\tau)=\sum_{n\in\mathds Z} (-1)^n\delta(\tau+n\beta)$  to ensure the validity of the equations beyond the time domain $\tau\in[0,\beta)$.
Equivalently, the SD equations can be derived from an expansion in Feynman diagrams as 
\begin{subequations}\label{eq:FeynmanSD}
\begin{align}
\twoptf{a [particle=$x$] -- [very thick] b [particle=$y$] } 
=&~ \twoptf{a [particle=$x$] -- b [particle=$y$]} 
+  \twoptf{a [particle=$x$]-- [edge label=~~~$z$] c [blob] -- [very thick, edge label=$z'$~~~] b [particle=$y$]} \\
x~ \twoptf{a [blob]} ~y
=&~ \ x~ \twoptf{a --[very thick,half left,looseness=1] b --[very thick,half left,looseness=1] a --[very thick] b --[scalar,half left,looseness=2] a} ~y \ 
+	\ x~ \twoptf{a [dot, label=$\mu$]} ~y 
\end{align} 
\end{subequations}
where the thin line represents the bare free propagator, while the thick line denotes the corresponding dressed propagator. Moreover, the circle indicates the one-particle-irreducible contributions, encoding the self-energy, the dashed line denotes the disorder average, and the dot represents the hopping. For clarity, the diagrams are shown for $q=4$, but their generalization to arbitrary values of $q$ is straightforward. The self-energy defined in \eqref{eq:SD_eqs_full_form_Sigma} enjoys the same symmetries of $G_{xy}(\tau)$ in \eqref{eq:symmetries_G}, namely
$\Sigma_{xy}(\tau)=\Sigma_{yx}(\tau)^*=\Sigma_{xy}(\beta-\tau)^*=-\Sigma_{xy}(-\tau)^*$.
Furthermore, to avoid clutter in later expressions, it is useful to define the integral matrix product $(A*B)_{xy}(\tau_{12})=\sum_z\int d\tau_3 A_{xz}(\tau_{13})B_{zy}(\tau_{32})$ such as to write the second term in \eqref{eq:SD_eqs_full_form_G} as $(\Sigma*G)_{xy}(\tau_{12})$. In matrix notation, \eqref{eq:SD_eqs_full_form_G} can be concisely written for $0\leq\tau<\beta$ as
\begin{equation}\label{eq:matrixeq}
    \partial G-\Sigma*G=I\,,
\end{equation}
where identity matrix $I$ is the matrix notation of $\delta(\tau)\delta_{xy}$ and $(\partial G)_{xy}(\tau_{12})=\sum_{z}\int d\tau_3 \delta'(\tau_{13})\delta_{xz}G_{zy}(\tau_{32})$.
We will often use this notation in the rest of the manuscript. Due to the symmetries of $G_{xy}(\tau)$ and $\Sigma_{xy}(\tau)$, the matrices $G$ and $\Sigma$, as well as the matrix equation \eqref{eq:matrixeq}, are invariant under the combination of spatial transposition $x\leftrightarrow y$ and complex conjugation.

At finite temperature $\beta<\infty$, the SD equations need to be complemented by suitable boundary conditions on the Green's function, which read
\begin{align}\label{eq:BoundaryCondition}
    G_{xx}(0)=G_{xx}(\beta)=1/2,\quad G_{x\neq y}(\beta)=-G_{x\neq y}(0)=G_{y\neq x}(0)\,.
\end{align}
When accessing the ground state at $\beta\to\infty$, these boundary conditions, in the absence of degeneracy, reduce to
\begin{align}
\label{eq:generalBC}
    G_{xx}(0)=1/2, \quad \lim_{\tau\to\infty} G_{xy}(\tau)=0\,.
\end{align}
Instead, if there is degeneracy present in the system, the right equation in \eqref{eq:generalBC} gets an additional constant contribution.

\subsection{Liouville equation in the large \texorpdfstring{$q$}{q} limit}
\label{subsec:Liouvillelargeq}

Although the SD equations \eqref{eq:SD_eqs_full_form} can be solved numerically for a reasonable choice of the parameters, they also admit analytical solutions in the limit where the parameter $q$ is taken to be large \cite{Maldacena:2016remarks,Maldacena:2018lmt}. The following analysis holds for general spatial distributions of the hopping parameters, although their explicit solutions may be rather complicated. We can consider the SD equations in the spirit of a $1/q$ expansion for large $q$, while keeping the parameters $\J,\hat{\mu}_x=q \mu_x, \tau$ and the hopping matrix $\hat{\mu}_{xy}=q\mu_{xy}$ finite. The above rescaled parameters are considered as $\order{1}$ in the large $q$ limit.
Therefore, we can see from \eqref{eq:SD_eqs_full_form_Sigma} that the self-energy obeys $\Sigma=0+\mathcal{O}\left(q^{-1}\right)$ and therefore vanishes at leading order in the large $q$ expansion.
Correspondingly, the SD equations \eqref{eq:SD_eqs_full_form_G} reduce to the free case
\begin{align}    
\label{eq:free_SD}
\partial_{\tau}\widehat{G}_{xy}(\tau)=\delta_{xy}\delta(\tau)\,.
\end{align}
Here, we use the superscript in $\widehat{G}_{xy}$ to denote the solution in the non-interacting Majorana case. The general solution to \eqref{eq:free_SD} reads
\begin{align}
\label{eq:defGcal}
    \widehat{G}_{xx}=\frac12 \sgn(\tau)\G_{xx},\quad
    \widehat{G}_{xy}(\tau)=\frac 12\G_{xy},\quad y\neq x,\quad \G_{xy}\in \mathds C,\ \abs{\G_{xy}}\leq 1\,,
\end{align}
where the undetermined matrix $\G_{xy}$ is Hermitian, namely $\G_{xy}=\G_{yx}^*$, with $\G_{xx}=1$, as follows from the symmetries of $G$ given in and below \eqref{eq:symmetries_G}. All values of $\G_{xy}$ in \eqref{eq:defGcal} solve the equation \eqref{eq:free_SD}. We notice that the periodic boundary conditions \eqref{eq:BoundaryCondition} yield $\G_{xy}=\delta_{xy}$. The reason why $G_{x\neq y}(0)$ are all vanishing is that the Hamiltonian itself vanishes at leading order in $1/q$ and therefore all the states become degenerate with an equal contribution to the correlations. We expect that the $1/q$ corrections will lift this degeneracy and lead to finite $G_{x\neq y}(0)$.

We now consider $1/q$ corrections to this behavior. For this, we make an ansatz for the Green's function of the form
\begin{align}\label{eq:qExpansion}
    G_{xy}(\tau)
    =\widehat{G}_{xy}(\tau)\left(1+\frac1{q}g_{xy}(\tau)+\cdots\right)
    \approx \widehat{G}_{xy}(\tau)e^{\frac{g_{xy}(\tau)}{q}
    }\,.
\end{align}
As mentioned above, our goal is to resolve the constant matrix $\G_{xy}$ by taking the $1/q$ corrections into account. Note that $g_{xy}^*(\tau)=g_{yx}(\tau)$ since $G_{xy}(\tau)$ is Hermitian. Moreover, from the condition $G_{xx}(0)=\frac{1}{2}$ we also have that $g_{xx}(0)=0$. Inserting this ansatz into (\ref{eq:SD_eqs_full_form_G}), in Appendix \ref{subapp:LiouvilleEq} we derive the equation that $g_{xy}(\tau)$ must satisfy. This equation, known as the \textit{Liouville equation}, reads
\begin{align}
    \label{eq:evolution_g_gen}&g_{xy}''(\tau) -2s_{xy}\J^2\G_{xy}^{q-2}e^{g_{xy}(\tau)}=0,\quad 0<\tau<\beta\,,
\end{align}
complemented with the boundary conditions
\begin{equation}
\label{eq:bc_gen g_R1}
    g_{xx}(0)=0\,,
\end{equation}
and
\begin{align}
\label{eq:bc_gen g_R23}
    &\G_{xy}g_{xy}'(0)+\sum_z \mathrm{i}\hat{\mu}_{xz}\G_{zy}=0,\quad x\neq y\,.
\end{align}
In the large $q$ limit, the factor $\G_{xy}^{q-2}$ in \eqref{eq:evolution_g_gen} takes the value $1$ or $0$ for $\abs{\G_{xy}}=1$ or $\abs{\G_{xy}}=e^{-\order{q^0}}<1$ respectively. The choice must be checked a posteriori, as we do for a chain of four sites in Sec.\,\ref{subsec:foursitechain} through an analytical self-consistency analysis and in Sec.\,\ref{subsec:numeric}  through a numerical computation.

Notice that the last equality in the ansatz \eqref{eq:qExpansion} holds 
until $g_{xy}(\tau)\sim\order{q}$ before some specific time scale. For now let us simply stress that for time scales where $g_{xy}(\tau)$ starts being of $\order{q}$, the solution to the SD has to be obtained via a different method. This resulting solution at late  times should then be matched with the solutions obtained from the ansatz \eqref{eq:qExpansion} in an intermediate overlapping regime. Within our analysis in the next section, we provide an example of such a computation, including the matching of the solutions.

\section{Real-space Decimation on SYK Chains}
\label{sec:foursite_SYK}

The central paradigm of this work is that of solving the SD equations \eqref{eq:SD_eqs_full_form} and their corresponding Liouville equation \eqref{eq:evolution_g_gen} via a general perturbative approach for inhomogeneous quantum chains. In particular, we will work within the framework of the so-called \textit{strong-disorder renormalization group} (SDRG) \cite{MDH79prl,MDH80prb}, which is a real-space renormalization procedure to access the low-energy properties of spatially inhomogeneous systems. This method works under the assumption that the strength of this spatial disorder can be considered large. More precisely, for our Hamiltonian \eqref{eq:Hamiltonian}, this approach requires to identify hopping parameters much larger than the other ones in the chain, decimate the degrees of freedom coupled by these strong hoppings, and finally treat the other weaker hoppings as perturbative parameters. In this way, the SDRG provides an analytical way of accessing the low-energy properties of strongly disordered systems. In particular, we are interested in accessing the ground state of the models under consideration. 
Although originally developed in terms of perturbation theory at the level of the Hamiltonian, in this work we present a novel take on SDRG by applying it directly to the SD equations for a chain of coupled SYK models. For the convenience of the reader, we review the Hamiltonian approach to SDRG in Appendix~\ref{apx:SDRG}. In this paper, we consider specific sequences of hopping parameters $\kc{\mu_1,\mu_2,\cdots,\mu_x,\cdots}$ such that the SDRG procedure on the corresponding system only involves the decimation of two sites connected by strong hoppings. For our analysis, we consider the following limits of the model (\ref{eq:Hamiltonian}), crucially taken in the specified order: 
\begin{senum}
\item \label{listitem:largeN_limit} the large $N$ limit $N\to\infty$\,;
    \item \label{listitem:zeroT_limit} the zero temperature limit $\beta \J,\,\beta \mu_x\to\infty$\,;
    \item \label{listitem:strong_disorder_limit} the strong disorder limit $\mu_\text{weak}/\mu_\text{strong}\to0$\,.  
\end{senum}
In the present section, we will first explain in general terms how a decimation step is applied to the SD equations of a four-site SYK chain using projection operators. We will show that this construction relies on the properties of the local ground state of two coupled SYK models, and we review how to solve this system following \cite{Maldacena:2018lmt}. This two-site system is then decimated, and we show how this decimation induces an effective self-energy in the four-site system, in the spirit of the SDRG.

\subsection{Decimation in the SD equations}
\label{subsec:setup_foursitechain}

Consider an open chain of Hamiltonian \eqref{eq:Hamiltonian} with $L=4$, $q \mod 4 =0$, and the hopping parameters such that
$\mu_1=\mu_a(1+\Delta)$, $\mu_2=\mu_b$ and $\mu_3=\mu_a(1-\Delta)$, where $-1<\Delta < 1$ such that the three $\mu_x$s are positive. This configuration is depicted in Fig.~\ref{fig:SYKChain}. Let us mention that a similar system of four coupled SYK dots has been previously considered in \cite{Numasawa:2020sty}. However, the crucial difference is that the author of  \cite{Numasawa:2020sty} considered a closed four-site chain, while we take it to have open boundary conditions, in view of applying our results to infinite chains. For later convenience, we introduce the hopping ratio $r\equiv \mu_a/\mu_b$. We will consider this four-site chain as its own standalone system for now, but in Sec.\,\ref{sec:wormholenetwork} it will be seen as a subsystem of longer chains.
Notice that if $\Delta=0$ the chain acquires reflection symmetry with respect to its center, and this constrains the entries of $G$ to satisfy
\begin{equation}
\label{eq:constraintonabab}
    G_{11}(\tau)=G_{44}(\tau)\,,\qquad
    G_{22}(\tau)=G_{33}(\tau)\,,\qquad
     G_{12}(\tau)=G_{34}(\tau)\,,\qquad
     G_{13}(\tau)=G_{24}(\tau)\,,
\end{equation}
where the same holds for the corresponding transposed entries. For the sake of generality, in this subsection, we will consider generically $\Delta\neq 0$ and therefore (\ref{eq:constraintonabab}) does not hold unless specified otherwise.

In this section, we adopt a perturbative approach, where we consider the hopping parameter $\mu_a$ as the smallest in the four-site chain, namely $\mu_a\ll\mu_b$, while the anisotropy $\Delta$ is assumed to be of $\order{1}$. In particular, this means that the hopping probability between the two central sites is much stronger than between central sites and left- and rightmost sites. In this regime, we can consider a perturbative expansion of the Green's function and the self-energy for small values of $\mu_a$.
Formally, they read
\begin{equation}
\label{eq:expansion_G_SDRG}
G=G^{(0)}+\mu_a G^{(1)}+\mu_a^2G^{(2)}+\order{\mu_a^3}\,,
\end{equation}
and 
\begin{equation}
\label{eq:expansion_Sigma_SDRG}
\Sigma=\Sigma^{(0)}+\mu_a\Sigma^{(1)}+\mu_a^2\Sigma^{(2)}+\order{\mu_a^3}\,.
\end{equation}
To identify the terms on the right-hand side of (\ref{eq:expansion_Sigma_SDRG}), we plug the expansion (\ref{eq:expansion_G_SDRG}) into (\ref{eq:SD_eqs_full_form_Sigma}) and we obtain
\begin{align}
   \Sigma_{xy} =&~\frac{\J^2}{q} s_{xy}(2G_{xy}^{(0)})^{q-1}-\mathrm{i}\mu_b M_{xy}^{(0)} \delta(\tau) \nn\\
    &~+\mu_a\kd{\J^22\kc{1-\frac{1}{q}}s_{xy}(2G_{xy}^{(0)})^{q-2}G_{xy}^{(1)}-\mathrm{i}M_{xy}^{(1)}\delta(\tau)}\\
    &~+\mu_a^2\kd{\J^22\kc{1-\frac{1}{q}}s_{xy}(2G_{xy}^{(0)})^{q-3}\kc{2G_{xy}^{(0)}G_{xy}^{(2)}+(q-2)(G_{xy}^{(1)})^2}}
    +\order{\mu_a^3}\,, \nn
\end{align}
from which we can identify $\Sigma^{(0)}$, $\Sigma^{(1)}$ and $\Sigma^{(2)}$ appearing in (\ref{eq:expansion_Sigma_SDRG})
\begin{subequations}
\begin{align}
 \label{eq:sigma0_def}
 \Sigma^{(0)}_{xy} =&~ \frac{\J^2}{q}s_{xy}(2G_{xy}^{(0)})^{q-1}-\mathrm{i}\mu_b M_{xy}^{(0)} \delta(\tau)\equiv \Sigma^{(0)}_{\J\,xy}+\Sigma^{(0)}_{\mu\,xy}  \,,
 \\
 \label{eq:Sigma1_def}
   \Sigma_{xy}^{(1)} =&~ \J^22\kc{1-\frac{1}{q}}s_{xy}(2G_{xy}^{(0)})^{q-2}G_{xy}^{(1)}-\mathrm{i}M_{xy}^{(1)}\delta(\tau)\equiv \Sigma^{(1)}_{\J\,xy}+\Sigma^{(1)}_{\mu\,xy} \,,
   \\
    \label{eq:Sigma2_def}
   \Sigma_{xy}^{(2)} =&~
  \J^22\kc{1-\frac{1}{q}}s_{xy}(2G_{xy}^{(0)})^{q-3}\kc{2G_{xy}^{(0)}G_{xy}^{(2)}+(q-2)(G_{xy}^{(1)})^2}\equiv \Sigma^{(2)}_{\J\,xy}\,.
\end{align}
\end{subequations}
In the previous equations, we have split the $4\times 4$ hopping matrix into the two dimensionless contributions, each multiplied by the corresponding energy scale, namely
\begin{equation}
 \label{eq:hoppingmatrix}
     \mu_{xy}= \mu_b M^{(0)}_{xy}+\mu_a M^{(1)}_{xy}\,,
 \end{equation}
 where
\begin{equation}
\label{eq:mu01_SDRG}
   M^{(0)}= \left(
\begin{array}{cccc}
 0  & 0 & 0 & 0 \\
 0 & 0  & 1 & 0 \\
 0 & -1 & 0  & 0 \\
 0 & 0 & 0 & 0  \\
\end{array}
\right) \,,
\qquad\quad
M^{(1)}= \left(
\begin{array}{cccc}
0  & 1+\Delta & 0 & 0 \\
 -1-\Delta & 0  & 0 & 0 \\
 0 & 0 & 0  & 1-\Delta \\
 0 & 0 & -1+\Delta & 0  \\
\end{array}
\right) \,.
\end{equation}

Plugging the expansions (\ref{eq:expansion_G_SDRG}) and (\ref{eq:expansion_Sigma_SDRG}) into (\ref{eq:SD_eqs_full_form_G}) and separately identifying the powers of $\mu_a$, we obtain the SD equations order by order in the expansion in $\mu_a$
\begin{subequations}
    \begin{align}
        \mu_a^0&:\quad\partial G^{(0)}=\Sigma^{(0)}*G^{(0)}+I\,, \label{eq:TruncatedSDeqs0} \\
        \mu_a^1&:\quad\partial G^{(1)}=\Sigma^{(0)}*G^{(1)}+\Sigma^{(1)}*G^{(0)}\,,\label{eq:TruncatedSDeqs1}\\
        \mu_a^2&:\quad\partial G^{(2)}=\Sigma^{(0)}*G^{(2)}+\Sigma^{(1)}*G^{(1)}\,,\label{eq:TruncatedSDeqs2}
    \end{align}
    \label{eq:TruncatedSDeqs}
\end{subequations}
where we have introduced the product $*$ indicating the convolution in Euclidean time and the matrix product, as defined in Sec.\,\ref{subsec:SDeqLargeN}.

In the spirit of implementing a real-space RG step on the short chain, we decimate the degrees of freedom associated with higher energy scales located on the central sites linked by a {\it strong} hopping parameter. To perform this decimation, we introduce the spatial projectors $P_{14}$ and $P_{23}$,
\begin{equation}
\label{eq:projectors}
    \begin{split}
        P_{14}=\left(
\begin{array}{cccc}
1  & 0 & 0 & 0 \\
 0 & 0  & 0 & 0 \\
 0 & 0 & 0  & 0 \\
 0 & 0 & 0 & 1  \\
\end{array}
\right)\,,\quad\quad
    P_{23}=\left(
\begin{array}{cccc}
0 & 0 & 0 & 0 \\
 0 & 1  & 0 & 0 \\
 0 & 0 & 1  & 0 \\
 0 & 0 & 0 & 0  \\
\end{array}
\right)\,,
    \end{split}
\end{equation}
which project onto the subspaces associated with the left- and rightmost sites and the central site, respectively.
From (\ref{eq:projectors}) we can deduce the following completeness and orthogonality relations,
\begin{align}
&P_{14}+P_{23}=I_{4\times 4} \,, \label{eq:completeness} \\ 
&P_{14}P_{23}=P_{23}P_{14}=0 \,.
\end{align}
The decimation of the central sites of the four-site chain is realized by projecting the SD equations by using $P_{14}$. The resulting {\it projected SD equations} account for the perturbative effect of the decimation on the remaining sites, i.e. the left- and the rightmost sites. 

In the spirit of perturbation theory, we start from the SD equations at zeroth order in $\mu_a$ reported in (\ref{eq:TruncatedSDeqs0}). These equations contain only $G^{(0)}$ and $\Sigma^{(0)}$, with the latter being given in terms of the former via (\ref{eq:sigma0_def}). According to the hopping matrix $M^{(0)}$ in \eqref{eq:mu01_SDRG}, only the two central sites are coupled with each other at the zeroth order. Thus, the zeroth-order SD equations (\ref{eq:TruncatedSDeqs0}) are projected into
\begin{subequations}\label{eq:SD0PP}
\begin{align}
    &\partial P_{23}G^{(0)}P_{23}=P_{23}\Sigma^{(0)}P_{23}*P_{23}G^{(0)}P_{23}+P_{23}\Sigma_{\J}^{(0)}P_{14}*P_{14}G^{(0)}P_{23}+P_{23}I\,,\label{eq:SD0P23P23}\\
    &\partial P_{14}G^{(0)}P_{23}=P_{14}\Sigma_{\J}^{(0)}P_{23}*P_{23}G^{(0)}P_{23}+P_{14}\Sigma_{\J}^{(0)}P_{14}*P_{14}G^{(0)}P_{23}\,,\label{eq:SD0P14P23}\\
    &\partial P_{23}G^{(0)}P_{14}=P_{23}\Sigma^{(0)}P_{23}*P_{23}G^{(0)}P_{14}+P_{23}\Sigma_{\J}^{(0)}P_{14}*P_{14}G^{(0)}P_{14}\,,\label{eq:SD0P23P14}\\
    &\partial P_{14}G^{(0)}P_{14}=P_{14}\Sigma_{\J}^{(0)}P_{23}*P_{23}G^{(0)}P_{14}+P_{14}\Sigma_{\J}^{(0)}P_{14}*P_{14}G^{(0)}P_{14}+P_{14}I\,,\label{eq:SD0P14P14}
\end{align}
\end{subequations}
where we have used the vanishing projections 
\begin{equation}
\label{eq:projectionSigma0G0mu}
    P_{14}\Sigma_\mu^{(0)}=\Sigma_\mu^{(0)}P_{14}=0\,,
\end{equation}
based on (\ref{eq:mu01_SDRG}).
Although \eqref{eq:SD0PP} are coupled equations, they allow for a solution with
\begin{equation}
\label{eq:conditionG_perturbative}
    P_{14}G^{(0)}P_{23}=P_{23}G^{(0)}P_{14}=0\,,
\end{equation} 
which is motivated by the solution at $\J=0$. Due to \eqref{eq:sigma0_def}, the condition \eqref{eq:conditionG_perturbative} leads to 
\begin{equation}
\label{eq:projectionSigma0G0J}
    P_{14}\Sigma_\J^{(0)}P_{23}=P_{23}\Sigma_\J^{(0)}P_{14}=0 \,.
\end{equation}
 Thus, \eqref{eq:SD0P14P23} and \eqref{eq:SD0P23P14} are fulfilled, while   \eqref{eq:SD0P23P23} and \eqref{eq:SD0P14P14} become decoupled
\begin{subequations}\begin{align}
    \partial P_{23}G^{(0)}P_{23}=P_{23}\Sigma^{(0)}P_{23}*P_{23}G^{(0)}P_{23}+P_{23}I\,,\label{eq:projectedSD0orderP23}\\
   \partial P_{14}G^{(0)}P_{14}=P_{14}\Sigma_\J^{(0)}P_{14}*P_{14}G^{(0)}P_{14}+P_{14}I \,.\label{eq:projectedSD0orderP14}
\end{align}\end{subequations}

Furthermore, due to the absence of the hopping contribution to the self-energy in \eqref{eq:projectedSD0orderP14}, the left- and rightmost sites are further decoupled at zeroth order. Thus, one might naively expect $P_{14}G^{(0)}P_{14} = 0$. However, each SYK cluster has a finite zero-temperature entropy $S_0 \sim N$, which reflects a high density of states near the ground state \cite{Maldacena:2016remarks}. Although this is a priori distinct from actual degeneracy, the average level spacing for these low-energy states scales as $\mathcal{J} e^{-S_0}$ in the large $N$ limit. Therefore, compared to any finite $\mu_a$, these low-energy states of the left- and rightmost sites at zeroth order in $\mu_a$ have negligible spacing and can be treated as effectively degenerate. Thus, we should apply degenerate perturbation theory with respect to $\mu_a$ to the left- and rightmost sites. Due to the order of limits \ref{listitem:largeN_limit}\,-\,\ref{listitem:strong_disorder_limit} we consider, this effective degeneracy should be lifted by the hopping at higher orders in $\mu_a$. For this reason, we expect the solution for $P_{14}G^{(0)}P_{14}$ to be determined at higher orders of $\mu_a$ rather than at zeroth order.
Summarizing, all the entries of $G^{(0)}$ are non-vanishing except the ones reported in (\ref{eq:conditionG_perturbative}). Due to the symmetries discussed in Sec.\,\ref{subsec:SDeqLargeN}, we thus have six independent non-vanishing entries.

As second step of our perturbative procedure, we project the first and second-order SD equations in (\ref{eq:TruncatedSDeqs1}) and (\ref{eq:TruncatedSDeqs2}), respectively, using $P_{14}$. For the first-order equation, exploiting \eqref{eq:projectionSigma0G0mu}, \eqref{eq:projectionSigma0G0J}, and 
\begin{equation}
\label{eq:projectionSigma1}
   P_{14}\Sigma_\mu^{(1)} P_{14}=0 \,,
\end{equation}
we have
\begin{equation}
\label{eq:projectedSD1storder}
    \partial P_{14}G^{(1)}P_{14}
    =
P_{14}\Sigma_\J^{(0)}P_{14}*P_{14}G^{(1)} P_{14}
    +P_{14}\Sigma_\J^{(1)}P_{14}*P_{14}G^{(0)} P_{14}\,.
\end{equation}
Studying this equation carefully, one can show that $P_{14}G^{(1)}P_{14}=0$ is indeed a solution. To see this, using (\ref{eq:Sigma1_def}) one first notices that
\begin{equation}
\label{eq:PSigmaJ1P}
\left(P_{14}\left(\Sigma_{\J}^{(1)}\right)P_{14}\right)_{xy}=\sum_{zs}(P_{14})_{xz}\left(\Sigma_{\J}^{(1)}\right)_{zs}(P_{14})_{sy}\propto \sum_{zs}\left(G^{(0)}_{zs}\right)^{q-2}(P_{14})_{xz}G_{zs}^{(1)}(P_{14})_{sy}\,.
\end{equation}
In this way, it becomes manifest that (\ref{eq:PSigmaJ1P}) and therefore the second term in \eqref{eq:projectedSD1storder} vanish when 
\begin{equation}
\label{eq:vanishingfirstorder}
    P_{14}G^{(1)}P_{14}=0.
\end{equation} 
Thus, (\ref{eq:vanishingfirstorder}) is in fact a solution to the projected SD equations (\ref{eq:projectedSD1storder}).
Having found this solution, we can check that $P_{14}\Sigma_{\J}^{(1)}P_{23}=P_{23}\Sigma_{\J}^{(1)}P_{14}=0$ upon exploiting \eqref{eq:Sigma1_def} and \eqref{eq:conditionG_perturbative}. Combining these findings with the fact that \eqref{eq:PSigmaJ1P} vanishes, we obtain the useful relations
\begin{equation}
\label{eq:projectionsigmaJ1}
P_{14}\Sigma_\J^{(1)}=\Sigma_\J^{(1)}P_{14}=0\,.
\end{equation}
Let us remark that even though $P_{14}G^{(1)}P_{14}=0$ is a solution, one may ask if there are any other possible solutions to \eqref{eq:projectedSD1storder}. 
However, since we know that $P_{14}G^{(1)}P_{14}$ cannot be different from zero in the free case $\J=0$ (see the discussion in Appendix~\ref{apx:freecase}), and we further expect a continuity property to the solutions in the interacting case, any other possible solutions are not of interest for us.

We then project the SD equations at $\order{\mu_a}$ by sandwiching them with $P_{23}$ and $P_{14}$ and simply them by use \eqref{eq:conditionG_perturbative}\eqref{eq:vanishingfirstorder}
\begin{subequations}
\begin{align}
    \partial P_{23}G^{(1)}P_{14}
    =&~P_{23}\Sigma^{(0)}P_{23}*P_{23}G^{(1)}P_{14}
    +P_{23}\Sigma_\mu^{(1)}P_{14}*P_{14}G^{(0)}P_{14}\\
    \partial P_{14}G^{(1)}P_{23}
    =&~P_{14}\Sigma_\J^{(0)}P_{14}*P_{14}G^{(1)}P_{23}
    +P_{14}\Sigma_\mu^{(1)}P_{23}*P_{23}G^{(0)}P_{23}\,.
\end{align}
\end{subequations}
Formally, we can write the solutions as
\begin{subequations}
\begin{align}
    P_{23}G^{(1)}P_{14}
    =&~P_{23}\frac1{\partial-\Sigma^{(0)}}P_{23}*P_{23}\Sigma_\mu^{(1)}P_{14}*P_{14}G^{(0)}P_{14} \nn \\
    =&~P_{23}G^{(0)}P_{23}*P_{23}\Sigma^{(1)}_\mu P_{14}*P_{14}G^{(0)}P_{14}\,,\label{eq:p23_g1_p14}\\
    P_{14}G^{(1)}P_{23}
    =&~P_{14}\frac1{\partial-\Sigma_\J^{(0)}}P_{14}*P_{14}\Sigma_\mu^{(1)}P_{23}*P_{23}G^{(0)}P_{23}\,. \nn 
\end{align}
\end{subequations}

Let us move to the second-order equation in (\ref{eq:TruncatedSDeqs2}) and project it using $P_{14}$. Using (\ref{eq:completeness}), (\ref{eq:projectionSigma0G0J}), (\ref{eq:projectionSigma1}) and (\ref{eq:projectionsigmaJ1}), we obtain the projected second-order SD equation
\begin{equation}
\label{eq:SDproject2ord}
  \partial P_{14}G^{(2)}P_{14}
   =
P_{14}\Sigma_\J^{(0)}P_{14}*P_{14}G^{(2)}P_{14}
    +P_{14}\Sigma_\mu^{(1)}P_{23}*P_{23}G^{(1)}P_{14}
    +P_{14}\Sigma_\J^{(2)}P_{14}*P_{14}G^{(0)}P_{14}\,.
\end{equation}
Now we combine the projected SD equations at leading, first- and second-order in $\mu_a$. Adding up (\ref{eq:projectedSD0orderP14}) and (\ref{eq:SDproject2ord}) (notice that the first order \eqref{eq:vanishingfirstorder} gives a vanishing contribution), we have
\begin{eqnarray}
\label{eq:totalSDequation_v1}
    &&\partial P_{14}(G^{(0)}+\mu_a^2G^{(2)})P_{14}=
    P_{14}I+\\
    &&
    P_{14}(\Sigma_\J^{(0)}+\mu_a^2\Sigma_\J^{(2)})P_{14}*P_{14}(G^{(0)}+\mu_a^2G^{(2)})P_{14}
+\mu_a^2P_{14}\Sigma_\mu^{(1)}P_{23}*P_{23}G^{(1)}P_{14}\,,
\nonumber
\end{eqnarray}
where we have added a term of order $\mu_a^4$ which is subleading in our approximation and therefore counts as a vanishing term.
Rewriting (\ref{eq:totalSDequation_v1}) into the form $G=G^{(0)}+\mu_aG^{(1)}+\mu_a^2G^{(2)}+O(\mu_a^3)$ and neglecting all the contributions of order larger than $\mu_a^2$, we finally obtain \begin{equation}
\label{eq:totalSDequation_v2}
   \partial P_{14}GP_{14}
   =P_{14}I+\kc{P_{14}\Sigma_\J P_{14}
    +\Sigma_\mu^{\rm eff}}*P_{14}GP_{14}+O(\mu_a^3) \,,
\end{equation}
where
\begin{equation}
   \label{eq:Sigma eff}
   \Sigma_\mu^{\rm eff}=\mu_a^2P_{14}\Sigma_\mu^{(1)}P_{23}*P_{23}G^{(0)}P_{23}*P_{23}\Sigma^{(1)}_\mu P_{14}\,,
\end{equation}
and
\begin{equation}
    \label{eq:Sigma J_ren}
    \Sigma_\J\equiv \Sigma_\J^{(0)}+\mu_a^2\Sigma_\J^{(2)} \,.
\end{equation}
We refer to the contribution $\Sigma_\mu^{\rm eff}$ as {\it effective self-energy} and to (\ref{eq:totalSDequation_v2}) as projected SD equations. 
The effective self-energy (\ref{eq:Sigma eff}) has a non-trivial time dependence that can be determined in terms of the leading contribution to the Green's function $G^{(0)}$. 
Using (\ref{eq:Sigma1_def}), (\ref{eq:mu01_SDRG}) and taking into account that only some entries of $G^{(0)}$ are non-vanishing, we find that the four entries of $\Sigma_\mu^{\rm eff}$ are given by (we choose here to label rows and columns with 1 and 4 to keep in mind that they are obtained by projections through $P_{14}$)
\begin{equation}
    \begin{split}
        \Sigma^{\rm eff}_{\mu\,11}&= \mu_a^2(1+\Delta)^2 G_{22}^{(0)}\,,
\qquad
\Sigma^{\rm eff}_{\mu\,44}= (1-\Delta)^2\mu_a^2G_{22}^{(0)}\,,\\
\Sigma^{\rm eff}_{\mu\, 14}&= -\Sigma^{\rm eff}_{\mu\, 41}=-\mu_a^2 (1-\Delta^2) G_{23}^{(0)}\,.
    \end{split}
    \label{eq:effSigma component}
\end{equation}
Notice that, when $\Delta=0$, the entries in (\ref{eq:effSigma component}) satisfy the expectation (\ref{eq:constraintonabab}) in the case of symmetry with respect to the center of the chain.
Eqs. (\ref{eq:totalSDequation_v2})-(\ref{eq:effSigma component}) perform a real-space RG decimation step on SD equations of the four-site chain with Hamiltonian (\ref{eq:Hamiltonian}) and consist a central result of this work. The effect of the decimation is reflected in the solution of (\ref{eq:totalSDequation_v2}), which provides a large $N$ two-point function renormalized according to this RG step. As discussed in Sec.\,\ref{sec:wormholenetwork}, (\ref{eq:totalSDequation_v2})-(\ref{eq:effSigma component}) are at the core of a generalization of the SDRG methods, originally developed in \cite{MDH79prl,MDH80prb}, to inhomogeneous chains with SYK-like on-site interactions.

Remarkably, the derivation of (\ref{eq:totalSDequation_v2}) and (\ref{eq:Sigma eff}) relies only on the perturbative assumption of having $\mu_a\ll\mu_b$, leading to (\ref{eq:conditionG_perturbative}). This fact implies that the effective SD equations are valid in any other regime of parameters of the model, including any value of $q$. These regimes influence the explicit expression of $\Sigma_\mu^{\rm eff}$ and the consequent solutions of the effective SD equation, but not the general form of (\ref{eq:totalSDequation_v2}).
Notice that in (\ref{eq:totalSDequation_v2}) the presence of on-site SYK terms in the self-energy is not modified by the projection onto the Hilbert space 1-4 (indeed $\Sigma_\J$ in (\ref{eq:Sigma J_ren}) is simply given by the initial self-energy after the projection (see (\ref{eq:sigma0_def})-(\ref{eq:Sigma2_def})), while the hopping contribution is affected by the presence of strong hopping between the central sites.

\subsection{Two-site chain: eternal traversable wormhole}
\label{subsec:two-sitechain}

From the analysis in the previous section it becomes apparent that the leading order contribution $G^{(0)}$ plays a key role, in particular the entries associated to the central sites labeled by the indices $x=2,3$. Within the perturbative approach, these are the sites coupled by the strong hopping parameter $\mu_b$.
To determine the two-point functions related to the central sites, we solve the SD equations in \eqref{eq:projectedSD0orderP23}, which are equivalent to \eqref{eq:SD_eqs_full_form} when $L=2$.
For this purpose, we now review the computation of $G_{22}^{(0)}$, $G_{33}^{(0)}$ and $G_{23}^{(0)}$, following the treatment of \cite{Maldacena:2018lmt}, which sets the stage for deriving the new results contained in this manuscript. The strategy is to cover the full time domain with two partially overlapping time regimes which we will qualitatively denote as \textit{early} and \textit{late} times, respectively. In each of these regimes, the total self-energy $\Sigma$ can be approximated such as to make the SD equations manageable. The quantitative time scales at which these approximations break down can be identified a posteriori once the solutions for $G_{22}^{(0)}$, $G_{33}^{(0)}$ and $G_{23}^{(0)}$ have been found, as we explain in the following.
As discussed in detail in \cite{Maldacena:2018lmt}, the ground state of the two-coupled SYK models with a hopping parameter of the same order of magnitude as the random on-site coupling describes the same physics as an eternal traversable wormhole in a nearly-AdS$_2$ spacetime.

Before considering the two time regimes separately, we notice that at leading order in perturbation theory, the two-site chain is symmetric under reflection with respect to its center and therefore $G^{(0)}_{22}(\tau)=G^{(0)}_{33}(\tau),\, \Sigma^{(0)}_{22}(\tau)=\Sigma^{(0)}_{33}(\tau)$. In the strict limit $q\to\infty$, using (\ref{eq:defGcal}), we have $G^{(0)}_{22}(\tau)=G^{(0)}_{33}(\tau)=\frac{1}{2}\sgn(\tau)+\order{q^{-1}}$, and $G_{23}^{(0)}(\tau)=-G_{23}^{(0)}(\tau)=\frac12\G_{23}+\order{q^{-1}}$, where the constant $\G_{23}$ is purely imaginary due to the symmetries of $G(\tau)$ discussed below \eqref{eq:symmetries_G}.

\subsubsection{Early time}

As discussed in Sec.\,\ref{subsec:Liouvillelargeq}, for early times and in the limit $q\to\infty$, it is convenient to consider the following ansätze for the components of the Green's functions
\begin{equation}
\label{ansatz_largeq_Gren_simpl_23}
\begin{split}
G^{(0)}_{22}(\tau)=&~\frac12\sgn\tau\kc{1+\frac1q g_{22}(\tau)+\cdots}
= \frac12\sgn\tau\, e^{\frac{g_{22}(\tau)}{q}}\ \,,
    \\
G^{(0)}_{23}(\tau)=&~\frac{\mathrm{i}}{2} \kc{1+\frac1q g_{23}(\tau)+\cdots}
= \frac{\mathrm{i}}{2}\, e^{\frac{g_{23}(\tau)}{q}} \,,
\end{split}\end{equation}
where the functions $g_{22}(\tau)$ and $g_{23}(\tau)$ are assumed to be $\order{q^0}$ and have no dependence on $q$. The constants defined in \eqref{eq:defGcal} are chosen to be $\G_{22}=-i\G_{23}=1$, as done in \cite{Maldacena:2018lmt}. Inserting these expressions into the Liouville equation (\ref{eq:evolution_g_gen}) with $s_{22}=s_{33}=1,\, s_{23}=s_{32}=i^q\kappa_{23}^2$ and $\kappa_{23}^2=e^{-1/\xi^2}$, and imposing the boundary conditions (\ref{eq:bc_gen g_R1}) and (\ref{eq:bc_gen g_R23}) applied to a two-site chain, we obtain solutions for the $g_{22}$ and $g_{23}$
\begin{equation}
\label{eq:Sol_LiouvilleEq_twosite}
e^{g_{22}}=\kc{\frac{\alpha_{22}}{\J\sinh(\alpha_{22}\abs{\tau}+\gamma_{22})}}^2\,,\qquad 
 e^{g_{23}}=\kc{\frac{\alpha_{23}}{\kappa_{23}\J\cosh(\alpha_{23}\abs{\tau}+\gamma_{23})}}^2\,,
\end{equation}
with coefficients 
\begin{equation}
\alpha_{22}=\J\sinh\gamma_{22}\,,\qquad
\alpha_{23}=\frac{\hat\mu_b}{2\tanh\gamma_{23}}\,.
\end{equation}
Plugging these solutions back into our ansatz \eqref{ansatz_largeq_Gren_simpl_23}, we find the Green's functions
\begin{equation}
\label{eq:G22G23smalltau}
\begin{split}
    G^{(0)}_{22}(\tau)&
    =\frac{\textrm{sgn}(\tau)}{2}\left(\frac{\alpha_{22}}{\J\sinh(\alpha_{22}\abs{\tau}+\gamma_{22})}\right)^{\frac{2}{q}}
    \,,\\
    G^{(0)}_{23}(\tau)&= \frac{\mathrm{i}}{2}\left(\frac{\alpha_{23}}{\kappa_{23}\J\cosh(\alpha_{23}\abs{\tau}+\gamma_{23})}\right)^{\frac{2}{q}}\,.
\end{split}
\end{equation}
These solutions are valid in the early-time regime, namely when $\abs{g_{22}(\tau)}/q,\, \abs{g_{23}(\tau)}/q \ll 1$ such that the initial assumptions in the ansätze \eqref{ansatz_largeq_Gren_simpl_23} are fulfilled.
More precisely, we see that when $|\tau|\gg \alpha_{22}^{-1}$, the solutions in \eqref{eq:Sol_LiouvilleEq_twosite} are $|g_{22}|\sim 2\alpha_{22} |\tau| \gg 1$, and similarly for $g_{23}$. Thus, for times such that $\frac{\alpha_{22} |\tau|}{q}\sim\order{1}$, our initial assumptions going into the ansatz \eqref{ansatz_largeq_Gren_simpl_23} are no longer fulfilled. The corresponding solutions in \eqref{eq:G22G23smalltau} are therefore only reliable as long as $|\tau|\ll q/\alpha_{22}$ and $|\tau|\ll q/\alpha_{23}$, respectively.

\subsubsection{Late time}
\label{subsec:largetauregime}

To find an alternative way to solve the SD equations for the two-site chain when $|\tau|\gtrsim q/\alpha_{22},\, q/\alpha_{23}$, we can consider a different approximation, exploiting the fact that $G(\tau)$ varies on a much larger time scale than $\Sigma(\tau)$. 
In the late time regime $\tau\gg \alpha^{-1}_{22}, \alpha_{23}^{-1}$, plugging (\ref{eq:G22G23smalltau}) into (\ref{eq:SD_eqs_full_form_Sigma}), we find the exponential damping $\Sigma_{\J\,22}(\tau)\propto e^{-\alpha_{22}\abs{\tau}}$ and $\Sigma_{\J\,23}(\tau)\propto e^{-\alpha_{23}\abs{\tau}}$ for the terms of the self-energy due to the SYK interactions.
Thus, we can approximate their contribution to the convolution integrals entering \eqref{eq:SD_eqs_full_form_G} by a Dirac-distribution \cite{Maldacena:2018lmt}.
More precisely, given that $q$ is chosen to be even, and since $G_{22}(-\tau)=-G_{22}(\tau)$ and $G_{23}(-\tau)=G_{23}(\tau)$, we can take the late-time approximation $\Sigma_{22}^{(0)}(\tau)\approx \rho_{22}\delta'(\tau)$ and $\Sigma_{23}^{(0)}(\tau)\approx-\mathrm{i}\nu_{23}\delta(\tau)$. The weights $\rho_{22}$ and $\nu_{23}$ are determined by the following integrals over the late time domain
\begin{align}
    \rho_{22}=&~-\int_{-\infty}^{+\infty}d\tau \Sigma_{22}^{(0)}(\tau)\tau
    =-\frac1q\log(1-e^{-2\gamma_{22}})\equiv \frac{\hat\rho_{22}}{q} \,,
    \label{eq:deltatrick_int}\\
    \nu_{23}=&~\mathrm{i}\int_{-\infty}^{+\infty}d\tau \Sigma_{23}^{(0)}(\tau)
         = \frac{\mu_b}{\tanh{\gamma_{23}}}
        \equiv \frac{\hat\nu_{23}}{q} \,,\label{eq:couplingunique} 
\end{align}
where we have inserted the self-energy obtained by plugging \eqref{eq:G22G23smalltau} into \eqref{eq:SD_eqs_full_form_Sigma}. Notice that the rescaled coefficients $\hat \nu_{23}$ and $\hat\rho_{22}$ are $\order{q^0}$.

Taking into account both the contributions from the diagonal and the off-diagonal entries of the self-energy, we can approximate the SD equations \eqref{eq:SD_eqs_full_form_G} in the late time regime, i.e. for $\tau\gg \alpha^{-1}_{22}, \alpha_{23}^{-1}$, as
\begin{equation}
    (1+\rho_{xx})\partial_\tau G_{xy}^{(0)}(\tau)+\mathrm{i}\sum_z\nu_{xz}G_{zy}^{(0)}(\tau)=0\,,
    \qquad\qquad
    x,y,z\in\{2,3\}\,,
    \label{eq:renormalized_SD_eq}
\end{equation}
where $\nu_{32}=-\nu_{23}$, $\nu_{22}=\nu_{33}=0$ and $\rho_{33}=\rho_{22}$. Since $\rho_{xx}\sim\order{q^{-1}}$, we will neglect $\rho_{xx}$ compared to $1$ in the large $q$ limit.
Then, Eq.\,(\ref{eq:renormalized_SD_eq}) is the same as the SD equations at $\tau\neq0$ of a free system given by two sets of free Majorana fermions coupled by a hopping term, i.e. described by the Hamiltonian (\ref{eq:Hamiltonian}) with $L=2$, $\J=0$ and $\mu_x=\nu_{23}$. The solutions are straightforwardly obtained, as reviewed for completeness in Appendix \ref{apx:freecase}, and read
\begin{equation}
\label{eq:G22G23largetau}
  G^{(0)}_{22}(\tau)
  =  \frac{\textrm{sgn}(\tau)}{2} A_1 e^{-\nu_{23}|\tau|}
  \,,
    \qquad
    G^{(0)}_{23}(\tau)=\frac{\mathrm{i}}{2} A_1 e^{-\nu_{23}|\tau|}\,.
\end{equation}

\subsubsection{Matching the solutions}
\label{subsec:matching_twosites}

The early and the late time regimes, where the solutions 
(\ref{eq:G22G23smalltau}) and (\ref{eq:G22G23largetau}) respectively hold, overlap when $\alpha^{-1}_{22},\, \alpha^{-1}_{23}\ll|\tau|\ll q/\alpha_{22},\, q/\alpha_{23}$. Thus, to fix the undetermined constants $A_1$, $\alpha_{22}$, $\alpha_{23}$, $\gamma_{22}$ and $\gamma_{23}$, we impose that the two solutions match in this common time domain.
Expanding the entries in (\ref{eq:G22G23smalltau}) for $\alpha_{22}\tau\gg 1$ and $\alpha_{23}\tau\gg 1$ and comparing them to the exponentials in (\ref{eq:G22G23largetau}), we find \cite{Maldacena:2018lmt}
\begin{equation}
   \label{eq: matching conditions 1}\alpha_{22}=\alpha_{23}=\frac{\hat\nu_{23}}{2}=\frac{\hat\mu_b}{2\tanh\gamma_{23}}=\J\sinh\gamma_{22} \,,
\end{equation}
where $\hat{\nu}_{23}$ is defined in (\ref{eq:couplingunique}), and
\begin{equation}
 \label{eq: matching conditions 2}
   \gamma_{22}=\gamma_{23}+\log\kappa_{23}\,,
   \qquad
   A_1=\left(1-e^{-2\gamma_{22}}\right)^{2/q}\to 1\,,
\end{equation}
with the limit of the last equation taken for $q\to\infty$.
Combining \eqref{eq: matching conditions 1} and \eqref{eq: matching conditions 2}, we can determine all the unknown constants $A_1$, $\alpha_{22}$, $\alpha_{23}$, $\gamma_{22}$, $\gamma_{23}$ in terms of the input parameters $\J$, $\hat{\mu}_b$, $q$ and $\kappa_{23}$ in the Hamiltonian. In particular, we only need to determine $\gamma_{23}$ in order to fix all other coefficients. The value of this parameter is obtained by solving the transcendental equation
\begin{equation}
    \sinh\kc{\gamma_{23}+\log\kappa_{23}}\tanh{\gamma_{23}}=\frac{\hat{\mu}_b}{2\J}\equiv \chi\,,
    \label{eq:transcendental_eq_gamma23}
\end{equation}
for given input values for the dimensionless combination $\chi$. Notice that, in the case of $\kappa_{23}=1$, we can find an explicit expression for $\tanh{\gamma_{23}}$, directly written in terms of the input parameters as
\begin{equation}
    \tanh{\gamma_{23}}=\frac{1}{\sqrt{\frac{1}{2}+\sqrt{\frac{1}{4}+\frac{1}{\chi ^2}}}}\,,\quad \text{for}\quad \kappa_{23}=1\,.
    \label{eq:tanh_gamma_23}
\end{equation}
Notice that $\tanh{\gamma_{23}}\to 1-2\J^2/\hat\mu_b^2$ when $\hat{\mu}_b\gg \J$, and $\tanh{\gamma_{23}}\to \sqrt{\hat\mu_b/2\J}$ when $\hat{\mu}_b\ll \J$.
These parameters allow us to fully match the solutions in the overlapping regime, ensuring a continuous passage from early to late times. For convenience, we summarize the expressions of the two-point functions $G^{(0)}_{22}$ and $G^{(0)}_{23}$, which read
\begin{equation}
\begin{split}
   G^{(0)}_{22}(\tau)=&~\frac12
   \textrm{sgn}(\tau)\times \begin{cases}
\left(\frac{\sinh\gamma_{23}}{\sinh(\hat{\nu}_{23}\abs{\tau}/2+\gamma_{23})}\right)^{\frac{2}{q}}\,,
         & |\tau| \ll  \frac{q}{\hat{\nu}_{23}}\,,
        \\
   \left(1-e^{-2\gamma_{23}}\right)^{2/q} e^{-\nu_{23}|\tau|}\,,
     & |\tau| \gg \frac{1}{\hat{\nu}_{23}} \,,
    \end{cases}     \,,\\
   G^{(0)}_{23}(\tau)=&~\frac{\mathrm{i}}{2}\times
    \begin{cases}
\left(\frac{\sinh\gamma_{23}}{\cosh(\hat{\nu}_{23}\abs{\tau}/2+\gamma_{23})}\right)^{\frac{2}{q}}\,,
         & |\tau| \ll  \frac{q}{\hat{\nu}_{23}}\,,
        \\
\left(1-e^{-2\gamma_{23}}\right)^{2/q} e^{-\nu_{23}|\tau|}\,,
     & |\tau| \gg \frac{1}{\hat{\nu}_{23}} \,,
    \end{cases}
     \label{eq:G23} \,,
\end{split}
\end{equation}
where we have re-expressed the relevant time scales in terms of $\hat{\nu}_{23}$ using (\ref{eq: matching conditions 1}). The exponent $\nu_{23}$ of the decay in \eqref{eq:G23} encodes the energy gap of the central sites given by the zero-order Hamiltonian.

In \cite{Maldacena:2018lmt}, it was argued that in the low-energy regime $\hat{\mu}_b\ll\J$ (also denoted as the conformal regime), the dynamics of the correlation functions given in \eqref{eq:G23} are governed by the same effective action that arises from a dilaton-gravity theory in AdS$_2$ with a non-local double-trace coupling between the two conformal boundaries of the AdS space in global coordinates. More precisely, the authors of \cite{Maldacena:2018lmt} considered Jackiw-Teitelboim gravity \cite{JACKIW1985343,TEITELBOIM198341} with nearly-AdS$_2$ boundary conditions. The bulk geometry is fixed to be AdS$_2$ after the dilaton field is integrated out and the exact position of both boundaries is seen as a fluctuating gravitational degree of freedom. The gravitational fluctuation is governed by the Schwarzian action plus the  the interaction between the two boundaries induced by the double-trace deformations. Such interactions have been shown to lead to a quantum matter stress-energy tensor which violates the averaged null energy condition, thus making wormholes traversable \cite{AriasPRD2011,SolodukhinPRD2005,Gao:2016bin,Maldacena:2017axo}. The ground state of such effective action is an eternal (i.e. static and time-independent) traversable wormhole (ETW) in AdS$_2$ \cite{Maldacena:2018lmt}. More precisely, in the conformal coordinates for AdS$_2$, the metric of the wormhole is given by
\begin{equation}\label{eq:ETW}\begin{split}
    &ds^2=\frac{d\sigma^2+d\theta^2}{\sin^2\sigma}, \quad \sigma\in (0,\pi),\\ 
    &\text{left boundary: } \ \sigma_L=\hat\nu_{23}\epsilon, \ 
    \theta_L=\hat\nu_{23} \tau, \\
    &\text{right boundary: } \ 
    \sigma_{R}=\pi-\hat\nu_{23}\epsilon, \ \theta_R=\hat\nu_{23}\tau,
\end{split}\end{equation}
with the UV cutoff $\epsilon\to0$.
The SYK clusters at sites $2$ and $3$ are described by the left and right boundaries, respectively, which are holographically parameterized by the imaginary time $\tau$.
In the light of the discussion in \cite{Maldacena:2018lmt}, by identifying the hopping term $\mathrm{i}\sum_j \mu_b\psi_2^j\psi_3^j$ in \eqref{eq:Hamiltonian} as an effective double-trace deformation which leads to the same interactions between the two boundaries, we may regard the coupled two-site system as enjoying a gravitational dual description in terms of such an ETW. 

An important consequence of the presence of the wormhole is that the rate at which the information can travel from one boundary to the other, i.e. from site 2 to site 3 or vice-versa, is enhanced by the SYK interactions. More precisely, such an information transfer is described by the retarded Green's functions
\begin{equation}\label{eq:RetardG}
    G^R_{xy}(t)\equiv \frac{\Theta(t)}{NZ}\sum_j\Tr[e^{-\beta H}\ke{\psi_x^j(it),\psi_y^j(0)}]
    = \Theta(t)\kc{G_{xy}(0^++it)+G_{yx}(0^+-it)}\,,
\end{equation}
where $Z$ is defined below \eqref{eq:Definition_G} and $t$ is the real time coordinate related to $\tau$ by the Wick's rotation $t=-{\rm i}\tau$.
Expressing the late-time Euclidean Green's function \eqref{eq:G23} in real time and using (\ref{eq:RetardG}), we obtain
\begin{equation}\label{eq:RetardG23}
\begin{split}
    &G^R_{22}(t)=\Theta(t)(1-e^{2\gamma_{23}})^{2/q} \cos(\nu_{22}t)\,,\\
    &G^R_{23}(t)=\Theta(t)(1-e^{2\gamma_{23}})^{2/q} \sin(\nu_{23}t)\,.
\end{split}
\end{equation}
where the decay rate in $G_{xy}(\tau)$ has now the physical meaning of frequency in the retarded Green's function $G^R_{xy}(t)$.
The times taken by information traveling from site $2$ to site $3$ and then coming back to site $2$ are $\frac{\pi}{2\nu_{23}}$ and $\frac{\pi}{\nu_{23}}$, respectively, which are determined by imposing $\abs{G^R_{23}(t)}\approx1$ and $\abs{G^R_{22}(t)}\approx1$. Due to \eqref{eq: matching conditions 1} and $\tanh\gamma_{23}\in[0,1)$, these time intervals are shorter than $\frac{\pi}{2\mu_b}$ and $\frac{\pi}{\mu_b}$ obtained in the non-interacting case with $\J=0$.
The information transfer is thus enhanced by the presence of the wormhole. For the interpretations of later results in this manuscript, let us stress that the aforementioned ETW description for coupled SYK dots is a direct consequence of the SYK contribution $\Sigma_{\J}$ to the total self-energy. Without this term, the system effectively behaves like a collection of $2N$ Majorana fermions which can freely hop from one site to the other, with no enhancement of the information transfer rate. Importantly, the system without SYK interactions does not admit a holographic wormhole description. In the next section, we will examine a four-site chain focusing on the self-energy $\Sigma_{14}$ between the left- and the rightmost sites in the large-$q$ limit. In the instance considered, the self-energy term $\Sigma_{\J\,14}$ coming from the SYK interactions is much smaller than the effective hopping term $\Sigma_{\mu\,14}^\text{eff}$ in \eqref{eq:Sigma eff}, indicating the absence of wormhole between the left- and the rightmost sites.

\subsection{Four-site chain: wormhole-induced effective hopping}
\label{subsec:foursitechain}

In this section, we exploit the solutions (\ref{eq:G23}) to obtain perturbative results for the two-point functions of the degrees of freedom on the left- and the rightmost sites of the four-site chain described in Sec.\,\ref{subsec:setup_foursitechain}. In particular, we will analyse the correlations arising as solutions to \eqref{eq:totalSDequation_v2}. The results can be interpreted as correlations induced by the wormhole geometry connecting the central sites of the chain.

\subsubsection{Projected SD equations}

We begin by inspecting the self-energy governing \eqref{eq:totalSDequation_v2}, which we recall here for convenience,
\begin{equation}
\label{eq:projectedSigma}\Sigma=P_{14}\Sigma_\J P_{14}+\Sigma_\mu^{\rm eff} \,.
\end{equation}
Notice that this self-energy contains two distinct pieces, $P_{14}\Sigma_\J P_{14}$ and $\Sigma_\mu^{\rm eff}$, whose behavior we can analyse individually.
As commented at the end of Sec.\,\ref{subsec:setup_foursitechain}, the on-site contribution to the self-energy is not modified by the decimation of the central sites through the projection, while the hopping contribution is renormalized to the effective self-energy (\ref{eq:Sigma eff}). For this reason, we do not make a change of notation for this term, even though after the projection through $P_{14}$, only four of the sixteen entries are non-vanishing. Using (\ref{eq:effSigma component}), the effective self-energy can be written in terms of \eqref{eq:G23} and read 
\begin{equation}
\begin{split}
   &\Sigma^{\rm eff}_{\mu\,11}(\tau)= \frac 12 \sgn\tau\mu_a^2 (1+\Delta)^2
   \times
   \begin{cases}
    \kc{\frac {\alpha_{22} }{\J\sinh(\alpha_{22} \abs{\tau}+\gamma_{22})}}^{2/q},& |\tau| \ll  \frac{q}{\hat{\nu}_{23}}\\
    e^{-\nu_{23}\abs{\tau}},& |\tau| \gg  \frac{1}{\hat{\nu}_{23}} \\   
    \end{cases}\\
   &\Sigma^{\rm eff}_{\mu\,44}(\tau)= \frac 12 \sgn\tau\mu_a^2 (1-\Delta)^2
   \times 
   \begin{cases}
    \kc{\frac {\alpha_{22} }{\J\sinh(\alpha_{22} \abs{\tau}+\gamma_{22})}}^{2/q},& |\tau| \ll  \frac{q}{\hat{\nu}_{23}}\\
    e^{-\nu_{23}\abs{\tau}},& |\tau| \gg  \frac{1}{\hat{\nu}_{23}} \\   
    \end{cases}\\
\label{eq:sigmaeff_14}
    &\Sigma^{\rm eff}_{\mu\, 14}(\tau)= -\Sigma^{\rm eff}_{\mu\, 41}(\tau)=
      -\frac{\mathrm{i}}{2} \mu_a^2(1-\Delta^2)\times\begin{cases}
     \kc{\frac {\alpha_{23} }{\J\cosh(\alpha_{23} \abs{\tau}+\gamma_{23})}}^{2/q},& |\tau| \ll  \frac{q}{\hat{\nu}_{23}} \\
     e^{-\nu_{23}|\tau|},& |\tau| \gg  \frac{1}{\hat{\nu}_{23}}    
    \end{cases}
\end{split}
\end{equation}
where the two time regimes characterizing $\Sigma^{\rm eff}_{\mu}$ reflect the ones of the two-point functions on the central sites. Similarly to \eqref{eq:couplingunique}, by integrating the late-time behavior of the effective self-energy, we obtain the approximate effective hopping parameter connecting the left- and the rightmost sites
\begin{equation}\label{eq:EffectiveHopping}
    \mathrm{i}\int_{-\infty}^{+\infty} d\tau\Sigma_{\mu\,14}^\text{eff}(\tau)
    \approx \frac{\mu_{a}^2}{\nu_{23}}(1-\Delta^2)\,,
\end{equation}
which is suppressed both by the on-site SYK interactions and by the inhomogeneity of the four-site chain 

Note that, crucially, the self-energy \eqref{eq:projectedSigma} with \eqref{eq:sigmaeff_14} reduces to an exponential dependence of the form $\Sigma=\Sigma_{\mu}^{\textrm{eff}}\propto \mu_a^2e^{-\mu_b\abs{\tau}}$ in the free limit $\J\to0$ independently of the time regime. As we show in Appendix \ref{apx:freecase}, this form of the effective self-energy $\Sigma_{\mu}^{\textrm{eff}}$ can be obtained by directly applying (\ref{eq:effSigma component}) with $G^{(0)}$ obtained in the free case.

Given the non-trivial time dependence, solving the problem for the exact form of (\ref{eq:projectedSigma}) is a formidable challenge. Therefore, in the spirit of the method of \cite{Maldacena:2018lmt} reviewed in Sec.\,\ref{subsec:two-sitechain}, we identified time regimes where $\Sigma$ in (\ref{eq:projectedSigma}) can be simplified, allowing to solve the corresponding SD equations. At the end of the computation, we match the solutions found in the various regimes. To parallel the discussion of Sec.\,\ref{subsec:two-sitechain}, we will also refer to these two regimes as early and late times, although the exact orders of magnitude will be different than those in Sec.~\ref{subsec:two-sitechain}. 

For simplicity, in the following, we solve the problem for $\Delta=0$. As discussed at the beginning of Sec.\,\ref{subsec:setup_foursitechain}, this choice induces on the four-site chain the symmetry under reflection with respect to its center. As a consequence, we have
\begin{equation}
\label{eq:symmetryDelta0}
    G_{11}(\tau)=G_{44}(\tau)\,.
\end{equation}
For this reason, in the rest of this section, we will report the results only for the entries $G_{11}(\tau)$ and $G_{14}(\tau)$, since the other two non-vanishing entries in $P_{14}GP_{14}$ can be obtained from them by (\ref{eq:symmetryDelta0}) and the properties of the Green's functions.
At the end of the computation, we will comment on how the solutions are modified in the presence of $\Delta\neq 0$.

\subsubsection{Early time}
\label{subsec:earlytime4sitechain}

As discussed in Sec.\,\ref{subsec:Liouvillelargeq}, in the early time regime and in the $q\to\infty$ limit, we can write the Green's function through the ansätze 
\begin{equation}
    \label{ansatz_largeq_Gren_simpl_14}
\begin{split}
G_{11}(\tau)=&~\frac12\sgn\tau\kc{1+\frac1q g_{11}(\tau)+\cdots}= \frac12\sgn\tau\, e^{\frac{g_{11}(\tau)}{q}}\ \,,
    \\
G_{14}(\tau)=&~
\frac{\mathrm{i}}{2} \E_{14}\kc{1+\frac1q g_{14}(\tau)+\cdots}
=\frac{\mathrm{i}}{2} \E_{14}e^{\frac{g_{14}(\tau)}{q}}\,,
\end{split}
\end{equation}
where $g_{11}(\tau),\,g_{14}(\tau)\sim\order{q^0}$. The prefactor $\E_{14}$ 
has to be determined by solving the projected SD equations at $\order{\mu_a^2}$.

We find worth it stressing the different notation employed here for the two-point functions $G_{11}$, $G_{44}$ and $G_{14}$ and in Sec.\,\ref{subsec:two-sitechain} for $G^{(0)}_{22}$, $G^{(0)}_{33}$ and $G^{(0)}_{23}$.
As highlighted in Sec.\,\ref{subsec:setup_foursitechain}, solving (\ref{eq:totalSDequation_v2}) leads to expressions for two-point functions that are of $\order{\mu_a^2}$ in perturbation theory while the solutions reviewed in Sec.\,\ref{subsec:two-sitechain} are of $\order{\mu_a^0}$. This is the reason for the different notations employed. Since we will not compute the $\order{\mu_a^2}$ corrections to the two-point functions of degrees of freedom on the central sites, we keep the distinction throughout the manuscript.

To obtain a Liouville equation for the projected SD equations \eqref{eq:totalSDequation_v2} with the ansätze \eqref{ansatz_largeq_Gren_simpl_14}, it is instructive to preliminarily discuss the orders of magnitude of $P_{14}\Sigma_\J P_{14}$ and $\Sigma_{\mu}^\text{eff}$ at large $q$. 
To guide the discussion, let us draw the leading Feynman diagrams in the small $\mu_a/\mu_b$ expansion of $\Sigma_{\J\,14}$ and $\Sigma_{\mu\,14}^\text{eff}$  
\begin{equation}\label{eq:FeynmanSigma14}
    \includegraphics[width=0.6\linewidth]{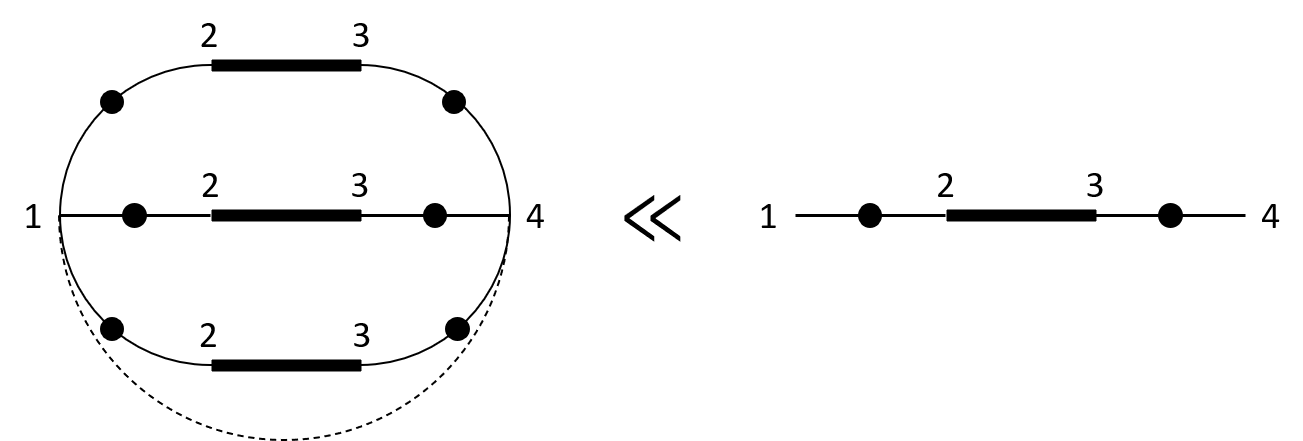}
\end{equation}
with the same notations as \eqref{eq:FeynmanSD} and with $q=4$ for illustration purposes.
We notice their scaling $\Sigma_{\J\,14}(\tau)\sim \frac{1}{q}\J^2\mu_a^{2(q-1)}$ and $\Sigma_{\mu\,14}^\text{eff}(\tau)\sim\mu_a^2$. Thus, we expect $\Sigma_{\J\,14}(\tau)\ll \Sigma_{\mu\,14}^\text{eff}(\tau)$ at large $q$, which is helpful for our large $q$ expansion of the projected SD equations. First, similarly to the original Liouville equation \eqref{eq:evolution_g_gen}, the limit of $\E_{14}^{q}$ at large $q$ has to be taken into account. Given that the order of the dynamical factor is $e^{g_{14}(\tau)/q}\sim e^{\order{q^{-1}}}$ in \eqref{ansatz_largeq_Gren_simpl_14}, each of the lines in \eqref{eq:FeynmanSigma14} contains a factor $\E_{14}$. Thus, the relation in \eqref{eq:FeynmanSigma14} suggests that $\E_{14}^q\ll\E_{14}$. As a consequence, we can assume that the prefactor $\E_{14}^{q}$ and the resulting non-local self-energy $\Sigma_{\J\,14}(\tau)$ behave as
\begin{equation}\label{eq:sigma14 approx}
    \E_{14}=e^{-\order{q^0}}<1\ \Rightarrow\ \Sigma_{\J\,14}(\tau)= -\mathrm{i}e^{-(3/\xi)^2}\frac{\J^2}{q}\E_{14}^{q-1}e^{g_{14}(\tau)}\sim e^{-\order{q}}\,,
\end{equation}
in the large $q$ limit.  On physical grounds, we are assuming that the correlation $\Im G_{14}(0)$ between sites $1$ and $4$ is smaller than the maximal value $1/2$ by a factor of $\O(q^0)$ so that their self-energy $\Sigma_{\J\,14}(\tau)$ is exponentially suppressed at large $q$. Later, we will provide an \textit{a posteriori} consistency check for this assumption once $G_{14}(\tau)$ is calculated.
The second remark is that, from \eqref{eq:sigmaeff_14}, $\Sigma_{\mu}^\eff(\tau)\sim \order{q^{-2}}$ and therefore it is much smaller than  $\Sigma_{\mu\,23}^{(0)}(\tau)\sim \order{q^{-1}}$.
Thus, since $\Sigma_{\J\,14}(\tau)$ and $\Sigma_{\mu}^\eff(\tau)$ are much smaller than $\order{q^{-1}}$, they will not contribute to the Liouville equation for the projected SD equations at early times. More precisely, the components of the self-energy reduce to $\Sigma_{11}(\tau)\approx\Sigma_{\J\,11}(\tau),\,\Sigma_{14}(\tau)\approx0$ and the Liouville equations read 
\begin{equation}
\label{eq:Liouville_shorttime_simple}\partial^2g_{11}(\tau)-2\J^2e^{g_{11}(\tau)}=0  \,,\qquad 
\partial^2g_{14}(\tau)=0\,,
\end{equation}
with boundary conditions
\begin{equation}
\label{eq:projected bc}
    g_{11}(0)=0\,,\qquad\qquad g'_{14}(0)=0\,.
\end{equation}
The first equation in  (\ref{eq:Liouville_shorttime_simple}) is formally the same as (\ref{eq:evolution_g_gen}) with $\G_{11}^q=1$ and therefore can be derived in the same way (reviewed in Appendix \ref{subapp:LiouvilleEq}). 
The solutions to \eqref{eq:Liouville_shorttime_simple} with \eqref{eq:projected bc} are 
\begin{equation}
\label{eq:shortime_sol_simple}
e^{g_{11}(\tau)}=\frac{\alpha_{11}^2}{\J^2\sinh^2(\alpha_{11}\abs{\tau}+\gamma_{11})} \,,\quad 
g_{14}(\tau)=c_{14} \,,
\end{equation}
with undetermined constants $c_{14}\sim \order{q^0}$ and
\begin{equation}
 \label{eq:bc_aba_simpl}
   \alpha_{11}=\J\sinh\gamma_{11}.
\end{equation}
These solutions are valid in the early-time regime when $\abs{g_{11}(\tau)}/q,\, \abs{g_{14}(\tau)}/q \ll 1$ such that the initial ansätze \eqref{ansatz_largeq_Gren_simpl_14} are fulfilled. Similarly to the analysis below \eqref{eq:G22G23smalltau}, when $\abs{\tau}\gg\alpha_{11}^{-1}$, we have $\abs{g_{11}(\tau)}\sim 2\alpha_{11}\abs{\tau}\gg1$, which becomes $\order{q^{1}}$ at $\abs{\tau}\sim q/\alpha_{11}$. Thus, the solutions in \eqref{eq:shortime_sol_simple} are only reliable as long as $|\tau|\ll q/\alpha_{11}$.

\subsubsection{Late time}
\label{subsec:largetauregime_4sites}

Due to the breakdown of the approximation (\ref{ansatz_largeq_Gren_simpl_14}) discussed above, we consider a different approach to solve the SD equations in the late time regime.
We expect new equations to determine the behavior of the Green's functions in the regime $\abs{\tau}\gg \alpha_{11}^{-1}$. The non-local self-energy $\Sigma_{\J\,14}(\tau)$ in \eqref{eq:sigma14 approx} is still negligible at late times due to the exponentially large $q$ suppression.
In addition, since the on-site self-energy $\Sigma_{\J\,11}(\tau)$ exhibits exponential damping as $e^{-\alpha_{11}\abs{\tau}}$ in the late time regime, we can consider the approximation $\Sigma_{\J\,11}(\tau)\approx\rho_{11}\delta'(\tau)$, where 
\begin{equation}\label{eq:sigma11}
    \rho_{11}=-\int_{-\infty}^{+\infty}d\tau \Sigma_{\J\, 11}^{(0)}(\tau)\tau
    =-\frac1q\log\kc{1-e^{-2\gamma_{11}}}\equiv \frac{\hat\rho_{11}}{q}\,.
\end{equation}
This approximation is similar to the one considered in \eqref{eq:deltatrick_int} while studying the two-site chain. For this reason, as discussed below \eqref{eq:renormalized_SD_eq}, the term $\rho_{11}\delta'(\tau)$ can be neglected when compared to the derivative term in the projected SD equations \eqref{eq:totalSDequation_v2} given that $\rho_{11}\sim\order{q^{-1}}$. Differently from the early time regime, since $\Sigma_\mu^{\rm eff}(\tau)$ decays slowly as $e^{-\nu_{23}|\tau|}$ in \eqref{eq:sigmaeff_14}, its late time contribution is no longer negligible. The crossover between the regime where $\Sigma_{\J\,11}(\tau)$ dominates in the self-energy and the one where $\Sigma_{\mu}^\text{eff}(\tau)$ gives the leading contribution occurs in the window where $\alpha_{11}^{-1}\ll\abs{\tau}\ll q/\alpha_{11}$. 
As a consequence, the self-energy reduces to
$P_{14}\Sigma P_{14}\approx \Sigma_{\mu}^{\rm eff}$ in the late time regime.
More precisely, when $|\tau|\gg \alpha_{11}^{-1}$, the effective self-energy $\Sigma_{\mu}^\text{eff}(\tau)$ is given by the second case of (\ref{eq:sigmaeff_14}) (with $\Delta=0$), whose Fourier transform is 
\begin{equation}
\label{eq:sigmaeff_omegaspace}
   \Sigma_\mu^{\rm eff}(\omega)= \frac{{\rm i}\mu_a^2}{\nu _{23}^2+\omega ^2}\begin{pmatrix}
 \omega  & -\nu_{23} \\
 \nu _{23} & \omega
\end{pmatrix}\,.
\end{equation}
Using (\ref{eq:sigmaeff_omegaspace}) in the projected SD equations \eqref{eq:totalSDequation_v2} in the frequency domain, we find
\begin{equation}
\label{eq:longtimesolution_simpl}
 P_{14}GP_{14}(\tau)= \frac{1}{2}A_+ e^{-E_+ \tau}\left(
\begin{array}{cc}
 1 & -\mathrm{i} \\
\mathrm{i}& 1 \\
\end{array}
\right)+\frac{1}{2}A_- e^{-E_- \tau}\left(
\begin{array}{cc}
 1 &\mathrm{i}\\
 -\mathrm{i} & 1 \\
\end{array}
\right)\,,
\end{equation}
where
\begin{equation}
\label{eq:Eplusminus_main}
    E_\pm=\frac{\sqrt{4 \mu _a^2+\nu _{23}^2}\pm \nu _{23}}{2}\,
\end{equation}
are given by the roots of $\det\kc{-\mathrm i\omega-\Sigma_\mu^{\rm eff}(\omega)}=0$ with $\omega=-\mathrm{i}E_\pm$.
When $\mu_a\ll \nu_{23}$,
\begin{equation}
\label{eq:Eplusminus-simple}
    E_+=\nu_{23}+\frac{\mu_a^2}{\nu_{23}}+\order{\mu_a^4}\,,
    \qquad\qquad
    E_-=\frac{\mu_a^2}{\nu_{23}}+\order{\mu_a^4}\,.
\end{equation}
A detailed derivation of the solution of the SD equations is reported for completeness in Appendix \ref{subapx:Fourier_expSigma}.

\subsubsection{Matching the solutions}
\label{subsec:matchingsolution_four sites}

To fix the undetermined constants in (\ref{eq:shortime_sol_simple}) and (\ref{eq:longtimesolution_simpl}), we use the same method employed in Sec.\,\ref{subsec:matching_twosites} for the two-point functions associated to degrees of freedom on the central sites.
Since the early time regime is defined by $|\tau|\ll q/\alpha_{11}$ while the long time regime by $|\tau|\gg \alpha_{11}^{-1}$, we impose that the solutions found in the two time domains match in the overlapping regime where
$\alpha_{11}^{-1}\ll|\tau|\ll q/\alpha_{11}$. 
Notice that, when $\alpha_{11}^{-1}\ll|\tau|\ll q/\alpha_{11}$, the early time solutions obtained by combining (\ref{ansatz_largeq_Gren_simpl_14}) and (\ref{eq:shortime_sol_simple}) can be approximated by a linear behaviour given by 
\begin{eqnarray}
\label{eq:G_11 linear_short_simple}
  2G_{11}(\tau) &=& 1-2\alpha_{11}\frac{\tau}{q}+\frac{2}{q}\left(\log\frac{2\alpha_{11}}{\J}-\gamma_{11}\right)\,,
    \\
   -2\mathrm{i}G_{14}(\tau) &=& \E_{14}\kc{1+\frac1q c_{14}}\,.
   \label{eq:G_14 linear_short_simple}
\end{eqnarray}
On the other hand, the late time solutions (\ref{eq:longtimesolution_simpl}) have an exponential behavior in time with an exponent of $\order{1/q}$. Thus, for $\alpha_{11}^{-1}\ll|\tau|\ll q/\alpha_{11}$, also these solutions can be expanded linearly in time as
\begin{eqnarray}
\label{eq:G_11 linear_long_simple}
 2G_{11}(\tau)  &=& A_-+A_+-( A_-E_-+A_+E_+)\tau\,,
    \\
   -2\mathrm{i}G_{14}(\tau) &=& A_--A_+-( A_-E_--A_+E_+)\tau\,.
   \label{eq:G_14 linear_long_simple}
\end{eqnarray}
Matching the coefficients of the linear functions in (\ref{eq:G_11 linear_short_simple}) with (\ref{eq:G_11 linear_long_simple}), and (\ref{eq:G_14 linear_short_simple}) with (\ref{eq:G_14 linear_long_simple}), we obtain four conditions. Together with the first equation in (\ref{eq:bc_aba_simpl}), these read 
\begin{eqnarray}
    1+\frac{2}{q}\left(\log\frac{2\alpha_{11}}{\J}-\gamma_{11}\right)&=&
    A_-+A_+\,,
    \\
   \frac{2\alpha_{11}}{q} &=& A_-E_-+A_+E_+\,,
    \\
    \E_{14}\kc{1+\frac1q c_{14}}&=&A_--A_+\,,
    \\
    0 &=&  A_-E_--A_+E_+\,,
    \\
   \alpha_{11} &=&\J\sinh\gamma_{11}\,.
\end{eqnarray} 
From these conditions, we can determine the expression of some unknown parameters at leading order in the large $q$ expansion
\begin{eqnarray}
    \alpha_{11}&=&q\frac{E_+E_-}{E_++E_-}
    \approx \frac{\hat{\mu}_a^2}{ \hat{\nu}_{23}}\,,
    \\
    \gamma_{11}&=&\sinh^{-1}\left(\frac{q}{\J}\frac{E_+E_-}{E_++E_-}\right)
    \approx\frac{\hat{\mu}_a^2}{ \J\hat{\nu}_{23}}\,,
    \\
   A_+ &=&\frac{E_-}{E_++E_-}
    \approx \frac{\hat{\mu}_a^2}{ \hat{\nu}^2_{23}}\,,
    \\
   A_- &=& \frac{E_+}{E_++E_-}
    \approx \exp\kc{-\frac{\hat{\mu}_a^2}{ \hat{\nu}^2_{23}}}\,,
    \\
    \E_{14}&=& \frac{E_+-E_-}{E_++E_-}
    \approx \exp\kc{-2\frac{\hat{\mu}_a^2}{ \hat{\nu}^2_{23}}}\,,
    \label{eq:E14}
\end{eqnarray}
where the last identities are valid up to the perturbative order $\hat{\mu}_a^2$ and are obtained using (\ref{eq:Eplusminus-simple}). The constant $c_{14}$ cannot be determined at the order in the large $q$ expansion considered here.

Before writing down the full solutions, let us discuss a self-consistency check of the assumption \eqref{eq:sigma14 approx}. The solution (\ref{eq:E14}) satisfies $\E_{14}=e^{-\order{q^0}}$ since $\mu_a^2/\nu^2_{23}\sim \order{q^0}$. 
Furthermore, from \eqref{ansatz_largeq_Gren_simpl_14}, \eqref{eq:sigma14 approx} and \eqref{eq:shortime_sol_simple}, we obtain, at $\tau=0$,
\begin{equation}
    G_{14}(0)\approx\frac{\mathrm i}{2}\exp\kc{-2\frac{\mu_a^2}{ \nu^2_{23}}}\,,
\end{equation}
which shows weaker correlation compared with $G_{23}(0)=\mathrm i/2$, and 
\begin{equation}
\label{eq:crosscheck_SigmaJ14}
    \Sigma_{\J\,14}(0)\approx -\mathrm i e^{-(3/\xi)^2}\frac{\J^2}{q}\exp\kc{-2q\frac{\mu_a^2}{\nu^2_{23}}}\,,
\end{equation}
which is exponentially suppressed in the large $q$ limit given that $\mu_a^2/\nu^2_{23}\sim \order{q^0}$. Thus, our assumption \eqref{eq:sigma14 approx} is justified when the $q\to\infty$ limit is taken before the perturbative limit $\mu_a/\mu_b\to 0$. This hierarchy of limits is also crucial for the forthcoming discussions.

Let us report the solutions for the two independent entries $G_{11}(\tau),\,G_{14}(\tau)$, which interpolate continuously between the early and late time regimes. The expressions are shown in terms of parameters common to both the regimes and read
\begin{subequations}
\label{eq:G14}
\begin{align}
    G_{11}(\tau)=&~\frac12\mathrm{sgn}(\tau)
    \times\begin{cases}
\left(\frac{q\mu_a^2}{ \J\nu_{23}}\left[\sinh\left(\frac{q\mu_a^2}{ \J\nu_{23}}(\J\abs{\tau}+1)\right)\right]^{-1}\right)^{2/q},
         & |\tau| \ll q \frac{\hat{\nu}_{23}}{\hat{\mu}^2_a}\,,
        \\
     \frac{\mu_a^2}{\nu^2_{23}}e^{-\left(\nu_{23}+\frac{\mu_a^2}{\nu_{23}}\right)|\tau|}
     +
     \left(1-\frac{\mu_a^2}{\nu^2_{23}}
     \right)e^{-\frac{\mu_a^2}{\nu_{23}}|\tau|},
     & |\tau| \gg \frac{\hat{\nu}_{23}}{\hat{\mu}^2_a} \,,
    \end{cases}
     \label{eq:G11_simplified}\\
    G_{14}(\tau)=&~\frac{\mathrm{i}}{2}\times
    \begin{cases}
       1 -2\frac{\mu_a^2}{\nu^2_{23}},
        & |\tau| \ll q \frac{\hat{\nu}_{23}}{\hat{\mu}^2_a}\,,
        \\
         -\frac{\mu_a^2}{\nu^2_{23}}e^{-\left(\nu_{23}+\frac{\mu_a^2}{\nu_{23}}\right)|\tau|}
     +
     \left(1-\frac{\mu_a^2}{\nu^2_{23}}
     \right)e^{-\frac{\mu_a^2}{\nu_{23}}|\tau|}, & |\tau| \gg \frac{\hat{\nu}_{23}}{\hat{\mu}^2_a}\,,
    \end{cases}
    \label{eq:G14_simplified}
    \end{align}
\end{subequations}
which are valid if the hierarchy $q^{-1}\ll\mu_a^2/\nu_{23}^2\ll1$ is satisfied. As we can read from the second case of \eqref{eq:G14}, the late-time Green's functions $G_{11}(\tau)$ and $G_{14}(\tau)$ have two exponentially decreasing modes with different decay rates $\nu_{23}+\frac{\mu_a^2}{\nu_{23}}$ and $\frac{\mu_a^2}{\nu_{23}}$. The decay rate $\nu_{23}+\frac{\mu_a^2}{\nu_{23}}$ is the decay rate of $G_{22}(\tau)$ and $G_{23}(\tau)$ \eqref{eq:G23} corrected at $\order{\mu_a^2}$ with respect to the original decay rate $\nu_{23}$. Since $\frac{\mu_a^2}{\nu_{23}}\ll \nu_{23}$, from the lessons of SDRG in spin chains \cite{vieira05,Vieira:2005PRB}, we can interpret the decay rate $\frac{\mu_a^2}{\nu_{23}}$ as the true energy gap for the four-site chain.

Let us compare the two decay rates $\nu_{23}$ and $\frac{\mu_a^2}{\nu_{23}}$ in the interacting case with the two decay rates $\mu_b$ and $\frac{\mu_a^2}{\mu_b}$ in the free case, given in Appendix \ref{apx:freecase}. Since $\nu_{23}>\mu_b$, the decay rate of $G_{23}$ is enhanced by the SYK interaction. On the other hand, $\frac{\mu_a^2}{\nu_{23}}<\frac{\mu_a^2}{\mu_b}$ and therefore the decay rate of $G_{14}$ is suppressed by the SYK interaction. 
Taking the analytic continuation in the late-time Euclidean Green's function \eqref{eq:G14} to express it in real time $t$ and using (\ref{eq:RetardG}), we obtain
\begin{equation}\label{eq:RetardG14}
\begin{split}
    &G^R_{11}(t)
    = \Theta(t)\ke{\frac{\mu_a^2}{\nu_{23}^2} \cos \kd{ \kc{\nu_{23}+\frac{\mu_a^2}{\nu_{23}}} t} 
    +\kc{1-\frac{\mu_a^2}{\nu_{23}^2}} \cos \kc{\frac{\mu_a^2}{\nu_{23}}t}}
    \,,\\
    &G^R_{14}(t)
    =\Theta(t)\ke{- \frac{\mu_a^2}{\nu_{23}^2} \sin \kd{ \kc{\nu_{23}+\frac{\mu_a^2}{\nu_{23}}} t}
    + \kc{1-\frac{\mu_a^2}{\nu_{23}^2}} \sin \kc{\frac{\mu_a^2}{\nu_{23}}t} }
    \,.
\end{split}
\end{equation}
The frequencies $\nu_{23}+\frac{\mu_a^2}{\nu_{23}}$ and  $\frac{\mu_a^2}{\nu_{23}}$ of the two oscillating modes in the retarded Green's functions correspond to the two decay rates identified in \eqref{eq:G14}.
Notice that propagation of fermions between sites $1$ and $4$ depends on the time scale.
On the one hand, at short time $t\ll \kc{\nu_{23}+\mu_a^2/\nu_{23}}^{-1}$, the propagation is described by $G^R_{14}(t)\approx \frac16 \mu_a^2\nu_{23}t^3$, whose growth rate is enhanced by the SYK interactions, given that $ \nu_{23} > \mu_b$. The short-time propagation between sites $1$ and $4$ can be understood as consisting of three factors as $\mu_a^2\nu_{23}t^3=\mu_at\times\nu_{23}t\times \mu_at$. The first factor corresponds to the propagation between sites $1$ and $2$, the second to the one between sites $2$ and $3$, and the third to the propagation between sites $3$ and $4$.
In this case, the propagation between sites $2$ and $3$ is the only one affected by the SYK interaction and its enhancement is the one discussed in Sec.\,\ref{subsec:matching_twosites}. 
On the other hand, the long time propagation is controlled by the second terms in \eqref{eq:RetardG14}. The time taken for fermions to propagate from site $1$ to site $4$ is $\frac{\pi\nu_{23}}{2\mu_a^2}$, determined by $G_{14}(t)\approx1$. Thus, the long-time propagation is suppressed by the SYK interactions, following the suppression of effective hopping parameter in \eqref{eq:EffectiveHopping}.  Physically, the suppression can be explained by the local SYK random interactions on sites $2$ and $3$ acting as impurities, which obstruct the transport of fermion over the longer distance between sites $1$ and $4$.

In Sec.\,\ref{subsec:setup_foursitechain}, we stressed that $G^{(0)}_{11}$, $G^{(0)}_{44}$ and $G^{(0)}_{14}$, i.e. the leading orders in the perturbative expansion for small $\hat{\mu}_a$ of the two-point functions $G_{11}$, $G_{44}$ and $G_{14}$, should be determined by the SD equations at higher orders of $\hat\mu_a$ according to degenerate perturbation theory.
The knowledge of the higher orders in this expansion allows to lift this degeneracy and determine $G^{(0)}_{11}$, $G^{(0)}_{44}$ and $G^{(0)}_{14}$ by computing the limit $\hat{\mu}_a\to 0$ of $G_{11}$, $G_{44}$ and $G_{14}$ in \eqref{eq:G14} respectively.
By taking the limit $\hat{\mu}_a\to 0$, the diagonal entries in the early time regime read
\begin{equation}
\label{eq:G11_muazero}
    G^{(0)}_{11}(\tau)= G^{(0)}_{44}(\tau)=\frac{\sgn(\tau)}{2}\left(\frac{1}{\J|\tau|+1}\right)^{2/q}
    \,,
\end{equation}
which is a remnant of the on-site correlations of isolated SYK models characterized by an algebraic decay \cite{Maldacena:2016remarks}. Given that the time scale $\frac{\hat{\nu}_{23}}{\hat{\mu}^2_a}$ becomes larger and larger in the perturbative limit, it is only at extremely large times that the effect of the new energy scale given by the hopping is observable. More precisely, at such long times,
the diagonal entries of  the Green's function behave like an exponential governed by this hopping energy scale. On the other hand, when $\hat{\mu}_a\to 0$, the off-diagonal entry is
\begin{equation}
\label{eq:G14_muazero}
   G^{(0)}_{14}(\tau) =\frac{\mathrm{i}}{2}\,,
\end{equation}
which is constant as predicted in Sec.\,\ref{subsec:setup_foursitechain} and equal to the analogous result in the free case (cf. Appendix \ref{apx:freecase}). Notice that the results (\ref{eq:G11_muazero}) and (\ref{eq:G14_muazero}) are equivalently obtained in the limit $\J\gg \hat{\mu}_b$.

\begin{figure}
    \centering
    \includegraphics[width=\textwidth]{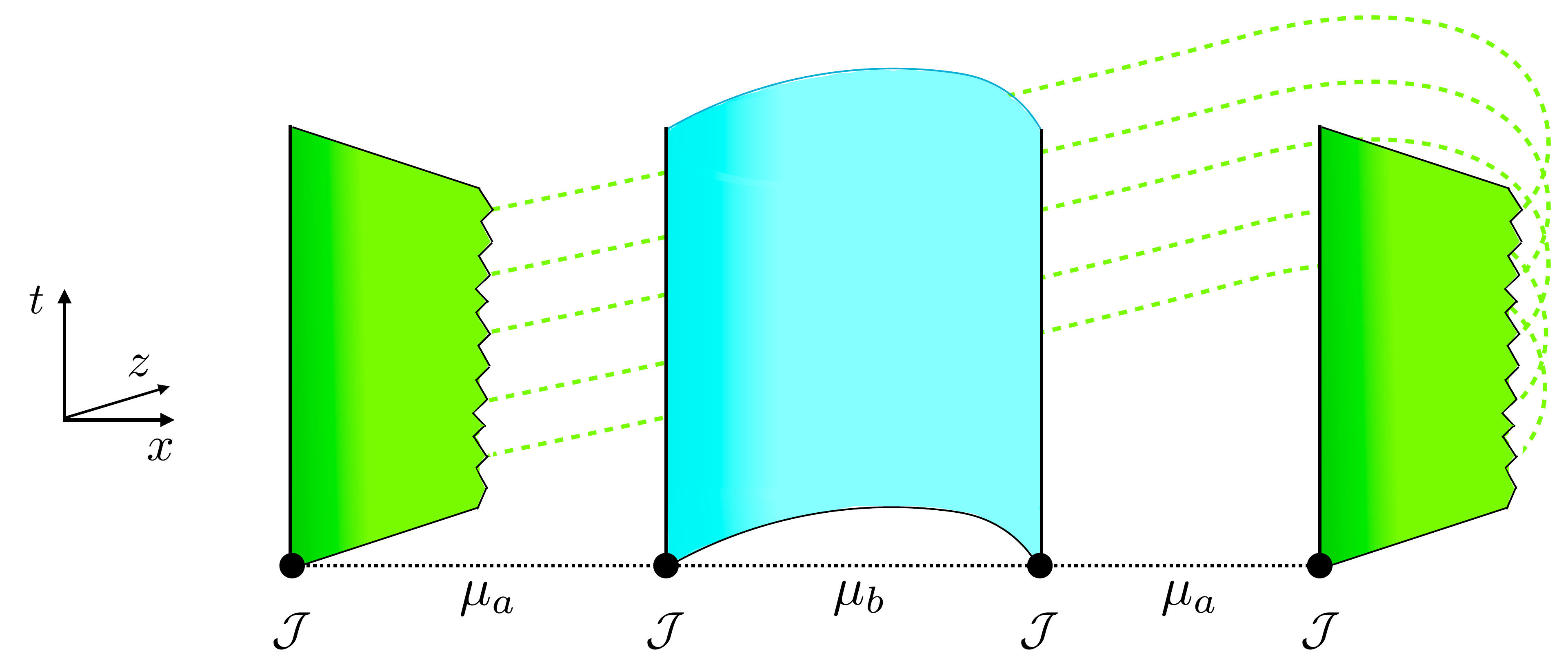}
    \caption{Pictorial representation of the geometry dual to the ground state of the four-site chain in the limit $\J\gg \hat{\mu}_b$ and the perturbative regime $\hat{\mu}_a\ll \hat{\mu}_b$. The vertical direction represents the real time direction parametrized by $t={\rm i}\tau$, the horizontal $x$ direction is the spatial direction along the chain, while the transverse direction is parametrized by the holographic coordinate $z$. The blue surface represents the ETW geometry in \eqref{eq:ETW}, while the green ones are the two nearly AdS$_2$ spaces describing the portions of spacetime around the left- and rightmost sites. These two patches are related by the correlations (green dashed lines) given in (\ref{eq:G14_simplified}), which do not have a geometric description.
    \label{fig:dualgeometry}}
\end{figure}

Let us comment on the interpretation of our results on the Green's functions in terms of dual geometries. First, notice that the consistency check provided by \eqref{eq:crosscheck_SigmaJ14} crucially implies that the total self-energy $\Sigma_{14}=\Sigma_{\J\,14}+\Sigma^{\textrm{eff}}_{\mu\,14}$ between left- and rightmost sites is always dominated by the effective hopping $\Sigma_{14}\simeq\Sigma^{\textrm{eff}}_{\mu\,14}$ and receives no contribution from the SYK interactions $\Sigma_{\J\,14}\simeq 0$ due to the exponential suppression at large $q$ in \eqref{eq:sigma14 approx}. This is drastically different from the situation encountered for the two-site chain in Sec.~\ref{subsec:two-sitechain}, where it was precisely the contribution from $\Sigma_{\J}$ which allowed for the interpretation of the two coupled SYK dots as an ETW \cite{Maldacena:2018lmt} in the regime $\J\gg\hat\mu_b$. Given the result \eqref{eq:crosscheck_SigmaJ14}, we conclude that even though the left- and rightmost sites of the four-site chain have non-vanishing correlations (as we show in \eqref{eq:G14_simplified}), these cannot be described by a dual ETW geometry in the large $q$ limit. Based on these considerations, we give a pictorial representation of the geometry dual to the ground state of the four-site chain is shown in Fig.\,\ref{fig:dualgeometry}. Let us stress that the represented spacetime emerges in the regime of our perturbative approximation $\hat{\mu}_a\ll \hat\mu_b$ and in the limit where the SYK interaction dominates, i.e. $\J\gg\hat{\mu}_b$. The ETW geometry is represented by the blue surface connecting the central sites. Interestingly, in the limit where $\J$ is much larger than the two hopping parameters, we can identify nearly AdS$_2$ geometries (depicted as green patches) along the holographic coordinate close to the left- and the rightmost sites. This is consistent with the fact that, when $\J\gg\hat{\mu}_b$, the expressions of $G_{11}$ and $G_{44}$ reduce to (\ref{eq:G11_muazero}), namely to the on-site correlation functions of decoupled SYK dots in the low-energy regime. The two nearly AdS$_2$ geometries are correlated as given by (\ref{eq:G14_simplified}) and pictorially represented by the green dashed curves. 

Finally, we qualitatively discuss the features of the two-point functions $G_{11}$, $G_{44}$ and $G_{14}$ in the case when the parameter $\Delta$ in the effective self-energy (\ref{eq:sigmaeff_14}) is different from zero. We recall that $\Delta$ is related to the anisotropy of the hopping parameters connecting the left- and the rightmost sites to the rest of the four-site chain (see the discussion at the beginning of Sec.\,\ref{subsec:setup_foursitechain}). The first remark is that, when $\Delta\neq 0$, $G_{11}\neq G_{44}$, and, therefore, in general, we have three independent entries of $G$ to determine. The qualitative behavior in $\tau$ of the two-point functions does not change due to $\Delta\neq 0$, namely, the correlations are characterized by exponential damping. The value of $\Delta$ enters in the constant prefactors of the exponentials and, most importantly, in the rate of decay of the Green's function. The general value of the decay rate is $\frac{\hat{\mu}_a^2(1-\Delta^2)}{q\hat{\nu}_{23}}$. This rate and the modifications to the constant prefactors are computed in Appendix \ref{subapx:Fourier_expSigma}, where the late-time SD equations are solved for generic values of $\Delta$.
The generalization to $\Delta\neq 0$ allows to apply our results in the context of infinite chains with on-site SYK-like interaction and random hoppings between nearest-neighbor sites. This application is discussed in detail in Sec.\,\ref{sec:wormholenetwork}.

\subsection{Numerical checks}
\label{subsec:numeric}

\begin{figure}
    \centering
    \includegraphics[width=\textwidth]{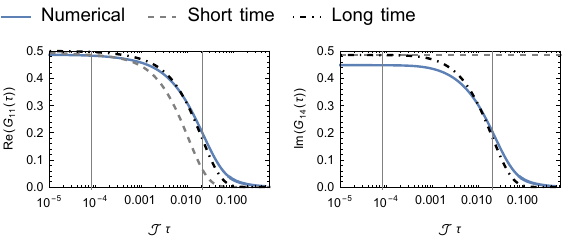}
     \caption{The nonzero parts of the components of Green's functions $G_{11},G_{14}$ as functions of the dimensionless Euclidean time $\J\tau$, obtained by solving the SD equations with the effective self-energy $\Sigma_{\mu}^{\eff}$ in \eqref{eq:sigmaeff_14} introduced artificially. The parameters considered in the numerical calculations are $q=256\,,\beta=5\times 10^5\,,\J=1\,,\hat\mu_a=0.2\,,\hat\nu_{23}=1.6$. The blue solid curves represent the numerical results. The gray dashed curves and black dot-dashed curves represent the analytical result at early time \eqref{eq:shortime_sol_simple} and at late time \eqref{eq:longtimesolution_simpl} respectively. The vertical gray lines represent the dimensionless time scales $\J/\alpha_{11}$ and $q\J/\alpha_{11}$.}
    \label{fig:GreenFunctionsEff}
\end{figure}

\begin{figure}
    \centering
    \includegraphics[width=\textwidth]{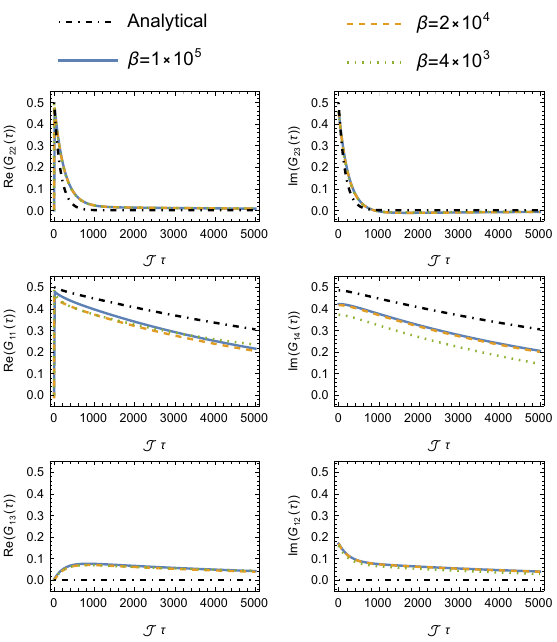}
    \caption{The nonzero part of the components of Green's functions $G_{22}$, $G_{23}$, $G_{11}$, $G_{14}$, $G_{12}$ and $G_{13}$ as functions of the dimensionless Euclidean time $\J\tau$. The curves are obtained for $q=256\,,\beta=5\times 10^5, 1\times 10^5,2\times 10^4\,,\J=1\,,\hat\mu_b=1\,,\hat\mu_a=0.2$ and $\xi=\infty$. The colored curves represent the numerical results at different $\beta$. The black dot-dashed curves represent the analytical approximation at late time \eqref{eq:longtimesolution_simpl}.
    \label{fig:GreenFunctionsBeta}}
\end{figure}

\begin{figure}
    \centering
    \includegraphics[width=\textwidth]{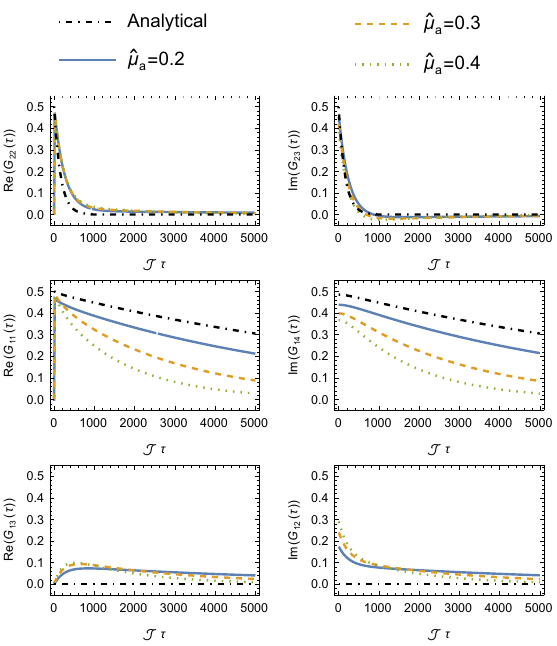}
    \caption{The nonzero part of the components of Green's functions $G_{22}$, $G_{23}$, $G_{11}$, $G_{14}$, $G_{12}$ and $G_{13}$ as functions of the dimensionless Euclidean time  $\J\tau$. The curves are obtained for $q=256\,,\beta=5\times 10^5\,,\J=1\,,\hat\mu_b=1\,,\hat\mu_a=0.2,0.3,0.4$. The colored curves represent the numerical results at different $\hat\mu_a$. The black dot-dashed curves represent the analytical approximation \eqref{eq:longtimesolution_simpl} at $\hat\mu_a=0.2$.
    \label{fig:GreenFunctionsMua}}
\end{figure}

We investigate the four-site chain considered in Sec.\,\ref{subsec:foursitechain} at low temperature by numerically solving the SD equations. As before, we assume that the hopping parameters follow the sequence $(\mu_a,\mu_b,\mu_a)$ along the chain. As discussed above in this section, from symmetry arguments, the independent components of Green's function are $G_{23},G_{23},G_{11},G_{14},G_{12},G_{13}$. 

Our numerical test consists of two levels. At the first level, we numerically solve the SD equations for sites $1$ and $4$ solely by artificially introducing the effective self-energy $\Sigma_{\mu\,11}^{\eff},\Sigma_{\mu\,44}^{\eff},\Sigma_{\mu\,14}^{\eff}$ in \eqref{eq:sigmaeff_14} with $\Delta=0$ and $\nu_{23}$ determined by \eqref{eq: matching conditions 1}. We compare the numerical and analytical results in Fig.~\ref{fig:GreenFunctionsEff}. Their deviations are mainly due to the effect of finite $q$ and finite $\beta$. 

At the second level, we numerically solve the full SD equations for the four sites. We compare the numerical and analytical results in Figs.~\ref{fig:GreenFunctionsBeta} and ~\ref{fig:GreenFunctionsMua}. We have not reported the analytical results for $G_{13}$ and $G_{12}$ since they are of the order $\mu_a$ and therefore subleading compared to the other analytical results obtained in perturbation theory. The deviations between the numerical and analytical results are reduced by increasing $\beta$ and decreasing $\hat\mu_a$. This observation supports the analytical results obtained before in this section.

In the middle-left panel of Fig.~\ref{fig:GreenFunctionsBeta}, the results at $\beta=2\times10^3$ are reliable only as long as $\tau\ll\beta$. In this regime, the numerical curves at higher $\beta$ are closer to the analytical curves, as expected.
We find it worth noticing that increasing $q$ and $\beta$, we observe smaller deviations and a better agreement between the numerical and the analytical curves in Fig.\,\ref{fig:GreenFunctionsEff}. However, in order to use the same parameters in all the figures of this section, we report only the outcomes obtained for the values of $q$ and $\beta$ displayed in the caption. 

\section{SDRG on SYK chains}
\label{sec:wormholenetwork}

The results obtained in the previous sections give analytical insights into the behavior of correlations in a four-site chain with on-site random interactions. In particular, in the regime where the bond connecting the central sites is much stronger than those connecting to the left- and rightmost sites, we find a hierarchy between the two gaps in the system, which leads to the interpretation of the correlations $G_{23}$ in terms of an ETW, while ruling out the emergence of such a geometry between the degrees of freedom correlated by $G_{14}$. We now follow the idea alluded to in the beginning of last section and think of this four-site chain as being embedded in a larger chain with a non-homogeneous hopping distribution. In the regime where 
the strongest hopping parameter along the chain can be identified, i.e. the strong disorder regime, the ground state of such non-homogeneous quantum systems can be investigated via a tool called \textit{strong-disorder renormalization group} (SDRG) \cite{MDH79prl,MDH80prb}.  In the following, we briefly review this approach and then we elucidate on its application to the SD equations considered above.

\subsection{SDRG and SD equations}
\label{subse:generaldecimation}

As a real-space renormalization procedure, the core idea of the SDRG approach is to first identify the sites of the chain are coupled by the stronger hopping and subsequently decimate them.
This decimation corresponds to the instance where the connection of these sites to the rest of the chain via weak couplings can be considered in perturbation theory. This decimation step can be shown to induce a renormalization of the hopping parameters between the sites neighboring the strongly coupled block.

More precisely, in the Hamiltonian formalism of SDRG, if we only consider two-site blocks coupled by a strong hopping with Hamiltonian $H_0$, which is given by the restriction of the general Hamiltonian to only these sites, we can study the influence of the neighboring sites to the left and right of the strongly-coupled block as a perturbation $\delta H_{LR}$, see Appendix~\ref{apx:SDRG}. 
Using the weaker hopping as a perturbative parameter, the second-order corrections to the energy of the local ground state give the renormalized hopping parameter in the effective Hamiltonian $H'_{LR}$ between the left and right sites. After this decimation step, the system has, in principle, fewer degrees of freedom than before. However, when considering an infinite chain, we can assume that the decimation step again yields an infinite chain, and the effect of the decimation is entirely captured by the renormalization of the hopping parameters. The Hamiltonian of the system after a decimation step can in principle contain new terms, but here we will focus on the cases where a decimation step leads to a Hamiltonian $H'_{LR}$ of the same form as the original $H_0$, but with renormalized hopping parameters. After a decimation step, one needs to check whether the perturbative assumption is still valid, since it may be that the renormalized hopping parameters no longer fall into such a regime. If the perturbative assumption is still present, we can iteratively repeat the decimation procedure to probe the system at low-energy. In particular, the RG flow induced by these decimation steps will drive the system into a disorder-induced fixed point, where the ground state properties of the chain can be extracted.

Having reviewed the Hamiltonian formalism of the SDRG procedure, we would like to investigate how a similar philosophy can be applied to the 
the infinite SYK chain \eqref{eq:Hamiltonian}. Upon inspection, we see that the basic setting for performing an SDRG decimation step is already given in the four-site chain considered in Sec.~\ref{sec:foursite_SYK}, with the central two sites being connected by a strong hopping, while the left- and right-most sites are only coupled to the central block via a weak hopping. However, the presence of many-body random interactions renders the Hamiltonian formalism of the SDRG procedure unapproachable, as the ground state of the central sites is extremely hard to obtain. Consequently, we are compelled to develop an alternative formalism for the SDRG step based on the SD equations in the large $N$ limit, as demonstrated in the four-site SYK chain in Sec.\,\ref{sec:foursite_SYK}. In the following, we show how to formally apply this novel approach to an infinite chain.\\
Recall that, in the large $N$ limit, the system is governed by the SD equations \eqref{eq:SD_eqs_full_form}, which are fully determined by the form of the self-energy $\Sigma$. Thus, instead of searching for an effective Hamiltonian after a decimation step, we can actually look for an effective self-energy induced by the decimation of the strong hopping parameter. However, the implementation of the SDRG has an inherent limitation in that it is only reliable at very large times. The reason for this is that the regime of validity of perturbation theory within the SDRG inherently introduces a new energy scale up to which the analysis can be trusted. This means that we can only probe energies below this new scale, $\Delta E$, which by the uncertainty principle means we can only resolve the dynamics of the system at times larger than $\frac{1}{\Delta E}$. Since $\Delta E$ is perturbatively smaller compared to the other energy scales in the system, the SDRG is expected to be valid at large time scales. Notably, however, it turns out that this large $\tau$ regime is also the regime where we can identify a renormalized self-energy of the same form as the original one, allowing us to iteratively perform the SDRG. Moreover, in Appendix~\ref{apx:freecase}, this physical intuition on the validity of the SDRG is checked in the free $\J=0$ case, where our take on the SDRG applied to the SD equations has to give the same results as the procedure based on the renormalization of the Hamiltonian \cite{MDH79prl,MDH80prb}.

Let us now apply this novel take on the SDRG procedure to provide the explicit form of a single decimation step in the model of our interest. For this, we focus on the four-site chain system studied in Sec.~\ref{sec:foursite_SYK}. Specifically, in Sec.\,\ref{subsec:setup_foursitechain}, we have realized a real-space RG decimation on the four-site system and we have found that the post-decimation dynamics are governed by the projected SD equations (\ref{eq:totalSDequation_v2}), where the self-energy is characterized by the original (non-renormalized) on-site SYK interaction \eqref{eq:Sigma J_ren} and a renormalized effective self-energy $\Sigma_{\mu}^{\rm eff}$ \eqref{eq:Sigma eff}. Exploiting (\ref{eq:sigmaeff_14}), we observe that for $\tau$ large enough, i.e. $\tau\gg \nu_{23}^{-1}$, the effective self-energy $\Sigma_{\mu}^{\rm eff}$ in \eqref{eq:sigmaeff_14} can be approximated by a Dirac-$\delta$ function
\begin{subequations}\label{eq:mu_prime}
\begin{align}
\Sigma^{\rm eff}_{\mu\,11}(\tau)
\approx -\mu_a^2(1+\Delta)^2\delta'(\nu_{23}\tau)
\,,\\
\Sigma^{\rm eff}_{\mu\,44}(\tau)
\approx -\mu_a^2(1-\Delta)^2\delta'(\nu_{23}\tau)
\,,\\
\Sigma^{\rm eff}_{\mu\, 14}(\tau)
= -\Sigma^{\rm eff}_{\mu\, 41}(\tau)
\approx -{\rm i} \mu'_{14}\delta(\tau)
\,,
\end{align}
\end{subequations}
with
\begin{equation}
\label{eq:mu_prime_expression}
 \mu_{14}'\equiv \frac{\mu_a^2(1-\Delta^2)}{2} \int_{-\infty}^{\infty} e^{-\nu_{23}\abs{\tau}}
=\frac{\mu_a^2(1-\Delta^2)}{\nu_{23}}=\frac{\mu_a^2}{\mu_b}(1-\Delta^2)\tanh\gamma_{23}
    \,.
\end{equation}
In the frequency domain, this effective self-energy reads
\begin{equation}
   \Sigma_\mu^{\rm eff}(\omega)= \frac{{\rm i}\mu_a^2}{\nu _{23}^2}\begin{pmatrix}
 (1+\Delta)^2\omega  & -(1-\Delta^2)\nu_{23} \\
 (1-\Delta^2)\nu_{23} & (1-\Delta)^2\omega
\end{pmatrix}\,.
\end{equation}
As discussed in Secs.\,\ref{subsec:largetauregime} and \ref{subsec:largetauregime_4sites}, due to \eqref{eq:sigma14 approx} and \eqref{eq:sigma11} in the large $q$ expansion, the components $\Sigma_{\J\,14}(\tau)$ and $\Sigma_{\J\,11}(\tau)$ are negligible when $\tau\gg \nu_{23}^{-1}$. By solving $\det\kc{-\mathrm i \omega - \Sigma_{\mu}^\text{eff}(\omega)}=0$, we find 
\begin{equation}
    E=-\mathrm i\omega
    = \frac{\mu_a^2 \left(1-\Delta ^2\right) \nu_{23} }{\sqrt{\mu_a^4 \left(1-\Delta ^2\right)^2+2 \mu_a^2 \left(1+\Delta ^2\right) \nu_{23} ^2+\nu_{23} ^4}}
    =\mu_{14}'+\order{\mu_a^4}\,,
\end{equation}
which, as explained in Appendix \ref{subapx:Fourier_expSigma}, controls the exponential decay of the projected Green's function $P_{14}GP_{14}\propto  e^{-E\abs{\tau}}$.
One can check that the leading expansion $E\approx\mu_{14}'$ is determined by the off-diagonal parts of the effective self-energy $\Sigma_{\mu\,14}^\text{eff}$ and $\Sigma_{\mu\,41}^\text{eff}$, and is unaffected by deleting the diagonal parts $\Sigma_{\mu\,11}^\text{eff}$ and $\Sigma_{\mu\,44}^\text{eff}$. In other words, we have $E=\mu_{14}'$ in the case of 
\begin{subequations}\label{eq:mu_prime_off}
\begin{align}
\Sigma^{\rm eff}_{\mu\,11}(\tau)
= \Sigma^{\rm eff}_{\mu\,44}(\tau)
= 0
\,,\\
\Sigma^{\rm eff}_{\mu\, 14}(\tau)
= -\Sigma^{\rm eff}_{\mu\, 41}(\tau)
\approx -{\rm i} \mu'_{14}\delta(\tau)
\,.
\end{align}
\end{subequations}
Thus, if we are only interested in the leading $\mu_a$ contribution, it suffices to consider the simplified self-energy \eqref{eq:mu_prime_off} instead of \eqref{eq:mu_prime}.

Let us stress that in the regime of validity of the SDRG, the projected SD equations (\ref{eq:totalSDequation_v2}) with \eqref{eq:mu_prime_off} after a single decimation step have the same form as before, with the same SYK coupling $\J$ but with a renormalized hopping parameter $\mu_{14}'$. Importantly, notice that $\tanh\gamma_{23}<1$, such that assuming $\mu_a\ll\mu_b$ and $\Delta^2<1$, we have that $\mu_{14}'\ll \mu_b$ and therefore the perturbative assumption still holds. We thus expect that (\ref{eq:totalSDequation_v2}) accurately describes the low-energy dynamics of the degrees of freedom on the left- and the rightmost sites of the four-site chain. Based on this result, we can expect to apply such decimation steps iteratively on an infinite chain for specific choices of inhomogeneous spatial distributions for the hopping parameters, as we will show in the next subsections. Furthermore, notice that $\tanh{\gamma_{23}}=1$ when $\J=0$, as discussed in Sec.\,\ref{subsec:matching_twosites}. In this case, and for $\Delta=0$, we see that \eqref{eq:mu_prime_expression} recovers the expression for the renormalized couplings of the disordered XX chain, which we review in Appendix~\ref{apx:SDRG}. Importantly, our results in the free case retrieve those obtained from the usual Hamiltonian approach to SDRG, providing a crucial consistency check for our SDRG application to the SD equations.

\subsection{SDRG on aperiodic SYK chains}
\label{subsec:sdrg_aperiodic}

We begin by considering a chain of interacting Majorana fermions with local random SYK interactions, governed by the Hamiltonian \eqref{eq:Hamiltonian}, where the hopping parameters $\{\mu_x\}$ are spatially inhomogeneous and are modulated according to so-called \textit{binary aperiodic sequences} \cite{Baake2013}. These are deterministically defined sequences with no finite-sized periods, meaning that no finite subsequence can be found to repeatedly appear at fixed intervals. In other words, a chain described by this sequence will not be translationally invariant. They are binary because they consist of only two letters, $a$ and $b$. Based on an aperiodic sequence, we will modulate the hopping parameters of our chain via the assignment $\mu_x=\mu_a=\hat{\mu}_a/q$ and $\mu_x=\mu_b=\hat{\mu}_b/q$ whenever the $x$-th letter of the sequence is $a$ or $b$, respectively. Note the similarities of this suggestively chosen notation with that of Sec.~\ref{sec:foursite_SYK}. We will explain these in more detail below. Then, the Hamiltonian \eqref{eq:Hamiltonian} describes a chain with \textit{spatial aperiodic disorder} in the hopping parameters, in addition to the weakly correlated random on-site interactions described in \eqref{eq:Average_SYK_coupling}. Aperiodic sequences can be constructed via iterative implementation of substitution rules, as explained in more detail in Appendix~\ref{apx:Aperiodic_Sequence}. In particular, we will focus on a specific instance of such aperiodic sequences, the so-called \textit{silver mean} sequence, whose substitution rule rule is given by
\begin{equation}
\label{eq:silvermeanrule}
 a\mapsto aba\,, 
 \qquad\qquad
 b\mapsto a\,.
\end{equation}
The infinite silver mean sequence is obtained from the iterative application of these substitution rules, and an exemplary snippet of it is given by $\dots abaabaaabaabaaa\dots$. Notably, the silver mean sequence does not exhibit consecutive $b$ letters, implying that our aperiodic SYK chain will not have strongly-coupled blocks of more than two consecutive sites.\\

We now review how the application of the SDRG to a general aperiodic chain with hopping parameters modulated by the silver mean sequence.
Under the assumption that $\mu_a\ll\mu_b$, we can identify $\mu_b$ as the strong hopping and $\mu_a$ as the weak hopping. Due to the absence of consecutive $b$ letters, all pairs of sites connected by a strong hopping are necessarily going to be connected to their right and left neighbors via weak hoppings. These four-site block chains coupled by the hopping parameters in the weak-strong-weak pattern are the fundamental objects in each step of the SDRG. We already show how to solve those four-site SYK chains in Sec.~\ref{sec:foursite_SYK}. On each of these four-site block chains, we can now perform the decimation step on their central sites simultaneously and generate a unified renormalized hopping parameter between their left and right sites, as explained in Sec.~\ref{subse:generaldecimation}. 
The decimation step effectively acts as the inverse of the substitution rule \eqref{eq:silvermeanrule}.
For this reason, it is natural to denote the renormalized hopping parameter obtained from the decimation $aba\to a$ as $\mu_a'$ and the remaining hopping parameter $\mu_a$ as $\mu'_b$ due to the second rule in \eqref{eq:silvermeanrule}. Due to the self-similarity of aperiodic sequences (cf.~Appendix \ref{apx:Aperiodic_Sequence}), the sequence of $\ke{\mu_a',\mu_b'}$ after the decimation is still a silver mean sequence and, as we show later in this section, the hierarchy $\mu_a'\ll\mu_b'$ is maintained after the decimation. Thus, we can iteratively apply the decimation steps and obtain new generations of the hopping parameters. Formally, the hopping parameters on the $n$-th and $(n+1)$-th step are related by
\begin{equation}\label{eq:mu_prime_general}
\mu_a^{(n+1)}=\mu_a^{(n)} R_a^{(n)}\kc{\frac{\mu_a^{(n)}}{\mu_b^{(n)}}},\qquad 
\mu_b^{(n+1)}=\mu_a^{(n)}\,,
\end{equation}
where the function $R_a^{(n)}(r)$ should be determined by the renormalization of the coupling at each decimation step. For the XX chain or the Majorana chain with $\J=0$, $R_a^{(n)}(r)=r$ \cite{Luck1993,HermissonGrimm97,Hermisson00,Vieira:2005PRB,vieira05}. We will provide  the function $R_a^{(n)}(r)$ for the SYK chain (\ref{eq:Hamiltonian}) in \eqref{eq:mu_prime_summary}.

\begin{figure}
    \centering
    \includegraphics[width=\textwidth]{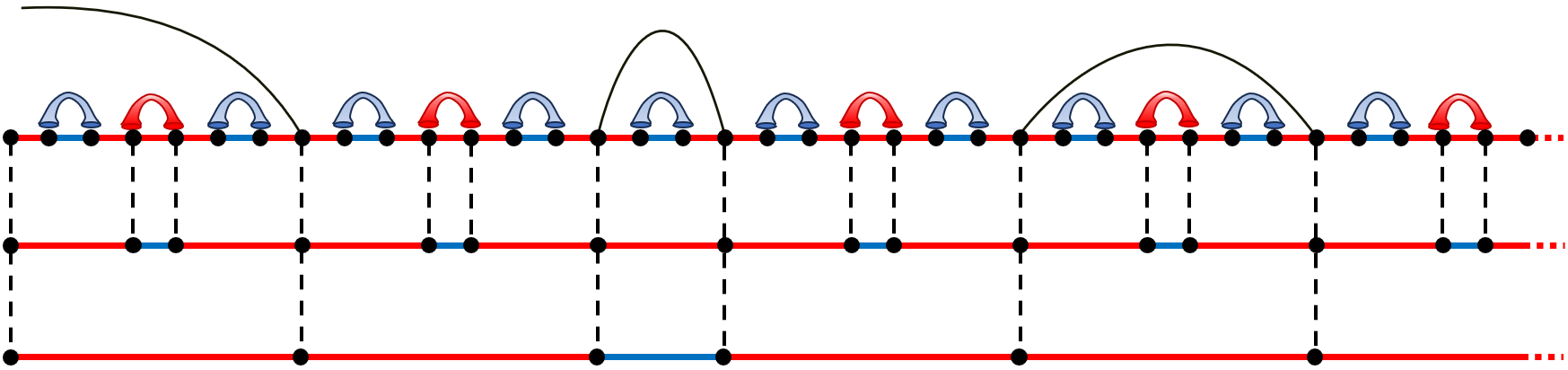}
    \caption{The schematic structure of the ground state of the aperiodic SYK chain, as given by \eqref{eq:factorised GS}, is shown in the top line. Contributions of the form $|{\rm ETW}_{s}(\mu_j)\rangle$ are depicted as wormholes, with blue and red ones corresponding to the $\mu_b$ and $\mu_a$ hoppings, respectively. These are obtained from the first two decimation steps in the SDRG, which we visualize in this finite portion of the infinite chain in the second and third lines, respectively. Non-wormhole contributions $|\Psi_s\rangle$ are shown as simple solid lines, and they arise from the third decimation step onward. The vertical dashed lines are a visual guide to those sites that are not decimated on a given SDRG step.}
    \label{fig:SYKgroundstate}
\end{figure}

Both the coherent time, {\it i.e.} the inverse of the gap, and the correlation length will grow along the SDRG. On the one hand, recall that this renormalized hopping can only be identified in the limit of large $\tau\gg q/\hat{\nu}_{23}$, with $\hat{\nu}_{23}$ defined in \eqref{eq: matching conditions 1}. For subsequent decimation steps, we will see below that this validity regime for $\tau$ will change throughout the SDRG as $\tau\gg 1/\mu_b^{(n)}$.
On the other hand, due to (\ref{eq:transcendental_eq_gamma23}), the renormalization of the couplings at each decimation step depends on the distance $\abs{x-y}$ between the sites which are connected by a strong coupling at that stage in the SDRG. For the aperiodic sequence generated by \eqref{eq:silvermeanrule}, the only decimation happen between the sites $x$ and $y$ with odd distance $\abs{x-y}$ so that $s_{xy}=\mathrm i^qe^{-|x-y|^2/\xi^2}$. The parameter $\log \kappa_{23}$ appearing in \eqref{eq:transcendental_eq_gamma23} generalizes to $\log \kappa_{xy}=-\frac{\abs{x-y}^2}{\xi^2}$, which explicitly accounts for the distance between decimated spins due to the correlation \eqref{eq:Average_SYK_coupling}. Initially, this distance is $\abs{x-y}=\eta$, where $\eta$ is an $\order{1}$ lattice constant, which can effectively amounts to a rescaling of the correlation length. As discussed later, this distance will generically grow as more and more sites are decimated throughout the SDRG.
For aperiodic chains with aperiodicity generated by the rule (\ref{eq:silvermeanrule}), the asymptotic distance between such sites after $n-1\gg1$ SDRG steps has been derived in \cite{Juh_sz_2007,Basteiro:2022zur} and reads
\begin{equation}
    \abs{x-y}=\eta(\lambda_+)^{\left\lfloor \frac{n-1}{2}\right\rfloor}\,,
    \label{eq:distance_aperiodic}
\end{equation}
where $\lfloor \cdot\rfloor$ denotes the floor function and $\lambda_+=1+\sqrt{2}>1$ is determined by the properties of the chosen aperiodic sequence, as we explain in detail in Appendix~\ref{apx:Aperiodic_Sequence}. For the purposes of the current analysis, it is sufficient to account for the scaling behavior \eqref{eq:distance_aperiodic} for any value of $n$. Due to the floor function, the distance between two decimated sites is the same if we consider the $2k-1$-th and  $2k$-th steps. This is a property of the SDRG procedure on aperiodic chains with disorder induced by the silver mean sequence. Indeed, as we can see from \eqref{eq:mu_prime_general}, one of the two post-decimation hopping parameters is not renormalized and is inherited from the pre-decimation chain. Given that this hopping parameter is the strong coupling in the following step, the distance $|x-y|$ remains the same for one more SDRG decimation. This effect is pictorially represented in Fig.\,\ref{fig:SYKgroundstate} for the first SDRG steps. In the remaining part of this section, we analyse the function $R_a^{(n)}(r)$ at each decimation step.\\

{\bf First step:} After the first decimation step, the remaining sites are coupled by one of two renormalized hoppings, namely 
\begin{equation}
    \mu_1'=\frac{\mu_a^2}{\mu_b}\tanh\gamma_{23}\quad \textrm{or}\quad \mu_2'=\mu_a\,,
  \label{eq:mu_prime_main_text}
\end{equation}
where $\gamma_{23}$ is the solution of (\ref{eq:transcendental_eq_gamma23}).
Notice how we momentarily switch labels since we do not know, a priori, which hopping is the strongest of both. We must therefore perform a check of our perturbative assumption, by noticing that
\begin{equation}
  \frac{\mu_1'}{\mu_2'}= 
  \frac{\mu_a}{\mu_b}\tanh\gamma_{23}<1\,.
\end{equation}
Thus, $\mu_a'=\frac{\mu_a^2}{\mu_b}\tanh\gamma_{23}$ and $\mu_b'=\mu_a$. If we define $r'\equiv \mu_a'/\mu_b'$, we have that $r'<r\ll 1$, meaning that this first SDRG step preserves, and actually strengthens, the initial perturbative assumption.\\

{\bf Second step:} To continue our analysis, we explicitly consider the renormalization of the couplings after two SDRG steps ($n=2$) . In this case, the decimated sites will be those coupled by the hopping $\mu_b'=\mu_a$. This is also shown in Fig.~\ref{fig:SYKgroundstate} when going from the second to the third line. 
After this second decimation step, we can again identify a renormalized self-energy only at $\tau\gg \tanh\gamma'_{23}/\mu'_b\equiv 1/\nu'_{23}$, where the parameter $\gamma_{23}'$ is given by
\eqref{eq:transcendental_eq_gamma23} 
evaluated with $\mu_b'$ instead of $\mu_b$, namely
\begin{equation}
\label{eq:trasc_eq_gamma_secondstep}
   \sinh\left(\gamma_{23}'-\frac{\eta}{\xi^2}\right)\tanh\left(\gamma_{23}'\right)=\frac{\mu_b'}{2\J} \,.
\end{equation}
Notice that, consistently with (\ref{eq:distance_aperiodic}) and the subsequent discussion, the distance between the decimated sites in this second SDRG step is $|x-y|=\eta$, i.e. the same considered for the first decimation. After this time scale, the renormalized self-energy is characterized by the renormalized hoppings
\begin{equation}
    \mu_b''=\frac{\mu_a^2}{\mu_b}\tanh\gamma_{23}\,,
    \qquad\qquad
    \mu_a''=\frac{\mu_a^3}{\mu_b^2}(\tanh\gamma_{23})^2\tanh\gamma_{23}'\,,
\end{equation}
 with a renormalized ratio given by $r''=\frac{\mu_a}{\mu_b}\tanh\gamma_{23}\tanh\gamma_{23}'$. 
\\

{\bf Third step:}
As seen for the previous two, also at the third SDRG step, the weak hopping parameter $\mu_a''$ becomes the strong one in the next generation, namely
\begin{equation}
\label{eq:mu_b3}
  \mu_b^{(3)}= \mu_a''=\frac{\mu_a^3}{\mu_b^2}(\tanh\gamma_{23})^2\tanh\gamma_{23}'  \,.
\end{equation}
To understand how to renormalize the hopping parameter after the decimation of the sites connected by $\mu_b''$, let us observe that the correlations within the pairs of sites that we aim at decimating are given by $G_{14}$, whose expression is given in (\ref{eq:G14_simplified}). Following the procedure described in Sec.\,\ref{subsec:setup_foursitechain}, to determine the effective self-energy accounting for the decimation of the strongly coupled sites, we have to plug $G_{14}$ (instead of $G^{(0)}_{23}$) into (\ref{eq:effSigma component}) with $\Delta=0$. Given that, at this step, the role of the weak hopping parameter is played by $\mu_a''$, the effective self-energy
$\Sigma_{\mu, {\rm dec}}^{\eff}$ across the bond to decimate reads
\begin{equation}
\label{eq:sigmamueff_nonMQ}
    \Sigma_{\mu, {\rm dec}}^{\eff}=-\frac{(\mu_a'')^2}{q^2}G_{14}\,,
\end{equation}
where $G_{14}$ is given by (\ref{eq:G14_simplified}).
The SDRG is valid at time scales $\tau\gg \nu_{23}/\mu_a^2=1/\mu_b''$. At this time scale, using the argument presented in Sec.\,\ref{subsec:largetauregime}, (\ref{eq:sigmamueff_nonMQ}) can be written as
\be
 \Sigma_{\mu, {\rm dec}}^{\eff}=-{\rm i}
 \frac{\mu^{(3)}_a}{q}
 \delta(\tau)
\,,
\ee
where the renormalized coupling reads
\be
\label{eq:mu_a3}
\mu_a^{(3)}=-2{\rm i}\frac{(\mu_a'')^2}{q}\int_0^\infty G_{14}d\tau
\simeq
\frac{(\mu_a'')^2}{q}
\int_0^{q\frac{\hat{\nu}_{23}}{\mu_a^2}}\left(1-2 \frac{\mu_a^2}{\hat{\nu}_{23}^2}\right)d\tau
\simeq 
\frac{\mu_a^4}{\mu_b^3}(\tanh\gamma_{23})^3(\tanh\gamma_{23}')^2\,,
\ee
and we have exploited that the solution $G_{14}$ decays quickly to zero as $\tau\gg \nu_{23}/\mu_a^2$.
Notice that the contribution from $G_{14}$ in (\ref{eq:mu_a3}) has to be evaluated at leading order for small values of the hopping ratio, given that the formula for $\Sigma_\mu^{\rm eff}$ has been derived at order $r^2$ and cannot predict higher orders. At this step, the ratio between the renormalized couplings reads
\begin{equation}
    r^{(3)}=
\frac{\mu_a^{(3)}}{\mu_b^{(3)}}
    =
    \frac{\mu_a}{\mu_b}\tanh\gamma_{23}\tanh\gamma_{23}'<r\ll 1\,.
\end{equation}
Given that the hopping parameter ratio remains within the perturbative assumptions, we can proceed with the SDRG procedure.
\\

{\bf $n$-th step, with $n>2$:}
Extending what already observed for the first three SDRG steps, after $n-1$ steps with $n>2$, the weak hopping parameter $\mu_a^{(n-1)}$ becomes the strong one in the $n$-th
generation, namely
\begin{equation}
\label{eq:mu_bn}
\mu_b^{(n)}=\mu_a^{(n-1)}\,.
\end{equation}
On the other hand, the sites connected by the strong hopping parameter $\mu_b^{(n-1)}$ after $n-1$ steps have to be decimated. 
The crucial observation is that all the hopping parameters remaining after two SDRG steps come from at least one renormalization procedure. In other words, no hopping parameters from the original chain are left at this point of the SDRG flow. This means that all the following steps involve decimations of sites with two-point correlations generically different from the one reviewed in Sec.\,\ref{subsec:two-sitechain}. 
To understand which is the corresponding effect on the SDRG flow of the hopping parameters along the chain, let us call $G_{\rm dec}$ the non-diagonal entries of the two-point functions connecting the two sites we want to decimate, i.e. the sites coupled by $\mu_b^{(n-1)}$. As seen in Sec.\,\ref{subsec:matchingsolution_four sites} for the entry $G_{14}$ in the context of the four-site chain, $G_{\rm dec}$ behaves perturbatively as
$G_{\rm dec}\simeq \frac{{\rm i}}{2}+\order{(r^{(n-1)})^2}$, where $r^{(n-1)}\equiv \mu_a^{(n-1)}/\mu_b^{(n-1)}$ is the hopping ratio after $n-1$ steps.
Thus, the renormalization works as in the third step by changing the index $3\to n$. More precisely,
at times scales $\tau\gg 1/\mu_{b}^{(n-1)}$,
the renormalized coupling $\mu_a^{(n)}$ obtained at the $n$-th step is given by
\begin{equation}
\label{eq:mu_an}
  \mu_a^{(n)} 
  =
  -2{\rm i}\frac{\left(\mu_a^{(n-1)}\right)^2}{q}\int_0^\infty G_{\rm dec}d\tau
\simeq
\frac{\left(\mu_a^{(n-1)}\right)^2}{q}
\int_0^{\frac{q}{\mu^{(n-1)}_b}}d\tau
=
\frac{\left(\mu_a^{(n-1)}\right)^2}{\mu_b^{(n-1)}}
\,.
\end{equation}
Using (\ref{eq:mu_bn}) and (\ref{eq:mu_an}), the hopping ratio $r^{(n)}$ after the $n$-th step reads
\begin{equation}
    r^{(n)}=\frac{\mu^{(n)}_a}{\mu^{(n)}_b}=\frac{\mu_a^{(n-1)}}{\mu_b^{(n-1)}}\,,
\end{equation}
for any $n>2$.
We can conveniently express these quantities in terms of the parameters in the original chain. When $n>2$, we have
\begin{equation}\begin{split}
\mu_b^{(n)}=\frac{\mu_a^{n}}{\mu_b^{n-1}}(\tanh\gamma_{23})^{n-1}(\tanh\gamma_{23}')^{n-2}\,,\\
    \mu_a^{(n)} =
\frac{\mu_a^{n+1}}{\mu_b^{n}}(\tanh\gamma_{23})^{n}(\tanh\gamma_{23}')^{n-1}
\,,
\end{split}\end{equation}
which leads to
\begin{equation}
r^{(n)}=\frac{\mu_a}{\mu_b}\tanh\gamma_{23}\tanh\gamma_{23}'
=
r \tanh\gamma_{23}\tanh\gamma_{23}'< r\ll 1\,.
\label{eq:fixed_point_r}
\end{equation}
The expression on the right-hand side of (\ref{eq:fixed_point_r}) implies that the hopping parameter ratio $r^{(n)}$ remains within the perturbative assumptions for any $n>2$ and, thus, we can iterate the SDRG procedure infinitely many times. 
\\

{\bf Summary and analysis of the fixed points:}
We summarize the renormalization rules of the hopping parameters $\mu_a^{(n)}$ and $\mu_b^{(n)}$ at the generic $n$-th SDRG step and their ratio $r^{(n)}$, which read
\begin{equation}\label{eq:mu_prime_summary}\begin{split}
   \mu_b^{(n)} =\mu_a^{(n-1)}  \,,\\
    \mu_a^{(n)} =\frac{\left(\mu_a^{(n-1)}\right)^2}{\mu_b^{(n-1)}}e^{-\zeta_n} 
    \,,
\end{split}\end{equation}
and
\begin{equation}
\label{eq:ratio_n_summary}
 r^{(n)}= \frac{\mu_a^{(n-1)}}{\mu_b^{(n-1)}}e^{-\zeta_n}\,,
\end{equation}
where
\begin{equation}
    e^{-\zeta_n}=
    \begin{cases}
       \tanh\gamma_{23} & \;\;{\rm if}\;\;n=1\,,
       \\
   \tanh\gamma_{23}' & \;\;{\rm if}\;\;n=2\,,
       \\
       1 & \;\;{\rm if}\;\;n>2\,.
    \end{cases}
\end{equation}
Notice that here we are adopting the convention according to which $\mu_a^{(0)}=\mu_a$ and $\mu_b^{(0)}=\mu_b$. 
Expressing the ratio (\ref{eq:ratio_n_summary}) in terms of the parameters of the original infinite chain, we have
\begin{equation}
\label{eq:ratio_n_originalchain}
   r^{(n)}=r\tanh\gamma_{23}\times\begin{cases}
   1
        & \;\;{\rm if}\;\;n=1\,,
       \\
  \tanh\gamma_{23}' & \;\;{\rm if}\;\;n>1\,.
    \end{cases} 
\end{equation}
Let us now discuss the behavior of the renormalized hopping ratio along the SDRG flow and, in particular, when $n\to\infty$, looking for aperiodicity-induced fixed points. The dependence of $r^{(n)}$ in (\ref{eq:ratio_n_originalchain}) on $n$  shows that the hopping ratio saturates to a constant value $r^*=r\tanh\gamma_{23}\tanh\gamma'_{23}$ after only two SDRG steps. Since its values stay constant for any $n$, and, therefore, for $n\to\infty$, $r^*$ is the SDRG fixed point value of the hopping ratio. Notice that this fixed point depends explicitly on the ratio $r$ of the original chain.
This means that the properties of the ground state in the IR depend on the chosen value of $r$. In this case, the aperiodic modulation is said to be\textit{marginal} \cite{Vieira:2005PRB}. 
The fixed point value of the hopping ratio depends only implicitly on the correlation length $\xi$ defined in (\ref{eq:Average_SYK_coupling}) via the parameters $\gamma_{23}$ and $\gamma'_{23}$. It is anyway insightful to notice that in the limit of totally correlated random on-site interactions, i.e. $\xi\to\infty$, $\tanh\gamma_{23}$ is given by (\ref{eq:tanh_gamma_23}). The same solution can be used for $\tanh\gamma_{23}'$, modulo the substitution $\mu_b\to\mu_b'=\mu_a$. In other words, in the regime $\xi\to\infty$, we can obtain an explicit expression of the fixed point hopping ratio $r^*$ in terms of the parameters of the original chain.

We find it particularly insightful to consider the limit $\J\to 0$. 
As stressed in Sec.\,\ref{subsec:themodel}, in this limit, the infinite chain with Hamiltonian (\ref{eq:Hamiltonian}) reduces to $N/2$ decoupled XX chains, namely to a free model. In this regime, (\ref{eq:transcendental_eq_gamma23}) has as a solution $\gamma_{23}\to\infty$, i.e. $\tanh\gamma_{23}\to 1$. The same conclusion is obtained from (\ref{eq:trasc_eq_gamma_secondstep}), i.e. $\tanh\gamma'_{23}\to 1$ as $\J/\hat{\mu}'_b\to 0$. Thus, (\ref{eq:ratio_n_originalchain}) becomes $r^{(n)}=r$ for any positive integer $n$. This SDRG flow for the hopping parameter ratio is the same found for aperiodic XX chains with the disorder generated by (\ref{eq:silvermeanrule}) \cite{Hermisson00,Vieira:2005PRB}. This finding provides a consistency check of our SDRG procedure given that it reduces to an existing result in the limit where the model becomes approachable via other methods (see Appendix \ref{apx:SDRG} for a review). According to the discussion above, in both the free model and the model with $\J\neq 0$, the modulation induced by the hopping parameters distributed following the silver mean sequence is marginal. 
As a consequence, the IR physics depends on the hopping ratio $r$ in the original chain. However, while in the free case $r^*=r$, in the presence of on-site random interactions, $r^*<r$, with the prefactor being dependent on the interaction strength $\J$. This means that the on-site random interactions change the properties of the line of fixed points parameterized by $r$, but do not affect the system so strongly to generate a new aperiodicity-induced fixed point independent of the original hopping ratio. This analysis and the SDRG marginality on the model (\ref{eq:Hamiltonian}) with hopping parameters distributed following the silver mean sequence constitute one of the main results of this work.

Let us comment on the structure of the ground state of the system at the SDRG fixed point described by $r^*$, which governs the IR physics of the system.
Given that $r^*\propto r$, only when $r\ll 1$ in the original chain we can make quantitative predictions on the ground state structure. Indeed, in this case, the large unbalance of the hopping parameters is enhanced by the SDRG flow, leading to the following factorised approximation for the ground state
\begin{equation}
\label{eq:factorised GS}
    |{\rm GS}\rangle=\bigotimes_{s_1\in G_1} |{\rm ETW}_{s_1}(\mu_b)\rangle \bigotimes_{s_2\in G_2} |{\rm ETW}_{s_2}(\mu_a)\rangle\bigotimes_{s\in G_{i>2}}|\Psi_s\rangle\,.
\end{equation}
Here, $|{\rm ETW}_{s}(\mu_j)\rangle$ are the ground states of two SYK clusters coupled by a hopping term with parameter $\mu_j$ \cite{Maldacena:2018lmt} located on pairs of sites labeled by $s$. Thus, each state $|{\rm ETW}_{s}(\mu_j)\rangle$ is the same as the one on the central sites of the four-site chain  with its corresponding hopping parameter $\mu_j$ and is also dual to an eternal traversable wormhole (see the correlation functions \eqref{eq:G23}). Let us stress that these are obtained exclusively from the two-site ground states decimated in the first $G_1$ and second $G_2$ generations in the SDRG.
The remaining contributions $|\Psi_s\rangle$ are complicated states in the Hilbert spaces attached to the sites decimated after the second SDRG step, with $G_i$ denoting the set of pairs of sites decimated at the $i$-th SDRG step. We expect that each state $\ket{\Psi_s}$ connecting the pair $s$ is close to the two-site state connecting the left- and rightmost sites of the four-site chain discussed in Sec.\,\ref{subsec:matchingsolution_four sites}. In particular, we expect that the correlations in $\ket{\Psi_s}$ are approximately given by \eqref{eq:G14} where the hopping parameters are substituted with the renormalized ones located between the pair $s$. Thus, each pair $s$ behaves as a holographic black hole on each site but with non-trivial and non-geometrical off-site correlations (see \eqref{eq:G14}).

A schematic depiction of this ground state structure is shown in Fig.~\ref{fig:SYKgroundstate} for an aperiodic distribution of the couplings following the silver-mean sequence.
In fact, we can exploit the letter statistics of the silver mean sequence to better characterize, in relative terms, what percentage of the ground state will consist of wormhole-like contributions, as we show in detail in Appendix~\ref{apx:Aperiodic_Sequence}. We find that the $83\%$ of the sites of the chain is pairwise coupled into the states $|{\rm ETW}_{s}(\mu_j)\rangle$, and therefore the vast majority of the factors in (\ref{eq:factorised GS}) are known.
Finally, we find it worth stressing that, due to the marginality of the aperiodic disorder, the approximation (\ref{eq:factorised GS}) for the actual ground state improves as the initial hopping ratio $r$ becomes smaller. This approximate structure of the ground state in the perturbative regime is another central finding of this manuscript.

The analysis presented above lends itself to drawing two main conclusions.
First, the introduction of on-site random disorder in the form of local SYK terms in \eqref{eq:Hamiltonian} has a significant physical consequence, modifying the properties of the SDRG fixed points. In particular, although the marginality of the modulation induced by the silver mean distribution of the hopping parameters is preserved, the value of the fixed point hopping ratio becomes smaller than the original ratio, i.e. $r^*<r$. Thus, the on-site random interactions modify the critical properties of the model compared to the free case but are not strong enough to drive the system into a new strong-disorder fixed point independent of the initial hopping ratio.
This finding is physically due to the fact that the on-site SYK terms in the Hamiltonian (\ref{eq:Hamiltonian}) affects the physics of the systems at short times, i.e. in the UV, while the SDRG captures the features of the system at low-energies, i.e. in the IR.
Second, at the SDRG fixed point, if the hopping ratio $r$ in the original chain is small enough, the ground state is well approximated by the factorised structure in \eqref{eq:factorised GS}. Although we cannot report the exact form of all its constituents, we can indeed identify the contributions to \eqref{eq:factorised GS} coming from the sites decimated in the first two steps of the SDRG since we have shown that these correspond to wormhole-like states $|{\rm ETW}_{s}(\mu_j)\rangle$, with $j=a,b$, as discussed in \cite{Maldacena:2018lmt}. For the remaining contributions in \eqref{eq:factorised GS}, denoted by $|\Psi_s\rangle$, we do not have an explicit form. Their on-site correlations suggest a description in term of black holes, while their off-site correlations do not have a geometrical description.

\subsection{SDRG on SYK chains with random hopping parameters}
\label{subsec:random_SDRG}

In this subsection, we consider another instance of an infinite inhomogeneous chain with Hamiltonian (\ref{eq:Hamiltonian}) characterized by on-site SYK interactions. More precisely, we consider the case where each hopping parameter $\mu_x>0$ with $x\in\mathds{Z}$ along the chain is independently drawn from a probability distribution $P(\mu_x)$. As we will comment more in what follows, the shape of the function $P(\mu_x)$ can, a priori, be taken to be quite generic. Later, however, we will elucidate certain conditions under which the shape of $P(\mu_x)$ is compliant with our framework. 
We find it worth noticing that, in this class of chains, there are two types of randomness. One is the on-site random SYK-like interaction, while the other is due to the randomly distributed hopping between nearest-neighbor sites. As mentioned in Sec.\,\ref{subsec:SDeqLargeN}, the on-site SYK disorder is integrated out in the path integral to obtain the large $N$ description of the model in terms of the SD equations \eqref{eq:SD_eqs_full_form}. In contrast, in the following analysis we consider a single realization of the disorder on the hopping parameters rather than integrating it out. Since the distribution of hopping parameters is locally independent, for a fixed realization, the probability density function of the hopping parameters along the whole chain is also given by the distribution $P(\mu)$.

\subsubsection{SDRG framework in random systems}
\label{eq:SDRGrandom}

Quantum chains with random hopping parameters can be treated with an SDRG approach, as originally done in \cite{MDH79prl,MDH80prb} (see also \cite{Fisher94SDRG}). Our goal is to extend this analysis to large $N$ chains with on-site SYK interaction, applying the SDRG method to the SD equations. For this purpose, we begin by identifying the largest among hopping parameters along the chain drawn from the distribution $P(\mu_x)$, namely $\mu_0\equiv\max_{x} \mu_x$. Notice that all hopping parameters in this section will refer to the unhatted quantities, which are related to the hatted ones used in Sec.~\ref{subsec:sdrg_aperiodic} by $\mu_x=\hat{\mu}_x/q$. We choose the former to avoid clutter, but all following arguments hold also for the hatted quantities. In the spirit of the SDRG, we work in a perturbative regime and we therefore assume that hopping parameters $\mu_{\textrm{l}}$ and $\mu_{\textrm{r}}$ neighboring $\mu_0$ satisfy $\mu_{\textrm{l}},\, \mu_{\textrm{r}}\ll \mu_0 $. The validity of this perturbative assumption will be fulfilled more or less easily according to the properties of the initial hopping distribution $P(\mu_x)$. In particular, we expect the perturbative condition to hold more easily if $P(\mu_x)$ is broad as a distribution. As discussed in Sec.\,\ref{subse:generaldecimation}, if $\mu_{\textrm{l}}, \mu_{\textrm{r}}\ll \mu_0 $ is satisfied, we can decimate the two set of degrees of freedom coupled by the strongest hopping parameter $\mu_0$. The decimation corresponds to the procedure described at the end of Sec.\,\ref{subsec:setup_foursitechain} and leads to the renormalization of the couplings discussed in Sec.\,\ref{subse:generaldecimation} and given by (see (\ref{eq:mu_prime})-(\ref{eq:mu_prime_expression}))
\begin{equation}   
\mu'=\frac{\mu_{\textrm{l}}\mu_{\textrm{r}}}{\mu_0}
   e^{-\zeta}\,,
\qquad\quad
\zeta\equiv -\ln\tanh \gamma_{23}\,,
   \label{eq:mu_prime_random}
\end{equation}
where we have the identifications with the previous notation as $\mu_b=\mu_0$, $\mu_{\textrm{l}}=(1+\Delta)\mu_a$, $\mu_{\textrm{r}}=\mu_a(1-\Delta)$ and $\gamma_{23}$ is the solution of (\ref{eq:transcendental_eq_gamma23}) with $\mu_b=\mu_0$.
The expression of $\zeta$ in (\ref{eq:mu_prime_random}) crucially depends on $\mu_0$ via $\gamma_{23}$ given that the first decimation necessarily involves sites which are nearest neighbors and, therefore, the computation of Sec.\,\ref{subse:generaldecimation} applies straightforwardly to this case.

To study how the decimation has modified the properties of the remaining degrees of freedom, we denote the hopping parameters along the post-decimation chain as $\{\mu'_x\}_{x\in\mathds{Z}}$, where we have removed $\mu_0$ from the pre-decimation hoppings $\{\mu_x\}_{x\in\mathds{Z}}$, we have replaced $\mu_{\textrm{l}}$ and $\mu_{\textrm{r}}$ with $\mu'$ and we have shifted the space labels $x$. Also in this example, if we decimate degrees of freedom from an infinite chain, we still obtain an infinite chain where some of the hopping parameters have been renormalized. To implement the second step of the RG procedure, given the set of new hopping parameters $\{\mu'_x\}_{x\in\mathds{Z}}$, we again identify the largest of them, $\mu_0'$, and, if the perturbative assumptions are still valid, i.e. $\mu'_{\textrm{l}}, \mu'_{\textrm{r}}\ll \mu'_0$, we can decimate the degrees of freedom coupled by $\mu_0'$, thus obtaining a set of renormalized hoppings $\{\mu''_x\}_{x\in\mathds{Z}}$ along the chain. Provided that at any step the perturbative assumptions are valid, we can iterate the decimation, implementing in this way the SDRG on the original chain. 

At this point, an important remark is in order. At the beginning of the SDRG procedure, the decimations involve sites that are nearest neighbours in the original chain. Deep enough in the IR direction of the SDRG flow, the strongest bonds along the chain connect sites that are no longer nearest neighbors in the original chain. Thus, after this threshold in the number of decimation steps (whose value does not alter the conclusion of our analysis), all the decimated sites lead to a renormalization of the coupling that parallels the computation shown in (\ref{eq:mu_a3}) and (\ref{eq:mu_an}). In other words, in the deep IR, the parameter $\zeta$ introduced in (\ref{eq:mu_prime_random}) becomes zero, and the hopping renormalization reduces to the same found for random XX chains \cite{Fisher94SDRG}.
This observation will be central to the forthcoming analysis.

\subsubsection{Distribution flow equations}

The decimation and the renormalization of the couplings have the effect of modifying the probability distribution of the hopping parameters along the post-decimation chain. After any RG step, the probability distribution changes. Following \cite{MDH79prl,MDH80prb,Fisher94SDRG}, we assume that the new distribution generated by the decimation remains locally independent. To label the RG step, we effectively introduce a dependence of $P$ on $\tilde{\mu}_0$, i.e. the strongest hopping parameter at a given step. The latter establishes the next-to-come decimation along the RG procedure. Thus, the distribution is denoted as $P(\mu_x;\tilde \mu_0)$.
The iterated decimations effectively induce a flow of the probability distribution according to an equation of the form
\begin{equation}
\label{eq:generalflowP}
    \frac{d P(\mu_x;\tilde{\mu}_0)}{d\tilde{\mu}_0}=R[P]\,,
\end{equation}
where $R$ is a suitable functional accounting for the effect of the SDRG transformation on $P$ and $\tilde{\mu}_0$ is now to be seen as setting the energy scale of the renormalization procedure. 

In the infinite quantum chain we consider in this section, the decimation steps have the neat effect of only renormalizing the hopping parameters, leaving all the other parameters, i.e.~the SYK couplings, unrenormalized.
The expression of the functional $R$ in (\ref{eq:generalflowP}) for this kind of systems was derived in \cite{MDH79prl,MDH80prb,Fisher94SDRG}. Following the convention in the previous literature, we introduce logarithmic energy scales for convenience, given by
\begin{equation}
\label{eq:logenergy}
 \beta\equiv \ln\frac{\tilde{\mu}_0}{\mu_x}\geq0  \,,
  \qquad
    \Gamma\equiv \ln\frac{\mu_0}{\tilde{\mu}_0}\geq0
\,.
\end{equation}
The notation $\beta$ for this quantity should not be confused with the one for the inverse temperature in the previous sections.
Upon a slight modification of the results in \cite{MDH79prl,MDH80prb,Fisher94SDRG}, which takes into account the specific detail of our model, we derive the following flow equation for the distribution of the logarithmic energy scales
\begin{equation}
\label{eq:RGequationP_XXX}
    \frac{\partial P(\beta;\Gamma)}{\partial\Gamma}=\frac{\partial P(\beta;\Gamma)}{\partial\beta}
    +
    P(0;\Gamma)\int_0^\infty d \beta_{\rm l} d \beta_{\rm r}\delta(\beta_{\rm l}+\beta_{\rm r}+\zeta(\Gamma)-\beta)P(\beta_{\rm l};\Gamma)P(\beta_{\rm r};\Gamma)\,.
\end{equation}
We review the derivation of the above equations in Appendix~\ref{apx:FlowEq}, where we highlight the appearance of $\zeta(\Gamma)$ inside the Delta function, in contrast to \cite{MDH79prl,MDH80prb,Fisher94SDRG} where $\zeta(\Gamma)=\textrm{const}$. The parameter $\mu_0$ in (\ref{eq:logenergy}) refers to the largest hopping in the original chain, in contrast to $\tilde{\mu}_0$, which runs with the RG flow. Accordingly, the logarithmic scale in \eqref{eq:logenergy} which plays the role of this RG coordinate is $\Gamma$. The introduction of logarithmic variables is convenient since they simplify the multiplicative renormalization rule \eqref{eq:mu_prime_random} to a more manageable additive one. Notice that the probability conservation $\int_0^\infty P(\beta;\Gamma)d\beta=1$ is guaranteed by the flow equation.\\
Let us explain the interpretation of the two terms in the right hand side of the flow equation \eqref{eq:RGequationP_XXX}. The second term embodies the contribution of two smaller couplings $\beta_1,\beta_2$ into one larger coupling $\beta>\beta_1+\zeta(\Gamma),\,\beta_2+\zeta(\Gamma)$, imposing \eqref{eq:mu_prime_random} in terms of logarithmic variables.
The first term then adjusts the form of the distribution by a left shift along $\beta$. Visualizing the effect of this mechanism on a distribution which is a monotonically decreasing function of $\beta$ makes manifest that it will become broader as $\Gamma$ increases. 

As anticipated in Sec.\,\ref{eq:SDRGrandom}, at the beginning of the SDRG procedure, the decimations involve sites that are nearest neighbors in the original infinite chain. This implies that the factor $e^{-\zeta}$ in (\ref{eq:mu_prime_random}) depends on $\tilde{\mu}_0$ and, via (\ref{eq:logenergy}), on $\Gamma$. On the other hand, from a certain threshold of SDRG coordinate, say $\Gamma>\Gamma_{\textrm{t}}$, $e^{-\zeta}=1$ since none of the decimated sites at those scales are nearest neighbors in the original chain anymore. In summary, we can write
\begin{equation}\label{eq:logfactor}
   \zeta(\Gamma)= -\ln\tanh [\gamma_{23}(\tilde{\mu}_0)]\, \Theta(\Gamma_{\textrm{t}}-\Gamma)\geq0 \,,
\end{equation}
where we have made explicit the dependence of $\gamma_{23}$ on $\tilde{\mu}_0$ which is given by solving (\ref{eq:transcendental_eq_gamma23}) with $\mu_b=\tilde{\mu}_0$ and $\Theta$ denotes the Heaviside step function.
Crucially, for $\Gamma> \Gamma_{\textrm{t}}$, $\zeta$ ceases to depend on the RG coordinate and vanishes. This implies that, since we are interested in evaluating the flow equation (\ref{eq:RGequationP_XXX}) deep in the IR along the SDRG, we can focus on the value of  $\zeta$ for $\Gamma> \Gamma_{\textrm{t}}$.
In this regime, $\zeta(\Gamma)= 0$, and the flow equation (\ref{eq:RGequationP_XXX}) becomes the one discussed in \cite{Fisher94SDRG} for the XX chain with random bond disorder. Importantly, this argument does not rely on the explicit value of $\Gamma_{\textrm{t}}$ (that we do not know) but only on its existence.
Notice that $\zeta(\Gamma)=0$ is also obtained when $\J\to 0$, which indeed corresponds to the free case where the Hamiltonian (\ref{eq:Hamiltonian}) reduces to $N/2$ decoupled XX chains.
A fixed point solution of (\ref{eq:RGequationP_XXX}) with $\zeta(\Gamma)=0$ was found in \cite{MDH79prl,MDH80prb,Fisher94SDRG}
and reads
\begin{equation}
    \label{eq:freeFPdistr}
    P_*(\beta;\Gamma)=\frac{1}{\Gamma}e^{-\frac{\beta}{\Gamma}}\,.
\end{equation}
Crucially, the distribution $P_*^{(0)}(\beta;\Gamma)$ becomes flat as a function of $\beta$ for increasing $\Gamma$. This broadening of the distributions is the feature that makes the application of the SDRG approach legitimate. In particular, this enhancement of the perturbative assumption makes the SDRG approach asymptotically exact for $\zeta(\Gamma)=0$. The fixed point distribution (\ref{eq:freeFPdistr}) attracts all the initial hopping distributions for which the SDRG approach is viable \cite{Fisher94SDRG}.
Remarkably, the attractive distribution (\ref{eq:freeFPdistr}) is an asymptotic solution of (\ref{eq:RGequationP_XXX}) for any finite initial value of $\xi$ defined in (\ref{eq:Average_SYK_coupling}), including the limit $\xi\to\infty$ of maximal correlation between the couplings, i.e. in the limit of identical on-site random interactions.

Let us comment on the finding in (\ref{eq:freeFPdistr}). The asymptotic hopping parameter distribution obtained in the presence of on-site random interactions is the same as found in the free case. This result is similar in spirit to the one discussed in Sec.\,\ref{subsec:sdrg_aperiodic} for the aperiodic chain with Hamiltonian (\ref{eq:Hamiltonian}). Although the on-site random interactions modify the early stages of the SDRG flow, their effect is not strong enough to alter the IR behavior, and, therefore, the long-distance correlation pattern of this model is essentially the same as in the free case.

\subsubsection{Ground state correlation}

Similarly to the situation discussed in Sec.\,\ref{subsec:sdrg_aperiodic} along the SDRG for infinite aperiodic chains, the ground state properties emerging at the end of the RG flow are captured by a factorised phase also in the case of randomly distributed hoppings. This means that, in the regime of validity of the SDRG, the ground state of the chain is well approximated by the tensor product of terms correlating pairs of sites, which can be separated by any length scale. This factorised phase differs from the one discussed in Sec.\,\ref{subsec:sdrg_aperiodic} by how the correlated pairs are spatially distributed, which now entirely relies on the random hopping distribution and not on any pre-existing spatial configuration of the couplings.
The ground state of the random SYK chain here is factorized into different correlated states involving pairs of sites. On the one hand, when the pair contains nearest-neighbour sites in the original infinite chain which are decimated during the SDRG, according to the discussion in Sec.\,\ref{subsec:matching_twosites}, we can associate the corresponding decimation with the emergence of a wormhole geometry and the state $|{\rm ETW}_{s}(\mu_s)\rangle$ as the one described in \cite{Maldacena:2018lmt}. The hopping parameters $\mu_s$ determining these wormhole geometries are different for any coupled nearest-neighbour pair and are randomly distributed.
On the other hand, the correlations between sites that are further apart than one lattice spacing in the original chain are characterized by the behavior discussed in Sec.\,\ref{subsec:matchingsolution_four sites}, and are encapsulated by the states $\ket{\Psi_s}$. Notice that these states are the same as those in \eqref{eq:factorised GS} and we therefore have no further insights on their possible dual geometry.  
As a consequence, the ground state of the random chain also has a factorised form as in \eqref{eq:factorised GS}, but, due to the random nature of the disorder, we can no longer precisely identify which degrees of freedom belong to which two-site ground state.
Schematically, the form of the ground state of the entire chain is given by
\begin{equation}
\label{eq:factorised GS_random}
    |{\rm GS}\rangle=\bigotimes_{s\in G_{\rm UV}} |{\rm ETW}_{s}(\mu_s)\rangle \bigotimes_{s'\in G_{\rm IR}}|\Psi_{s'}\rangle\,,
\end{equation}
where $G_{\rm UV}$ includes the pair of nearest-neighbour sites connected by the randomly drawn hopping parameter $\mu_s$ and decimated throughout the SDRG process, while $G_{\rm IR}$ contains the pair of decimated sites along the SDRG which are at a distance larger than a single lattice spacing.
\\
Summarizing, the analysis of this section shows the applicability of the SDRG methods to the SD equations governing the dynamics of a class of chains with randomly distributed hopping parameters and correlated on-site SYK interactions. Through this approach, we shed light on the properties of the ground state of these models. In particular, the structure of the ground state factorised into the contributions from the spatially distributed pairs of sites correlated at different length scales encodes the physical properties of the system at low energies. 
Since the SYK interaction only contributes via a bounded logarithmic factor $\zeta(\Gamma)$ in \eqref{eq:logfactor}, for any choice of the correlation length $\xi$ defined in (\ref{eq:Average_SYK_coupling}), the SDRG-attractive hopping parameter distribution is the same as the free case. This distribution is attractive with respect to any initial hopping distribution compatible with the SDRG perturbative assumptions. We can interpret this result by saying that the on-site SYK random interactions characterizing the model in (\ref{eq:Hamiltonian}) are unable to influence the low-energy physics of the system, which is, therefore, dominated by the hopping part of the Hamiltonian and is the same as the free XX chains.
\\
Finally, it is worth commenting on similar systems previously considered in the literature on spatial generalizations of SYK models \cite{Gu:2016local,Jian:2017unn,Song:2017pfw,Chowdhury:2018sho,Patel:2017mjv,Patel:2018rjp,Patel:2019qce,Cha:2019zul,Zhang:2017jvh}. It is therefore desirable to position our results in this context for completeness. The main differences of previous works with respect to our current setup are that these models typically consider random hopping parameters specifically drawn from Gaussian distributions, i.e. $\avg{\mu_x\mu_y} \propto \mu^2\delta_{xy}$ or $\mu^2$,
while here we deal with general probability distributions $P(\mu_x;\tilde\mu_0)$ with $\mu_x>0$.
In addition, previous analyses mostly consider totally uncorrelated or correlated on-site random SYK interactions, i.e.
$\avg{J_{j_1j_2\cdots j_q,x}J_{j_1j_2\cdots j_q,y}}\propto \J^2\delta_{xy}$ or $\J$.
In this work, we have extended these cases by introducing a correlation length $\xi$ to interpolate between them.
Therefore, based on our results explained above in this section, the effect of weakly correlated random disorder, i.e.~finite $\xi$, is washed away along the SDRG, leading to the asymptotically attractive distribution \eqref{eq:freeFPdistr} corresponding to the free model. Thus, for the case of initially totally uncorrelated random couplings for the original chain, we expect to observe the same asymptotic SDRG properties. 
    
\section{Conclusions}
\label{sec:conclusions}

\begin{figure}
    \centering
    \includegraphics[width=\linewidth]{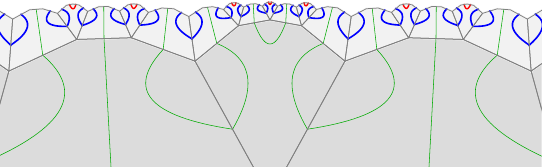}
    \caption{Embedding the sequence of wormholes of Fig.~\ref{fig:SYKgroundstate} on the $\ke{6,4}$ tiling of hyperbolic disk. The sites are embedded on the edges of hexagon. The red curves and blue curves represent the eternal traversable wormholes connected by $\mu_a$-hopping interactions and $\mu_b$-hopping interactions. The green curves represent correlations between the two sites at the boundary that do not have a two-dimensional holographic dual in the sense that there is no metric description. The grayscale labels the layers in SDRG.}
    \label{fig:TN}
\end{figure}

We studied the ground state of a Majorana chain with on-site $q$-body SYK interactions and nearest-neighbor hopping interactions, where the hopping parameters are modulated by either an aperiodic sequence or a random distribution. We assume the spatial distribution of the hopping parameters to be strongly inhomogeneous and such that we can always identify a hopping parameter along the chain much stronger than the others. This assumption allows to treat the weaker hopping terms perturbatively within the paradigm of the strong disorder renormalization group (SDRG). Our analytical results have been obtained from a novel merging of techniques in which we combine this SDRG procedure with the Schwinger-Dyson (SD) equations governing the dynamics of large-$N$ Majorana chain. The SDRG method involves iteratively projecting pairs of nearest-neighbor SYK dots with strong hopping interactions onto their ground state, which is dual to an eternal traversable wormhole (ETW). This decimation step induces an effective hopping interaction between the remaining left and right sites, enabling further decimations along the SYK chain and allowing the SDRG to proceed. Our work advances the discrete holography program by exploring spacetime descriptions of inhomogeneous quantum chains viewed as boundary models.

To implement the SDRG decimation, we first consider a toy model, an open four-site chain where the hopping parameter connecting the central sites is much larger than the ones connecting these central sites to the left- and rightmost sites. Leveraging this perturbative assumption on the SD equations, we have realized a block decimation of the two strongly coupled central sites by solving their SD equations at zeroth order and applying perturbation theory to the SD equations on the left- and rightmost sites, given in \eqref{eq:totalSDequation_v2}. At zeroth order, the correlations \eqref{eq:G23} for the central two sites are described by an ETW, as previously investigated in \cite{Maldacena:2018lmt}. The wormhole between the central sites perturbatively induces an effective interaction between the left- and rightmost sites, described by \eqref{eq:effSigma component}.
Thus, we find that the self-energy governing these projected SD equations has two terms: \textbf{i)} the original SYK self-energy, meaning that the on-site interaction is not renormalized under block decimation, and \textbf{ii)} the effective self-energy (\ref{eq:effSigma component}) resulting from the renormalization of the hopping parameters which describes an effective induced coupling between the left- and rightmost sites. The realization of the real-space decimation at the level of the SD equations and the resulting renormalized self-energy represent the first main finding of this work.

Furthermore, we computed the leading perturbative correction to the two-point correlation functions in the limit $q\to\infty$. 
Solving the SD equations in two different regimes and continuously matching the solutions, we obtain the expressions (\ref{eq:G11_simplified}) and (\ref{eq:G14_simplified}) for the correlations on the left- and rightmost sites and between them. Differently from the correlations on the central sites, which are known to admit a dual description in terms of an ETW, the correlations \eqref{eq:G14_simplified} between the edge sites do not admit an interpretation as an ETW, since the SYK contributions to the total self-energy between these sites is negligible in the perturbative regime of validity of the SDRG, cf.~\eqref{eq:crosscheck_SigmaJ14}.  
Our analytical findings are supported and complemented with a numerical analysis, whose results are reported in Figs.\,\ref{fig:GreenFunctionsEff},\,\ref{fig:GreenFunctionsBeta} and \ref{fig:GreenFunctionsMua}.
This analysis and solutions of the SD equations for a four-site SYK chain represents the second main result of this work.

The aforementioned two-site block decimation constitutes the fundamental operation to iteratively implement the SDRG procedure on infinite chains with Hamiltonian (\ref{eq:Hamiltonian}). The decimation is allowed only up to the perturbative energy scale, i.e. at large enough values of the Euclidean time. In this regime, the effective self-energy, and therefore the SD equation, reduces to the one before the decimation, up to a renormalization of the hopping parameter.
Thus, upon identifying the largest among the hopping parameters along the chain, we have iteratively applied the SDRG procedure and unveiled that the induced flow of the hopping parameter is perturbatively consistent, making the method asymptotically exact. We have employed this technique in two instances of bond inhomogeneities: aperiodic disorder and random disorder. In both cases, after infinitely many steps, we argue that the ground state properties in the regime of validity of the SDRG are captured by the factorised states (\ref{eq:factorised GS}) and (\ref{eq:factorised GS_random}), where the disposition of the pairs of coupled sites is determined either by the aperiodic or by the random distribution of the couplings, respectively. A fraction of the two-site states in (\ref{eq:factorised GS}) and (\ref{eq:factorised GS_random}) is given by the ground state of two identical coupled SYK dots, which is dual to an ETW.
This ground state, depicted in Fig.\,\ref{fig:SYKgroundstate}, generalizes the aperiodic singlet phase \cite{vieira05,Vieira:2005PRB,Juh_sz_2007,Iglói_2007} in the sense that the two-site states factorised in (\ref{eq:factorised GS}) are not spin singlets. 
The ground states (\ref{eq:factorised GS}) and (\ref{eq:factorised GS_random}) and its geometric description embody the last central result of this work.\\ 
For infinite chains with aperiodic bond disorder with generic correlation length $\xi$ between on-site random SYK interactions, we find that the aperiodic bond disorder is a marginal modulation. This means that the ratio of hopping parameters $r$ flows under the SDRG to a fixed point $0<r^*(r)<r$. A similar marginal behavior of aperiodic modulations was observed in the free case of our chain corresponding to XX chains \cite{Hermisson00,vieira05,Vieira:2005PRB}. 
For the case of chains with random bond disorder, we have shown that RG-attractive hopping distributions governing the low-energy physics of the model exist for any value of $\xi$. In particular, we find that the attractive distribution is given in \eqref{eq:freeFPdistr} and is the same as the distribution of free XX chains \cite{Fisher94SDRG}.
Based on its expression, we argue that this distribution becomes broader along the RG flow. These results provide evidence that the SDRG becomes asymptotically exact also in the presence of on-site SYK interactions. This is due to the fact that the broadening of the distribution enhances the fulfillment of the perturbative assumption for the SDRG. 
The validity of our low-energy predictions for a broad class of initial disorder distributions constitutes a main difference with respect to previous literature \cite{Gu:2016local,Jian:2017unn,Song:2017pfw,Chen:2020wiq}. In these references, specific choices of the random disorder were made, such as to integrate these distributions when obtaining the effective action of the models. 

Our results advance the discrete holography programme by providing a spacetime description of inhomogeneous chains beyond the ones considered in \cite{Jahn:2021kti,JahnCentralCharge,Basteiro:2022zur,Basteiro:2022xvu}. In this literature, aperiodic spin chains were taken as boundary theories in potential realizations of holographic dualities based on hyperbolic tilings, which are discretizations of the Poincaré disk. As was shown in \cite{Basteiro:2022zur}, the aperiodic modulations considered in this work (cf.~\eqref{eq:silvermeanrule}) have the special property that their SDRG decimations are in one to one correspondence with the construction of an instance of hyperbolic tilings. According to this mapping, the first layer of tiles (light grey) in Fig.~\ref{fig:TN} is generated by all the pairs of nearest-neighbor sites decimated by the SDRG. On the other hand, the inner layers of the tiling (corresponding to the deeper part of the bulk, depicted in dark grey) are constructed from the decimation of sites which are separated by more than one lattice spacing. The progress made in this work is to consider infinite SYK chains with Hamiltonian given in \eqref{eq:Hamiltonian} and with hopping parameters modulated by the silver mean aperiodic sequence as boundary theories. As discussed in detail in Sec.~\ref{subsec:sdrg_aperiodic}, the emerging wormhole geometries connect only nearest-neighbor sites, while the other ones do not possess a clear gravitational interpretation. In this sense, we can say that the spacetime description of the inhomogeneous SYK chains considered here is a sequence of wormhole geometries which covers the first layer of a tiling of the Poincaré disk represented in Fig.~\ref{fig:TN}. In fact, as we have shown in Appendix~\ref{apx:Aperiodic_Sequence}, the percentage of of two-site states appearing in \eqref{eq:factorised GS} that correspond nearest-neighbour sites, i.e. have a wormhole description, is approximately 83\%.
To further progress towards establishing discrete holographic dualities, it will be insightful to generalize the model considered in this manuscript by allowing for emergent dual geometries connecting sites at large boundary distances, corresponding to exploring deeper regions of the corresponding bulk.

A further extension of our analysis that can be of interest for the discrete holography programme concerns considering other aperiodic modulations for the hopping parameters. Such aperiodic sequences can be generated as infinite families associated to inflation rules for tessellations of hyperbolic space \cite{FlickerBoyleDickens,JahnCentralCharge,Basteiro:2022zur,Basteiro:2022xvu}. The main challenge for these extensions is perturbatively solving the model (\ref{eq:Hamiltonian}) for finite chains with $L>4$ sites. Such an extension would involve the generalization of the SDRG procedure of Sec.\,\ref{sec:wormholenetwork} to include the decimation of blocks of more than two consecutive SYK sites \cite{Vieira:2005PRB,Basteiro:2022zur}. The first generalization pertains SYK chains with the distribution of hopping parameters given by the sequence $a(ba)^{L/2-1}$ with even $L$. The decimation of the central $L-2$ SYK sites involves the projection of the SD equations on local ground states dual to multiple ETWs, each of which connecting sites coupled by a strong hopping parameter $\mu_b$. The resulting effective self-energy between the edge sites, which generalizes \eqref{eq:Sigma eff}, is given at order $\mu_a^{L/2}$ in perturbation theory. The induced effective hopping parameter between the two edges scales as $\mu_a^{L/2}/\nu_{23}^{L/2-1}$, which is suppressed by the SYK interaction. 
Thus, in the large $L$ limit, we anticipate the emergence of a mass of Majorana edge modes \cite{SSH:1979,Kitaev:2000nmw,Meier:2016SSH} with on-site SYK interactions on the two edges of this SYK chain.  
The second possible generalization concerns SYK chains with the hopping parameter sequence given by $ab^{L-3}a$ with even $L$. In this case, the decimation of the central $L-2$ SYK dots, all coupled by the hopping parameter $\mu_b$, involves the projection of the SD equations onto the ground state of this homogeneous SYK chain. It would be interesting to understand the form of the Hamiltonian governing more than two SYK clusters necessary to obtain emergent multi-throat traversable wormholes as dual geometries \cite{Emparan:2020ldj,Shen:2024uja}.

Another question to address in future investigations concerns the solution of the aperiodic or random SYK chain away from the large $q$ limit. The large $q$ limit plays an important role in this work for deriving the correlation and the suppression of the off-site self-energy contribution from the SYK interaction. The random charged SYK chain at $q=4$ is known to be a heavy Fermi liquid at low temperature \cite{Song:2017pfw}, where the SYK interaction highly renormalizes the effective mass of the fermion. Investigating similar effects in our setup is a promising strategy to gain insights away from the regime $q\gg 1$. 

Recall that the Hamiltonian (\ref{eq:Hamiltonian}) considered in this work has two terms: the on-site SYK interaction and a hopping term with either aperiodic or random bond disorder. Exploring the effect of more general interaction terms subject to these types of disorder, as treated respectively in Sec.~\ref{subsec:sdrg_aperiodic} and Sec.~\ref{subsec:random_SDRG}, provides an intriguing avenue for future research. It has been previously established that such perturbations can lead to interesting transition phenomena \cite{Jian:2017unn,Jian:2017tzg,Song:2017pfw,Altland:2019PRL} when they are governed by Gaussian distributions like that of the original SYK model. Moreover, recent investigations have shown that multi-hopping interactions lead to a rich phase diagram emerging in the large $N$ limit \cite{Basteiro:2024cuh}. In this case, different regions of these phase diagrams are characterized by different types of von Neumann operator algebras. Given the interest of these aspects in the context of quantum gravity and holography \cite{Leutheusser:2021qhd,Leutheusser:2021frk,Witten:2021jzq,Witten:2021unn,Chandrasekaran:2022eqq,Chandrasekaran:2022cip,Banerjee:2023eew,Aguilar-Gutierrez:2023odp,Engelhardt:2023xer,Gesteau:2024dhj,Xu:2024hoc,Kudler-Flam:2023qfl}, we aim at extending these analyses in the presence of an on-site SYK interaction, which, combined with the hopping, is the origin of the dual wormhole geometry between coupled SYK dots.
The presence of phase transitions in the generalization of the model (\ref{eq:Hamiltonian}) would correspond in the gravitational dual picture to transitions between wormhole and black hole geometries. A similar effect has been found in fermionic chains coupled to ancillary spin systems in view of describing a transition between a Fermi liquid and  a fractionalized Fermi liquid \cite{Nikolaenko:2021vlw}. 
Finally, an algebraic approach to detect and describe horizons and causal geometries was recently proposed in \cite{Gesteau:2024rpt}. This method applies to theories where the large $N$ factorisation allows for a complete characterization of the correlations in terms of the two-point functions. Given that the models considered in this work fall in this class, extending this new analysis to the short chains studied in Sec.\,\ref{sec:foursite_SYK} could be a first step towards a more precise characterization of their gravity duals.\\

\noindent {\bf \large Acknowledgements}
\\
We are grateful to Sergio E. Aguilar-Gutierrez, Igor Boettcher, Anffany Chen, Elliott Gesteau, Shao-Kai Jian, Juan Maldacena, Alexey Milekhin, Sarthak Parikh, Cheng Peng, Subir Sachdev, Gautam Satishchandran, Konstantin Weisenberger, Pengfei Zhang for fruitful discussions and comments. This work  was supported by Germany's Excellence Strategy through the W\"urzburg‐Dresden Cluster of Excellence on Complexity and Topology in Quantum Matter ‐ ct.qmat (EXC 2147, project‐id 390858490), and by the Deutsche Forschungsgemeinschaft (DFG) through the Collaborative Research Center ToCoTronics, Project-ID 258499086—SFB 1170, as well as a German-Israeli Project Cooperation (DIP) grant ”Holography and the Swampland”. ZYX also acknowledges support from the National Natural Science Foundation of China under Grant No.~12075298. GDG is supported by the ERC Consolidator grant (number: 101125449/acronym: QComplexity). Views and opinions expressed are however those of the authors only and do not necessarily reflect those of the European Union or the European Research Council. Neither the European Union nor the granting authority can be held responsible for them. We are grateful to the long term workshop YITP-T-23-01 held at the Yukawa Institute for Theoretical Physics, Kyoto University, where a part of this work was done. PB would also like to acknowledge the hospitality of the Theoretical Physics Institute of the University of Alberta, where part of this work was done.

The authors’ names are listed in alphabetical order.

\appendix
\addtocontents{toc}{\protect\setcounter{tocdepth}{1}}

\section{Insights and review from the free case}
\label{apx:freecase}

In this appendix, we discuss the case where $\J=0$ in (\ref{eq:Hamiltonian}), namely where our model reduces to a free system governed by $N/2$ decoupled XX spin chain Hamiltonians, as given in (\ref{eq:genericNandp1_spinDOFform}).

\subsection{Free chains with alternating hopping parameters}

We focus our discussion on the case of a finite open chain consisting of an even number $L$ of sites connected through alternating hopping parameters
$\mu_x=\mu_a$ if $x$ is odd, and $\mu_x=\mu_b$ if $x$ is even.
According to (\ref{eq:SD_eqs_full_form_Sigma}), the self-energy for this chain is
\begin{eqnarray}
\label{eq:freeselfenergy}
\Sigma_{xy}(\tau_1,\tau_2)
    &=&-\mathrm{i} \delta(\tau_1-\tau_2) \big[\frac{(1-(-1)^x)\mu_a+(1+(-1)^x)\mu_b}{2}\delta_{x+1,y}
    \\
    &-&\frac{(1-(-1)^y)\mu_a+(1+(-1)^y)\mu_b}{2}\delta_{x,y+1}\big]\,.
    \end{eqnarray}
The SD equations (\ref{eq:SD_eqs_full_form_G}) with self-energy (\ref{eq:freeselfenergy}), governing the dynamic of the system when $N\to\infty$, can be solved by performing a Fourier transform and working in frequency space. After Fourier transform, (\ref{eq:SD_eqs_full_form_G}) reads
\begin{align}
   -\mathrm{i}\omega G_{xy}(\omega)-\sum_z \Sigma_{xz}(\omega)G_{zy}(\omega)=\delta_{xy}\,,
\label{eq:SDFT}
\end{align}
where the convolution in the second term of (\ref{eq:SD_eqs_full_form_G}) is replaced by the standard matrix product due to the convolution theorem.
The strategy is therefore to determine
\begin{equation}
\label{eq:SDFT_sol}
  G(\omega) = \frac{1}{ -\mathrm{i}\omega-\Sigma(\omega)} \,,
\end{equation}
and to compute its inverse Fourier transform. This computation can be carried out explicitly, as shown in the next subsection, for a $L=4$ site chain with a sequence of alternating hopping parameters $(\mu_a,\mu_b,\mu_a)$. For larger values of $L$ even, the computation becomes cumbersome, but insights can be obtained by solving the equation numerically. 

Before discussing the solution in the case of the four-site chain, let us report general considerations on the entries of the solution (\ref{eq:SDFT_sol}) after the inverse Fourier transform.
First, plugging the Fourier transform of (\ref{eq:freeselfenergy}) into (\ref{eq:SDFT_sol}), we can argue by inspection that the entries of $G(\omega)$ are rational functions of $\omega$ with $L/2$ poles in the upper half complex plane and $L/2$ in the lower one.
In particular, the poles of the entries of $G(\omega)$ are always purely imaginary. To check this statement more formally, we can observe that the poles of the entries of $(-\mathrm{i}\omega -\Sigma)^{-1}$ are the zeroes of the determinant of $-\mathrm{i}\omega -\Sigma$. To evaluate such a determinant, we use its Laplace expansion, showing that it can be deconstructed into the sum of at most  $2^{L-2}$ determinants of $2\times2 $ matrices (since $L$ is always even). This is because every term in every step of the Laplace expansion will only give rise to two non-vanishing terms, given the tridiagonal structure of $\Sigma$. Specifically, the only matrices with non-vanishing determinant appearing in the final expansion are $\begin{pmatrix}\omega & -\mu_a\\\mu_a & \omega\end{pmatrix}$ and $ \begin{pmatrix}\mu_b & -\mu_a\\0 & \omega\end{pmatrix}$, with coefficients which can be tracked via a diagrammatic argument. We can convince ourselves that the determinant is then always a polynomial with even powers and real coefficients. Every polynomial can be factored over the real numbers into linear factors and irreducible quadratic factors, and we know that only even powers appear in our determinant. As a consequence, the determinant only consists of irreducible quadratic factors of the form $(\kappa^2+\omega^2)$ with $\kappa^2$ real, leading to purely imaginary roots.

When integrating the entries of $G(\omega)$, we can apply the residue theorem by closing the contour on the lower half-plane, thus picking up the corresponding $L/2$ poles. Therefore, the entries of $G(\tau)$ can be written as sums of terms of the form
\begin{align}
\label{eq:residues}
    -\mathrm{i} {\rm Res}\kc{\frac{\mathrm{i} \omega ^{\alpha }}{\omega ^2+\kappa^2}f(\omega),\,\omega=-\mathrm{i}\kappa}=\frac{1}{2} \mathrm{i}^{1-\alpha }\frac{f(-\mathrm{i}\kappa)}{\kappa}\,,
\end{align}
with $\kappa$ real function of the parameters of the model and $f$ rational function which depends on the entry of $G_{xy}$ we consider and such that $f(-\mathrm{i}\kappa)$ is real. By inspecting many instances of chains with $L>4$ sites with alternating hopping parameters, we can argue that the polynomials at the numerator of the entries of $G_{xy}(\omega)$ have always odd powers if $|x-y|$ is even and even powers otherwise. This fact implies that $\alpha$ in (\ref{eq:residues}) is odd for entries with $|x-y|$ even and even otherwise. 
 Thus, the final implication is that $ \arg G_{x,x+j}= \arg \mathrm{i}^{j\mode2}$ for $j\geq0$.
 This result is consistent with the explicit computation performed in the next subsection for the free four-site chain and, non-trivially, with the results obtained in Sec.\,\ref{sec:foursite_SYK} for the four-site chain with on-site SYK interactions.

\subsection{Explicit solutions and renormalization of the hopping parameter}

Let us begin the discussion of some explicit solutions with the case $L=2$, where we denote the only hopping parameter as $\mu_b$. For this instance, the self-energy reads
 \begin{equation}
 \label{eq:selfeenrgy_Bchainfree}
   \Sigma(\tau)=
   \left(
\begin{array}{cccc}
 0 & -\mathrm{i}\mu_b \\
 \mathrm{i}\mu_b & 0
        \end{array}\right)\delta(\tau)
        \,,
 \end{equation}
$G(\omega)$ is given in (\ref{eq:SDFT_sol}) and its Fourier transform can be computed straightforwardly leading to
\begin{equation}
   G(\tau)=\frac{e^{- \mu_b |\tau| }}{2}\left(
\begin{array}{cccc}
 \sgn(\tau) &\mathrm{i} \\
 -\mathrm{i} &  \sgn(\tau)
        \end{array}\right)
        \label{eq:Gbchain} \,.
\end{equation}  
If the case of $L=4$, we can compute the $4\times 4$ matrix of the two-point correlation functions $G(\tau)$ by evaluating explicitly (\ref{eq:SDFT_sol}) and its inverse Fourier transform, after exploiting the explicit form of the self-energy which is given by
 \begin{equation}
 \label{eq:selfeenrgy_4sites_free}
   \Sigma(\tau)=
   \left(
\begin{array}{cccc}
 0 & -\mathrm{i}\mu_a & 0 & 0 \\
 \mathrm{i}\mu_a & 0 & -\mathrm{i}\mu_b & 0 \\
 0 & \mathrm{i}\mu_b & 0 & -\mathrm{i}\mu_a \\
 0 & 0 & \mathrm{i}\mu_a & 0
        \end{array}\right)\delta(\tau)
        \,.
 \end{equation}
We refer the reader to the supplemental material of \cite{Basteiro:2024cuh} for a detailed derivation, but we report the final solution here for completeness. It reads
\begin{align}\label{Gfreeaba}
&G_{xy}(\tau/\mu_b)=\frac{e^{-\frac{1}{2} (s+1) |\tau| } }{4s}\times \\
&\left(
\begin{array}{cccc}
 (s+1) e^{|\tau| }+s-1 & \mathrm{i} \sqrt{s^2-1} \left(e^{|\tau| }+1\right) & \sqrt{s^2-1} \left(e^{|\tau| }-1\right) & \mathrm{i} \left((s+1) e^{|\tau| }-s+1\right) \\
 -\mathrm{i} \sqrt{s^2-1} \left(e^{|\tau| }+1\right) & (s-1) e^{|\tau| }+s+1 & -\mathrm{i} \left((s-1) e^{|\tau|u }-s-1\right) & \sqrt{s^2-1} \left(e^{|\tau| }-1\right) \\
 \sqrt{s^2-1} \left(e^{|\tau| }-1\right) & \mathrm{i} \left((s-1) e^{|\tau| }-s-1\right) & (s-1) e^{|\tau| }+s+1 & \mathrm{i} \sqrt{s^2-1} \left(e^{|\tau| }+1\right) \\
 -\mathrm{i} \left((s+1) e^{|\tau| }-s+1\right) & \sqrt{s^2-1} \left(e^{|\tau|}-1\right) & -\mathrm{i} \sqrt{s^2-1} \left(e^{|\tau| }+1\right) & (s+1) e^{|\tau|}+s-1 \\
\end{array}
\right) \nn\,,
\end{align}
where $r=\mu_a/\mu_b=\frac12\sqrt{s^2-1}$.
At $\tau=0$, we have
\begin{align}
    \label{eq:G0_v1}G_{11}=G_{22}=\frac12\sgn0,\quad
    G_{23}=G_{14}=\frac{\mathrm{i}}{2} \cos\theta,\quad 
    G_{12}=G_{34}=\frac{\mathrm{i}}{2} \sin\theta\,.
\end{align}
We stress that when $r\to 0$, i.e. $\theta\to 0$, the only surviving correlations are the ones between central sites and the ones between the left- and rightmost sites.

Although the explicit solution for the free four-site model is known, it is instructive to follow the perturbative approach developed in Sec.\,\ref{subsec:setup_foursitechain} by assuming $\mu_a\ll \mu_b$. In particular, we want to compute the effective self-energy (\ref{eq:effSigma component}) (with $\Delta=0$) coupling the left- and the rightmost sites upon decimation of the central sites. According to the procedure explained in Sec.\,\ref{subsec:setup_foursitechain},  $G^{(0)}$ in (\ref{eq:effSigma component}) contains the two-point functions of the degrees of freedom on the strongly coupled central sites, which in the case of interest is given by (\ref{eq:Gbchain}). Therefore, the entries of the effective self-energy read
 \begin{equation}
 \label{eq:renormalizeG_abafree}
     \Sigma^{\rm eff}_{\mu \,11}(\tau)=  \Sigma^{\rm eff}_{44}(\tau)=\frac{\mu_a^2}{2}e^{-\mu_b|\tau|}\sgn(\tau)\,,\qquad\qquad
       \Sigma^{\rm eff}_{14}(\tau)=\mathrm{i}\frac{\mu_a^2}{2}e^{-\mu_b|\tau|}\,.
 \end{equation}
Notice that the $2\times 2$ self-energy matrix with entries given by $\Sigma^{\rm eff}_{\mu \,11}$, $\Sigma^{\rm eff}_{\mu \,14}$, $\Sigma^{\rm eff}_{\mu \,41}$ and $\Sigma^{\rm eff}_{\mu \,44}$ has not the same time dependence as (\ref{eq:selfeenrgy_Bchainfree}). We retrieve the same behavior only in the limit $|\tau|\gg \mu_b^{-1}$, where the diagonal entries can be neglected and a $\delta$-function arises in the off-diagonal ones, as discussed in Sec.\,\ref{subsec:largetauregime}. In particular, comparing (\ref{eq:renormalizeG_abafree}) with (\ref{eq:selfeenrgy_Bchainfree}) when
$|\tau|\gg \mu_b^{-1}$, we obtain the renormalization of the hopping parameter after an SDRG step
\begin{equation}
    \mu'=\frac{\mu_a^2}{\mu_b}\,,
    \label{eqSM:mu_prime_ApxA}
\end{equation}
consistent with (\ref{eq:mu_prime_expression}) when $\Delta=0$ and $\J\to 0$, i.e. $\tanh\gamma_{23}\to 1$.

The fact that, for $|\tau|\gg \mu_b^{-1}$, we retrieve the same self-energy as in \eqref{eq:selfeenrgy_Bchainfree} but with the renormalized coupling \eqref{eqSM:mu_prime_ApxA} supports our understanding that the SDRG decimation on SD equations works only for times larger than the inverse of the energy scale of the four-site block. Indeed, as we review in Appendix \,\ref{apx:SDRG}, it is known that the free XX chain (or $N/2$ decoupled copies of it) preserves the same Hamiltonian (and therefore self-energy) after an SDRG step, albeit with a renormalized hopping given precisely by \eqref{eqSM:mu_prime_ApxA}.

\section{Details on Schwinger-Dyson equations}
\label{apx:Details_SD_Eqs}

In this Appendix, we provide some details on the derivation of the Liouville equation \eqref{eq:evolution_g_gen}, as well as on the validity of the large $q$ ansatz for the Green's function given in \eqref{eq:qExpansion}. Moreover, we provide a detailed derivation of the late-time solutions \eqref{eq:longtimesolution_simpl} to the SD equations.

\subsection{Derivation of the Liouville equation}
\label{subapp:LiouvilleEq}

To derive the Liouville equation \eqref{eq:evolution_g_gen}, we consider the large $q$ ansatz for the Green's function \eqref{eq:qExpansion}, which we recall here for the convenience of the reader,
\begin{align}\label{eq:qExpansion_Apx}
    G_{xy}(\tau)
    =\widehat{G}_{xy}(\tau)\left(1+\frac1{q}g_{xy}(\tau)+\cdots\right)
    \approx\widehat{G}_{xy}(\tau)e^{\frac{g_{xy}(\tau)}{q}
    }\,.
\end{align}
Inserting \eqref{eq:qExpansion_Apx} into the SD equations \eqref{eq:SD_eqs_full_form_G}, we obtain for the derivative term
\begin{equation}
\label{eq:SDequationG_subleadingq}
    \begin{split}
        \partial_{\tau} G_{xy}&= \partial_{\tau}  \widehat{G}_{xy}+\frac{1}{q} \partial_{\tau} (\widehat{G}_{xy} g_{xy}(\tau)) \\
        &= \delta_{xy} \delta(\tau) + \delta_{xy} \delta(\tau) \frac{g_{xy}(\tau)}{q}+\widehat{G}_{xy} \frac{g'_{xy}(\tau)}{q}\,,
    \end{split}
\end{equation}
where, in the second line, we have used the SD equations for the non-interacting case \eqref{eq:free_SD}. Simultaneously, we can insert the ansatz \eqref{eq:qExpansion_Apx} into the expression for the self-energy (\ref{eq:SD_eqs_full_form_Sigma}) and subsequently into \eqref{eq:SD_eqs_full_form_G} while using (\ref{eq:free_SD}) and (\ref{eq:SDequationG_subleadingq}), in order to find the SD equations at $\mathcal{O}\left(q^{-1}\right)$
\begin{align}
\label{eq:1overqSDequation}
    0=\delta(\tau_{12})\delta_{xy}g_{xy}(\tau_{12})+\widehat{G}_{xy}(\tau_{12})g_{xy}'(\tau_{12})
    +\sum_z \mathrm{i}\hat{\mu}_{xz}\widehat{G}_{zy}(\tau_{12})\\ 
    -\sum_z\int_{\tau_3} s_{xz}\J^2(2\,\widehat{G}_{xz}(\tau_{13}))^{q-1} e^{g_{xz}(\tau_{13})}\widehat{G}_{zy}(\tau_{32})\,,
    \nonumber
\end{align}
with $\tau_{ij}\equiv \tau_i-\tau_j$. Crucially, the integral over $\tau_3$ for the hopping term can be evaluated thanks to the simple  $\delta$-function time-dependence of the self-energy \eqref{eq:SD_eqs_full_form_Sigma}. We consider the above equation in the cases $0<\tau<\beta$ and $\tau=0$ separately, where the latter will play the role of boundary conditions. For $0<\tau<\beta$, $\delta(\tau)$ vanish, which implies that the first term of \eqref{eq:1overqSDequation} drops out. By taking a derivative with respect to $\tau_2$, and exploiting \eqref{eq:free_SD} on the hopping term above, we obtain a second order differential equation for $g_{xy}(\tau>0)$ as follows
\begin{align}
    0&=\begin{aligned}[t]\partial_{\tau_2}\Bigg(\widehat{G}_{xy}(\tau_{12})g_{xy}'(\tau_{12})&+\sum_z \mathrm{i}\hat{\mu}_{xz}\widehat{G}_{zy}(\tau_{12})\\
    &-\sum_z\int_{\tau_3} s_{xz}\J^2(2\,\widehat{G}_{xz}(\tau_{13}))^{q-1} e^{g_{xz}(\tau_{13})}\widehat{G}_{zy}(\tau_{32})\Bigg)\end{aligned}\label{eq:Liouville_derivative1}\\
    &=\begin{aligned}[t] \left(\widehat{G}_{xy}(\tau_{12})\right)'g_{xy}'(\tau_{12})+&\widehat{G}_{xy}(\tau_{12})g_{xy}''(\tau_{12})+\sum_z \mathrm{i}\hat{\mu}_{xz}\left(\widehat{G}_{zy}(\tau_{12})\right)'\\
    &-\sum_z\int_{\tau_3} s_{xz}\J^2(2\,\widehat{G}_{xz}(\tau_{13}))^{q-1} e^{g_{xz}(\tau_{13})}\left(\widehat{G}_{zy}(\tau_{32})\right)'\end{aligned}\label{eq:Liouville_derivative2}\\ 
    &=\widehat{G}_{xy}(\tau_{12})g_{xy}''(\tau_{12})-2^{q-1}s_{xy}\J^2 \widehat{G}_{xy}^{q-1}e^{g_{xy}(\tau)}\,,
    \label{eq:Liouville_derivative3}
\end{align}
where primes denote a derivative with respect to $\tau_2$. We used the solution $\widehat{G}_{xy}$ for the non-interacting case \eqref{eq:free_SD} twice when going from \eqref{eq:Liouville_derivative2} to \eqref{eq:Liouville_derivative3}. More precisely, for the first and third terms of \eqref{eq:Liouville_derivative2}, we use $\left(\widehat{G}_{xy}(\tau_{12})\right)'=\delta_{xy}\delta(\tau_{12})$, which vanishes due to $\delta(\tau_{12})\equiv\delta(\tau)=0$ for $\tau>0$. However, for the last term in \eqref{eq:Liouville_derivative2}, we have $\left(\widehat{G}_{zy}(\tau_{32})\right)'=\delta_{zy}\delta(\tau_{32})$, where this $\delta$-function does not vanish. In fact, since $\tau_3$ is integrated over, this simplifies the sum and the integral in the last term of the second line of \eqref{eq:Liouville_derivative2}, yielding the final expression in \eqref{eq:Liouville_derivative3}. From this result, recalling that $\widehat{G}_{xy}=\frac{1}{2}\G_{xy}$ when $\tau>0$ and writing $\tau=\tau_{12}$, we find 
\begin{align}
    &g_{xy}''(\tau) -2s_{xy}\J^2\G_{xy}^{q-2}e^{g_{xy}(\tau)}=0,\qquad \textrm{for}\quad  0<\tau<\beta\,,
    \label{eq:Final_Liouville_Apx}
\end{align}
which is precisely the Liouville equation given in (\ref{eq:evolution_g_gen}).
\\
As boundary conditions, we first consider the SD equations at $\tau=0$ and $x\neq y$. The latter immediately imposes $\delta_{xy}=0$ in \eqref{eq:1overqSDequation}, thus dropping the first term once again. Using again that $\widehat{G}_{xy}=\frac{1}{2}\G_{xy}$ as per \eqref{eq:defGcal}, and evaluating \eqref{eq:1overqSDequation} at $\tau=0$, we find the boundary condition (\ref{eq:bc_gen g_R23}). Notice that, for the off-diagonal components of the Green's function $G_{xy}(0)$ with $x\neq y$, we do not expect that the on-site SYK interaction plays a role. Rather, the physical expectation is that these off-diagonal terms should depend solely on the hopping parameters at $\tau=0$, and therefore the condition \eqref{eq:bc_gen g_R23} does not include any terms from the SYK part. This assumption can indeed be checked \textit{a posteriori} for the explicit cases of interest in this manuscript, namely $g_{23}$ in the two-site chain and $g_{14}$ for the four site-chain. Finally, given the structure of $\G$ in \eqref{eq:defGcal}, the diagonal components may also be analysed individually, recalling that $g_{xx}(0)=0$, now to be taken as boundary condition which imposes $G_{xx}(0)=1/2$, thus yielding the condition (\ref{eq:bc_gen g_R1}).\\

\subsection{SD equations with self-energy exponential in time}

\subsubsection{Exponential ansatz}
\label{subapx:exp_ansatz_expSigma}
In this Appendix, we report the details on the derivation of the Liouville equation in the presence of a self-energy with an exponential time-dependence, as we encounter in the case of \eqref{eq:sigmaeff_14} for the renormalized self-energy. In particular, we would like to explain the steps in deriving the second equation in \eqref{eq:Liouville_shorttime_simple} for the $g_{14}$ component, since the first equation in \eqref{eq:Liouville_shorttime_simple} follows the same steps as shown in the previous Appendix. In order to derive the second Liouville equation in \eqref{eq:Liouville_shorttime_simple}, recall that we assume that the SYK contribution to the off-diagonal components of the self-energy is negligible at short times (cf.~\eqref{eq:sigma14 approx}), i.e. $\Sigma_{\J\,14}\approx 0$. This is because we expect that at short times, the off-diagonal entries are only due to the presence of hopping terms, and not a consequence of the on-site interactions. This assumption needed to be checked \textit{a posteriori}, as we have done in \eqref{eq:crosscheck_SigmaJ14}.\\ 
The expression \eqref{eq:1overqSDequation}, now with an exponential self-energy given by \eqref{eq:sigmaeff_14}, reads
\begin{align}
\label{eq:1overqSDequation_exp}
    0=\delta(\tau_{12})\delta_{xy}g_{xy}(\tau_{12})+\widehat{G}_{xy}(\tau_{12})g_{xy}'(\tau_{12})
    -q\sum_z \int_{\tau_3}\Sigma_{\mu\,xz}^{\textrm{eff}}(\tau_{13})\widehat{G}_{zy}(\tau_{32})\\ 
    -\sum_z\int_{\tau_3} \Sigma_{\J\,xz}(\tau_{13})\widehat{G}_{zy}(\tau_{32})\,.
    \nonumber
\end{align}
Here, the hopping matrix in \eqref{eq:1overqSDequation} is now replaced by the effective self-energy $\Sigma_{\mu\,xy}^{\textrm{eff}}$ given in \eqref{eq:sigmaeff_14}, which implies that the integral over $\tau_3$ cannot be carried out immediately. The factor of $q$ is to match the order of the rest of the terms, meaning that we still consider the SD equations at $\order{q^{-1}}$. Moreover, note that now the indices run only over the values $x,y,z\in\{1,4\}$, since we are dealing with the projected SD equation. Taking a derivative with respect to $\tau_2$ as done in \eqref{eq:Liouville_derivative1}-\eqref{eq:Liouville_derivative3}, we can derive the corresponding Liouville equations of the system. The Liouville equation for the diagonal component $g_{11}$, under the assumption  $\Sigma_{11}(\tau)\approx\Sigma_{\J\,11}(\tau)$ (see the discussion in Sec.\,\ref{subsec:earlytime4sitechain}), is derived via the same steps as in Appendix \ref{subapp:LiouvilleEq}, leading to the first equation in \eqref{eq:Liouville_shorttime_simple}. As for the off-diagonal component $g_{14}$, we may drop the SYK term in \eqref{eq:1overqSDequation_exp} under the assumption \eqref{eq:sigma14 approx}, and also the first term vanishes identically due to the Kronecker-delta. Then, we have
\begin{align}
    0&=\partial_{\tau_2}\Bigg(\widehat{G}_{14}(\tau_{12})g_{14}'(\tau_{12})
    -q\sum_z \int_{\tau_3}\Sigma_{\mu\,1z}^{\textrm{eff}}(\tau_{13})\widehat{G}_{z4}(\tau_{32}))\Bigg)\label{eq:Liouville_derivative1_exp}\\
    &= \widehat{G}_{14}(\tau_{12})g_{14}''(\tau_{12})-\sum_z \int_{\tau_3}\Sigma_{\mu\,1z}^{\textrm{eff}}(\tau_{13}) \left(\widehat{G}_{z4}(\tau_{32})\right)'\label{eq:Liouville_derivative2_exp}\\
    &=\widehat{G}_{14}(\tau_{12})g_{14}''(\tau_{12})-q\Sigma_{\mu\,14}^{\textrm{eff}}(\tau_{12})\,,
\end{align}
where we have again used the free SD equations \eqref{eq:free_SD} as $\left(\widehat{G}_{xy}(\tau_{32})\right)'=\delta_{xy}\delta(\tau_{32})$. Now, recalling that  we have assumed (and consistently verified) that $\widehat{G}_{14}=\frac{\mathrm{i}}{2}\mathcal{E}_{14}\neq 0$, and using the renormalized effective self-energy given in \eqref{eq:sigmaeff_14} with $\Delta=0$, we find 
\begin{equation}
\mathcal{E}_{14}\partial^2g_{14}+\frac1q\hat\mu_a^2 e^{-\nu_{23}\tau}=0\,.
\end{equation}
Notice that, since $g_{14}$ is $\order{1}$, the second term in the expression is negligible in the limit $q\to\infty$, thus retrieving  precisely the second Liouville equation reported in \eqref{eq:Liouville_shorttime_simple}. This equation needs to be complemented with the boundary condition \eqref{eq:projected bc}. In contrast to the boundary conditions \eqref{eq:bc_gen g_R23}, this condition imposes that $g'_{14}(0)=0$ and reflects the fact that sites 1 and 4 were not originally coupled by a hopping parameter.

\subsubsection{Solution in the frequency space}
\label{subapx:Fourier_expSigma}

In this Appendix, we report the derivation of the solutions \eqref{eq:longtimesolution_simpl} in the long time regime $\tau\gg1/\hat{\nu}_{23}$. We will restore the parameter $\Delta$ and perform the general derivation for generic values $\Delta$, restricting only at the end to the case $\Delta=0$ which recovers \eqref{eq:longtimesolution_simpl}. Recall that the self-energy in this time regime is given by \eqref{eq:sigmaeff_14}, which we write collectively in the matrix
\begin{equation}
    \Sigma_{\mu}^{\textrm{eff}}(\tau)=\frac{1}{2}\sgn\tau \mu_a^2 \begin{pmatrix}
        \;(1+\Delta)^2\;\; & \;-\mathrm{i} (1-\Delta^2) \\[5pt]
        \;\mathrm{i} (1-\Delta^2)\;\;  & \;(1-\Delta)^2 
    \end{pmatrix}\times e^{-\nu_{23}\abs{\tau}}\,.
    \label{eq:Sigma_eff_matrix_apx}
\end{equation}
In the time domain, the SD equations with this self-energy are given by
\begin{align}
    \partial_{\tau} G(\tau)-\Sigma^{\textrm{eff}}_{\mu}(\tau)*G(\tau)=0\,,
\label{eq:SD_eq_exp_self_energy_apx}
\end{align}
where the $\delta(\tau)$ on the right-hand side is not present since we consider long times. We solve these equations by Fourier transforming \eqref{eq:SD_eq_exp_self_energy_apx} to frequency space, where they take on the form 
\begin{equation}
\label{eq:kernel_SD_eq}
    (-\mathrm{i}\omega-\Sigma^{\textrm{eff}}_{\mu}(\omega))G(\omega)\equiv M(\omega)G(\omega)=0\,.
\end{equation}
In this expression, $\Sigma_{\mu}^{\textrm{eff}}(\omega)$ is the Fourier transform of \eqref{eq:Sigma_eff_matrix_apx}, whose explicit form is given in \eqref{eq:sigmaeff_omegaspace}. Notice that the products in \eqref{eq:SD_eq_exp_self_energy_apx} are now standard matrix products due to the convolution theorem. Moreover, we have defined the matrix
\begin{equation}
    M(\omega)=\begin{pmatrix}
 -\frac{\mathrm{i} \omega (1+\Delta)^2 \mu _a^2}{\nu _{23}^2+\omega ^2}-\mathrm{i} \omega  & \frac{\mathrm{i} \nu _{23} (1-\Delta^2)\mu _a^2}{\nu _{23}^2+\omega ^2} \\
 -\frac{\mathrm{i} \nu _{23} (1-\Delta^2) \mu _a^2}{\nu _{23}^2+\omega ^2} & -\frac{\mathrm{i} \omega (1-\Delta)^2 \mu _a^2}{\nu _{23}^2+\omega ^2}-\mathrm{i} \omega
\end{pmatrix}\,.
\label{eq:Momega}
\end{equation}
As a general ansatz for solving \eqref{eq:kernel_SD_eq}, we may expand the Green's function in a generic linear combination of the eigenvectors of $M$, which we denote by $\ket{V_j(\omega)}$ with $j=1,2$, as
\begin{equation}
\label{eq:G_fourier_ansatz}
    G(\omega)=\sum_{j=1}^2 \Tilde{c}_j(\omega)\ket{V_j(\omega)}\bra{V_j(\omega)}\,,
\end{equation}
with frequency-dependent coefficients $\Tilde{c}_j(\omega)$.
Since \eqref{eq:kernel_SD_eq} is a kernel equation, we are specifically interested in the null eigenspace of $M$, i.e. those values of the frequency for which $M$ maps its eigenvectors to zero. We can find the null eigenspace by computing the zeroes of the determinant of $M$, which we denote by $\omega_i$. That is, we have
\begin{equation}
    \det(M(\omega))\overset{!}{=}0 \longrightarrow \omega=\omega_i\,,
\end{equation}
such that
\begin{equation}
    M(\omega_i)\ket{V_j(\omega_i)}=0\,.
\end{equation}
The determinant of $M$ has four zeroes $\omega_i$ with $i=\{1,2,3,4\}$ and $j=\{1,2\}$ which are purely imaginary and come in two pairs with opposite sign. That is, we have $\omega_i=\pm\omega_{\pm}$, with
\begin{equation}
\label{eq:omega_plusminus}
    \omega^2_\pm
    =~-\frac{1}{4} \left(\sqrt{4\Delta^2\mu_a^2+\nu_{23}^2}\pm\sqrt{4 \mu _a^2+\nu _{23}^2}\right)^2\,.
\end{equation}
The assignment of signs is based on the absolute value and the sign of the imaginary part of the frequencies, and reads $\omega_1=-\omega_+,\,\omega_2=+\omega_+,\,\omega_3=-\omega_-,\,\omega_4=+\omega_-$.
We can associate them to their respective null eigenvectors $\ket{V_j(\omega_i)}$ which will have an explicit $\Delta$ dependence in general. The assignment is as follows
\begin{equation}
    \begin{split}
        M(\omega_1&=-\omega_+)\ket{V_2(\omega_1=-\omega_+)}=0\,,\\
        M(\omega_2&=+\omega_+)\ket{V_1(\omega_2=+\omega_+)}=0\,,\\
        M(\omega_3&=-\omega_-)\ket{V_1(\omega_3=-\omega_-)}=0\,,\\
        M(\omega_4&=+\omega_-)\ket{V_2(\omega_4=+\omega_-)}=0\,,\\
    \end{split}
    \label{eq:Null_relations}
\end{equation}
Based on this assignment, we can now improve our ansatz \eqref{eq:G_fourier_ansatz} based on the frequencies we found, and write
\begin{equation}
    \begin{split}
        G(\omega)=& A_1\delta(\omega-\omega_1)\ket{V_2(\omega_1)}\bra{V_2(\omega_1)}+A_2\delta(\omega-\omega_2)\ket{V_1(\omega_2)}\bra{V_1(\omega_2)}\\
        &+A_3\delta(\omega-\omega_3)\ket{V_1(\omega_3)}\bra{V_1(\omega_3)}+A_4\delta(\omega-\omega_4)\ket{V_2(\omega_4)}\bra{V_2(\omega_4)}\,,
    \end{split}
\end{equation}
where the proportionality sign implicitly contains any numerical factors from the expansion coefficients, for which we have only fixed the frequency dependence.

We recall at this point that our final goal is to solve the SD equations \eqref{eq:totalSDequation_v2} in real time, which can be obtained from \eqref{eq:kernel_SD_eq} by Fourier transformation and by means of the convolution theorem
\begin{equation}
    M(\tau)*G(\tau)=\int_{-\infty}^\infty M(\omega)G(\omega)e^{-\mathrm{i}\omega \tau} d\omega = 0\,,
    \label{eq:Fourier_kernel_SD}
\end{equation}
where $M(\tau)$ is the inverse Fourier transform of the operator in (\ref{eq:Momega}) and is the actual differential operator appearing in the time-domain SD equation. 
By now plugging our ansatz \eqref{eq:G_fourier_ansatz} into \eqref{eq:Fourier_kernel_SD}, and exploiting \eqref{eq:Null_relations}, we can show that it automatically fulfills the last equality,
\begin{equation}
    \begin{split}
        \int_{-\infty}^\infty& M(\omega)G(\omega)e^{-i\omega \tau} d\omega =\\ 
        &A_1e^{-\mathrm{i}\omega_1\tau}M(\omega_1)\ket{V_2(\omega_1)}\bra{V_2(\omega_1)}+A_2e^{-\mathrm{i}\omega_2\tau}M(\omega_2)\ket{V_1(\omega_2)}\bra{V_1(\omega_2)}\\
        &+A_3e^{-\mathrm{i}\omega_3\tau}M(\omega_3)\ket{V_1(\omega_3)}\bra{V_1(\omega_3)}+A_4e^{-\mathrm{i}\omega_4\tau}M(\omega_4)\ket{V_2(\omega_4)}\bra{V_2(\omega_4)}=0\,.
    \end{split}
    \label{eq:FT transform_ansatz}
\end{equation}
We can therefore conclude that the l.h.s.~of \eqref{eq:Fourier_kernel_SD} must also vanish, thus implying that the Green's function in real time can be obtained by Fourier transformation of our ansatz and reads,
\begin{equation}
    \begin{split}
        G(\tau)=\,\int_{-\infty}^\infty&G(\omega)e^{-\mathrm{i}\omega \tau} d\omega= \\
        & A_1e^{-\mathrm{i}\omega_1\tau}\ket{V_2(\omega_1)}\bra{V_2(\omega_1)}+A_2e^{-\mathrm{i}\omega_2\tau}\ket{V_1(\omega_2)}\bra{V_1(\omega_2)}\\
        &+A_3e^{-\mathrm{i}\omega_3\tau}\ket{V_1(\omega_3)}\bra{V_1(\omega_3)}+A_4e^{-\mathrm{i}\omega_4\tau}\ket{V_2(\omega_4)}\bra{V_2(\omega_4)}\,.
    \end{split}
\end{equation}
In order for the Green's function to not diverge for $\tau\to\infty$, we are forced to set the coefficients in front of the exponentials whose frequency has a positive imaginary part to zero. These are $\omega_2$ and $\omega_4$, 
and thus Green's function reads (with normalized null eigenvectors),
 \begin{equation}
    G(\tau)=A_1 e^{-\mathrm{i}\omega_1 \tau}\ket{V_2(\omega_1)}\bra{V_2(\omega_1)}+A_3 e^{-\mathrm{i}\omega_3 \tau}\ket{V_1(\omega_3)}\bra{V_1(\omega_3)}
\end{equation}
\begin{equation}
    \begin{split}
    =A_+ e^{-E_+ \tau}&
  \left(
\begin{array}{cc}
 \frac{\Delta  \left(\Delta ^2-1\right)^2 \mu_a^2}{\left(\Delta ^2+1\right)^2 \nu_{23}^2}+\frac{\Delta }{\Delta ^2+1}+\frac{1}{2}\quad & \frac{\mathrm{i} \left(\Delta ^2-1\right)}{2 \left(\Delta ^2+1\right)}-\frac{2\mathrm{i}\Delta ^2 \left(\Delta ^2-1\right) \mu_a^2}{\left(\Delta ^2+1\right)^2 \nu_{23}^2} \\[10pt]
 \frac{2\mathrm{i}\Delta ^2 \left(\Delta ^2-1\right) \mu_a^2}{\left(\Delta ^2+1\right)^2 \nu_{23}^2}-\frac{i \left(\Delta ^2-1\right)}{2 \left(\Delta ^2+1\right)}\quad & -\frac{\Delta  \left(\Delta ^2-1\right)^2 \mu_a^2}{\left(\Delta ^2+1\right)^2 \nu_{23}^2}-\frac{\Delta }{\Delta ^2+1}+\frac{1}{2}
\end{array}
\right)\\[10pt]
&+A_- e^{-E_- \tau}\left(
\begin{array}{cc}
 \frac{1}{2}-\frac{\Delta  \mu_a^2}{\nu_{23}^2} & \frac{i}{2} \\
 -\frac{i}{2} & \frac{\Delta  \mu_a^2}{\nu_{23}^2}+\frac{1}{2} \\
\end{array}
\right)
\,,
    \end{split}
\label{eq:G_long_tau}
\end{equation}
where $E_\pm$ are the real and positive parameters given by $|\omega_\pm|$ in (\ref{eq:omega_plusminus}). More precisely, $E_+=\mathrm{i}\omega_1$, and $E_-=\mathrm{i}\omega_3$. We have also renamed $A_1=A_+$ and $A_3=A_-$ to match the notation of their respective exponents, and that used in the main text. 
From (\ref{eq:omega_plusminus}), we can write the parameters $E_\pm$ as
\begin{equation}
   E_\pm= \frac{1}{2} \left(
   \sqrt{4 \mu _a^2+\nu _{23}^2}
   \pm
   \sqrt{4\Delta^2\mu_a^2+\nu_{23}^2}\right)\,.
\end{equation}
For later convenience, let us expand these quantities for $\mu_a\ll 1$, finding
\begin{equation}
    \begin{split}
        E_+&=\nu _{23}+\frac{(1+\Delta^2)\mu _a^2}{\nu _{23}}+O\left(\mu _a^4\right)\,,\\
        E_-&=\frac{(1-\Delta^2)\mu _a^2}{\nu _{23}}+O\left(\mu _a^4\right)\,.
    \end{split}
    \label{eq:Energies_long_tau}
\end{equation}
In the case $\Delta=0$, we obtain the following expression for the projected Green's function
\begin{equation}
    G(\tau)=\frac{1}{2}A_+ e^{-E_+ \tau}\left(
\begin{array}{cc}
 1 & -\mathrm{i} \\
\mathrm{i}& 1 \\
\end{array}
\right)+\frac{1}{2}A_- e^{-E_- \tau}\left(
\begin{array}{cc}
 1 &\mathrm{i}\\
 -\mathrm{i} & 1 \\
\end{array}
\right)
\,,
\end{equation}
which recovers \eqref{eq:longtimesolution_simpl} of the main text, with the associated energies
\begin{equation}
        E_+=\nu _{23}+\frac{\mu _a^2}{\nu _{23}}+O\left(\mu _a^4\right)\,,\quad\quad
        E_-=\frac{\mu _a^2}{\nu _{23}}+O\left(\mu _a^4\right)\,.
\label{eq:Energies_long_tau_expanded}
\end{equation}
which are those given in \eqref{eq:Eplusminus-simple}.

\section{Hamiltonian SDRG of Majorana chain with singlets}
\label{apx:SDRG}

In this Appendix, we review the formulation of SDRG at the level of the Hamiltonian, as introduced originally in \cite{MDH79prl,MDH80prb} for quantum spin chains with random spatial disorder. We consider general Hamiltonians with nearest neighbor interactions of the form 
\begin{equation}
    H=\sum_{x\in\mathds{Z}}\mu_x h(\Vec{\sigma}_x\cdot \Vec{\sigma}_{x+1};\theta_x)\,,
    \label{eqSM:General_Hamiltonian}
\end{equation}
where the vector $\Vec{\sigma}_i$ contains our choice of local spin degrees of freedom which span the local Hilbert space of our model, whose dimension we label by $d$. The sets $\{\mu_x\}$ and $\{\theta_x\}$ in \eqref{eqSM:General_Hamiltonian} denote hopping and anisotropy parameters, respectively, which are spatially distributed along the chain. These could be distributed according to an aperiodic sequence, as considered in Sec.~\ref{subsec:sdrg_aperiodic}, or be drawn randomly at each site, as was the case in Sec.~\ref{subsec:random_SDRG}. For concreteness, note that the XX spin chain given in \eqref{eq:genericNandp1_spinDOFform}, which is equivalent to our model \eqref{eq:Hamiltonian} in the free case of $\J=0$, is of the form given in \eqref{eqSM:General_Hamiltonian} after choosing $\theta_x=0$, $\Vec{\sigma}_x=\left(X_x,Y_x\right)^{\textrm{T}}$, and $h(\Vec{\sigma}_x\cdot \Vec{\sigma}_{x+1})=X_xX_{x+1}+Y_xY_{x+1}$. In this case, the local Hilbert space has dimension $d=2$.\\
For the general description of the SDRG, the explicit nature of the disorder is not relevant, as long as we can make the assumption that the chain has one hopping parameter $\mu_{\textrm{\tiny strong}}$ which is much larger than all the other ones. This is the central strong-disorder assumption at the core of the SDRG procedure. The idea is that the low-energy features of the system can be accessed perturbatively by decimating the degrees of freedom which are coupled by $\mu_{\textrm{\tiny strong}}$. This is because if $\mu_{\textrm{\tiny strong}}$ is large, we expect these degrees of freedom to couple among themselves to a local state with a large associated energy of $\order{\mu_{\textrm{\tiny strong}}}$. At energies much lower than $\mu_{\textrm{\tiny strong}}$, the contribution of such states to the ground state of the entire chain  decouple and can be treated separately in perturbation theory. In particular, for the purposes of the current work, we restrict our attention to the cases where such strong hopping parameters appear isolated throughout the chain. It then suffices to consider a block of 4 sites $x=1,2,3,4$, where the central two-sites are coupled by a strong hopping parameter, while sites 1-2 and 3-4 are coupled by weaker hopping parameters, which we denote by $\mu_{\textrm{l,r}}\ll \mu_{\textrm{\tiny strong}}$. The Hamiltonian $H^{(4)}$ governing this 4-site block is a $(d^4\times d^4)$-dimensional matrix of the form \eqref{eqSM:General_Hamiltonian}, which we split into three terms,
\begin{equation}
    H^{(4)}=H_{\textrm{l}}+H_{\textrm{\tiny strong}}+H_{\textrm{r}}\,,
\end{equation}
where each of these terms corresponds to the specific site pairs 1-2, 2-3, and 3-4, with hopping parameters $\mu_{\textrm{l}},\mu_{\textrm{\tiny strong}},\mu_{\textrm{r}}$, respectively. These act non-trivially only on their respective sites, and a tensor product with the $d$-dimensional identity operator on the remaining two sites is implied. The local Hamiltonian of the two central sites, which itself is only a $(d^2\times d^2)$-dimensional matrix, will have a set of eigenstates $\{\ket{g},\ket{e_i}\}$, containing the local two-site ground state $\ket{g}$ (which we assume to be non-degenerate) and local excited states $\ket{e_i}$, with $i=0,\dots,d^2-2$. The perturbative assumption that $\mu_{\textrm{\tiny strong}}\gg \mu_{\textrm{l},\textrm{r}}$ implies that we can see $H_{\textrm{l}}+H_{\textrm{r}}$ as perturbations of the central Hamiltonian $H_{\textrm{\tiny strong}}$, with $\mu_{\textrm{l}},\, \mu_{\textrm{r}}$ as perturbation parameters. The action of decimating the central two sites is realized by projecting $H^{(4)}$ onto the local ground state $\ket{g}$, which we denote as the singlet state. This nomenclature originates from the fact that for $s=1/2$ SU$(2)$ spins, as is the case for the XX chain \eqref{eq:genericNandp1_spinDOFform}, $\ket{g}$ is precisely the singlet state of two spins. We can however identify this singlet state for more general choices of local degrees of freedom \cite{Basteiro:2022pyp}. In perturbation theory, the effect of the decimation is to induce a renormalized effective coupling and anisotropy $(\mu',\theta')$ between sites 1 and 4. In particular, we perform this real-space decimation on the Hamiltonian operators themselves, such that the second-order perturbative corrections are given by
\begin{equation}
    H'_{14}=\sum_{i=0}^{d^2-2}\frac{\bra{g}H_{\textrm{l}}+H_{\textrm{r}}\ket{e_i}\bra{e_i}H_{\textrm{l}}+H_{\textrm{r}}\ket{g}}{E_g-E_{e_i}}\equiv \mu'h(\Vec{\sigma}_1\cdot\vec{\sigma}_4;\theta')\,,
    \label{eqSM:renormalized_Ham}
\end{equation}
which is itself a Hamiltonian operator. It turns out that the first-order corrections vanish \cite{MDH79prl,MDH80prb,Vieira:2005PRB}. Crucially, this renormalized Hamiltonian, which is now a $(d^2\times d^2)$ matrix, is of the same form as the original one \eqref{eqSM:General_Hamiltonian}, albeit with renormalized parameters $(\mu',\theta')$. Though this need not be the case in general, it is indeed the behavior of the XX Hamiltonian \eqref{eq:genericNandp1_spinDOFform}.
Notice that it is very reasonable that the matrix \eqref{eqSM:renormalized_Ham} is now $(d^2\times d^2)$ since it acts on the two remaining sites of the original 4-site block which have survived the decimation. In general, after computing $\mu'$, we have to check that our decimation procedure is indeed driving the system towards lower energies, meaning that $\mu'$ we should check if $\mu'<\mu_{\textrm{\tiny strong}}$ to verify the validity of our perturbative assumption. If this is the case, we can iterate these decimation steps simultaneously throughout the chain, proceeding in this way along the RG trajectory. The low-energy properties of the system are then encoded in the strong-disorder fixed point reached after an infinite number if decimations.\\
Let us return to the case of disordered XX quantum spin chains, which describe the free limit $\J=0$ of our Hamiltonian \eqref{eq:Hamiltonian}. More precisely, let us focus on XX chains with Hamiltonian \eqref{eq:genericNandp1_spinDOFform}, $L=4$ and disordered hopping parameters $\mu_{\textrm{l}},\mu_{\textrm{\tiny strong}},\mu_{\textrm{r}}$. We can readily apply the formula \eqref{eqSM:renormalized_Ham}, as was done in \cite{vieira05,Vieira:2005PRB}, to find the explicit form of the renormalized hopping. Of course, the results for a single XX chain straightforwardly generalize to the case of $N/2$ decoupled copies, as we have in \eqref{eq:genericNandp1_spinDOFform}. For the XX chain, the renormalized hopping appearing in \eqref{eqSM:renormalized_Ham} reads
\begin{equation}
   \mu'=\frac{\mu_{\textrm{l}}\mu_{\textrm{r}}}{\mu_{\textrm{\tiny strong}}}\,.
   \label{eqSM:mu_prime_XX}
\end{equation}
The equation above provides a positive consistency check for our results derived in presence of SYK interactions in their free limit of $\J\to0$. More precisely, for the case of aperiodic SYK chains considered in Sec.~\ref{subsec:sdrg_aperiodic}, we have $\mu_{\textrm{\tiny strong}}=\mu_b$, $\mu_{\textrm{l}}=\mu_a(1+\Delta)$ and $\mu_{\textrm{r}}=\mu_a(1-\Delta)$, such that \eqref{eq:mu_prime_main_text} reduces to \eqref{eqSM:mu_prime_XX} when $\J=0$ since then $\tanh\gamma_{23}=1$. Similarly, for the random SYK chains considered in Sec.~\ref{subsec:random_SDRG}, we have $\mu_{\textrm{\tiny strong}}=\mu_0$, $\mu_{\textrm{l}}=\mu(1+\Delta)$, and $\mu_{\textrm{r}}=\mu(1-\Delta)$, and we can again confirm that \eqref{eq:mu_prime_random} reproduces \eqref{eqSM:mu_prime_XX} in the free limit of $\tanh\gamma_{23}=1$.

\section{Aperiodic sequence}\label{apx:Aperiodic_Sequence}

This Appendix is devoted to the review of binary aperiodic sequences used as spatial modulations for the hopping parameters of quantum chains in the main text (see Sec.\,\ref{subsec:sdrg_aperiodic}), where $a\equiv \mu_a$ and $b\equiv \mu_b$. We will first briefly cover the general treatment of general aperiodic sequences generated from substitution or inflation rules, and later focus on the case of the silver mean sequence. We refer the reader to \cite{FlickerBoyleDickens,JahnCentralCharge,Basteiro:2022xvu,Basteiro:2022zur} for more detailed descriptions and for context on the wide range of areas where these aperiodic sequences are of interest.

Binary aperiodic sequences are infinite sequences consisting of two letters which have no repeating finite-sized sub-sequences at fixed intervals. When thought of as modulating disorder pattern in quantum systems, they pose an interesting middle ground between homogeneous and completely random disorder. They can be obtained from so-called \textit{substitution rules} \cite{Juh_sz_2007}, which generate aperiodic sequences upon iterative application. In the case of binary aperiodic sequences, general substitution rules have the form $a\mapsto w_a(a,b),\,b\mapsto w_b(a,b)$, where $w_a,w_b$ are words consisting of both $a$ and $b$ letters. By applying to any seed sequence the same substitution rule an arbitrarily large number of times, we generate a sequence of $a$ and $b$ which has no patterns repeating at any fixed periods, but where a given sub-sequence can be found again at any scale, albeit with newly defined constituents. For example, if we have the sub-sequence $abaa$, an infinite aperiodic sequence is guaranteed to contain a larger sub-sequence of the form $w_a\circ w_b\circ w_a\circ w_a$, where the operation $\circ$ denotes simple sequence concatenation. The latter sub-sequence will, of course, be defined on a larger scale (i.e. will consist of more letters), but the original structure can be identified.

Aperiodic sequence obtained in this way have properties regarding the relative distribution of $a$ and $b$ letters which can be derived from their associated \textit{inflation matrix}, defined as
\begin{equation}
M_\sigma
=\,
\bigg( 
\begin{array}{cc}
\#_a(w_a)  &  \#_a(w_b) \\
\#_b(w_a)  &  \#_b(w_b)  \\
\end{array}   \bigg)\,.
\label{eq:general_inflation_matrix}
\end{equation}
Here, the entries $M_{\alpha\beta}=\#_\alpha(w_\beta)$, with $\alpha,\beta\in\{a,b\}$, denote the frequency of each letter $\alpha$  in the corresponding substitution word $w_\beta$. These entries are clearly real and positive, and thus, by the Perron-Frobenius theorem, the inflation matrix has a unique largest eigenvalue, denoted by $\lambda_+$. This eigenvalue $\lambda_+$ describes the asymptotic scaling factor between strings of subsequent inflation steps after a large number of applied substitutions. In other words, the length of the sequence after $n\gg1$ substitutions will be approximately $\lambda_+$ times larger than the length of the sequence after $(n-1)$ steps. To this eigenvalue, we can associate a statistically normalized right eigenvector, denoted as $\boldsymbol{v}_+=(p_a, p_b)^{\textrm{T}}$ with $p_a+p_b=1$, which is also guaranteed to be positive. The entries of $\boldsymbol{v}_+$ describe the relative frequency of the letters $a$ and $b$, respectively, in the asymptotic sequence. Thus, given a set of words $w_a,w_b$ defining the substitution rules, one can characterize the associated infinite aperiodic sequence by studying the eigenvectors and eigenvalues of the  substitution matrix \eqref{eq:general_inflation_matrix}.

For certain families of aperiodic sequences, so-called \textit{singlet-producing self-similar sequences} \cite{Juh_sz_2007}, the decimation steps associated with the SDRG procedure can be seen as inverse substitution steps from the point of view of the substitution matrix. Thus, many properties such as the distribution of decimated pairs arising in the SDRG procedure can be derived from the distribution of letters determined by the right eigenvector to the Perron-Frobenius eigenvalue $\lambda_+$. We will exploit this in the following.

Let us now focus on the case of the silver mean sequence considered in Sec.~\ref{subsec:sdrg_aperiodic}. In particular, let us use the properties of aperiodic sequences mentioned above to quantify the constituents of the factorised ground state reported in \eqref{eq:factorised GS}. The substitution rules for the silver mean sequence are given by  \cite{Juh_sz_2007,Basteiro:2022zur}
$a\mapsto aba, b\mapsto a$, and so the corresponding inflation matrix reads
\begin{equation}
    M=\begin{pmatrix}
        2 & 1\\
        1 & 0
    \end{pmatrix}\,,
    \label{eq:Inflation_Matrix_silvermean}
\end{equation}
with eigenvalues given by $\lambda_{\pm}=1\pm \sqrt{2}$. The silver mean sequence is indeed a singlet-producing self-similar sequence \cite{Juh_sz_2007}, and so the decimation steps used in the SDRG description of a chain with this aperiodic modulation can be seen as inverse substitution steps $aba\mapsto a$ and $a\mapsto b$. In view of the ground state \eqref{eq:factorised GS}, we are interested in the relative frequency of $b\equiv \mu_b$ hoppings in the asymptotic sequence, since spins connected by such strong hoppings in the original chain are going to couple into singlets after one decimation step. Although in the large $N$ chain considered in Sec.\,\ref{subsec:sdrg_aperiodic}, the local ground states of the decimated pairs are not singlets, in this appendix we will refer to them with this nomenclature borrowed from the application of SDRG to XXZ spin chains \cite{Vieira:2005PRB}. From the substitution matrix \eqref{eq:Inflation_Matrix_silvermean} we can readily compute its statistically normalized right eigenvector, which is given by $\boldsymbol{v}_+\equiv(\boldsymbol{v}_a,\boldsymbol{v}_b)^{\textrm{T}}=(\frac{1}{\sqrt{2}},1-\frac{1}{\sqrt{2}})^{\textrm{T}}$. Thus, we can immediately deduce that $\boldsymbol{v}_b=1-\frac{1}{\sqrt{2}}$ is exactly the size of the set $G_1$ quantifying the relative fraction of the spins in the chain that will give contributions of the form $|{\rm ETW}_{s}(\hat{\mu}_b)\rangle$ to the ground state \eqref{eq:factorised GS}. Furthermore, exploiting the fact that for this choice of aperiodic modulation, the SDRG decimation pattern follows exactly the inverse of the inflation procedure to generate the sequence, a relation for the concentration $\rho_n$ of singlets after $n\geq 1$ decimation steps was found in \cite{Juh_sz_2007}. It reads
\begin{equation}
    \rho_n= \boldsymbol{v}_b(\lambda_+)^{-(n-1)}\,.
    \label{eq:singlet_density}
\end{equation}
In particular, $\rho_1$ is the concentration of singlets after the first decimation step, which by construction is the relative fraction of strong bonds in the original chain, i.e.~$\boldsymbol{v}_b$.
Moreover, for $n=2$, the quantity $\rho_{2}$ is precisely the relative size of the set $G_2$ used in \eqref{eq:factorised GS}, namely the number of sites which couple into singlets after two decimation steps, giving contributions of the form $|{\rm ETW}_{s}(\hat{\mu}_a)\rangle$ to the ground state \eqref{eq:factorised GS}. Thus, using the relation \eqref{eq:singlet_density} and the explicit values of $\lambda_+$ and $\boldsymbol{v}_b$ obtained for the silver mean sequence, we find that the relative size of the $G_2$ set in \eqref{eq:factorised GS} is given by $\rho_2=\boldsymbol{v}_b\lambda_+^{-1}=\frac{3}{\sqrt{2}}-2$. Since the concentrations $\rho_n$ are normalized in such a way that $\sum_{n=1}^{\infty}\rho_n=\frac{1}{2}$, the obtained relative sizes can be expressed as true percentages of the total number of sites in the chain upon multiplication by two. Therefore, the total percentage of wormhole-like contributions to the ground state \eqref{eq:factorised GS}, i.e. the combined relative sizes of the sets $G_1$ and $G_2$, is $2\left(\frac{2}{\sqrt{2}}-1\right)$, i.e. approximately $83\%$. Indeed, this can be already qualitatively corroborated with the snippet of the aperiodic singlet phase shown in Fig.~\ref{fig:SYKgroundstate}, where 30 out of 35 sites are connected by a singlet after two decimation steps, yielding a percentage of approximately $85\%$.

\section{Flow equation of probability distributions along SDRG}
\label{apx:FlowEq}

In this appendix, we derive the flow equation (\ref{eq:RGequationP_XXX}) for the SDRG evolution of the hopping parameter probability distribution in infinite quantum chains with Hamiltonian (\ref{eq:Hamiltonian}) and random bond disorder. This model has been studied in Sec.\,\ref{subsec:random_SDRG} of this manuscript. Our derivation of the flow equation follows closely the ones in \cite{MDH79prl,MDH80prb,Fisher94SDRG}, but differs from them due to the presence of the factor $e^{-\zeta}$ in the hopping parameter renormalization rule (\ref{eq:mu_prime_random}). In general, this factor might depend on the strong hopping parameter $\mu_0$ of the bond decimated along the SDRG. As explained in Sec.\,\ref{subsec:random_SDRG}, this happens when the decimated sites are nearest neighbors in the original infinite chain.
Notice that, when $\J\to 0$, i.e. $e^{-\zeta}\to 1$, our setup and, consequently, the derivation of the flow equation, reduces to the one of \cite{Fisher94SDRG} for XX spin chains with random disorder in the hopping parameters.
Throughout this appendix, we will use the conventions introduced in Sec.\,\ref{subsec:random_SDRG}, namely we will denote with $\tilde\mu_0$  the strongest hopping parameter running along the SDRG and with $\mu_0$ the one in the original chain.

To derive the flow equation, we start by observing what happens if we perform a decimation that infinitesimally changes the value of the maximal hopping $\tilde{\mu}_0$ at a certain stage of SDRG to $\tilde{\mu}_0-d\tilde{\mu}_0$.
In particular, we ask what happens to the probability distribution $P(\mu_x)$ of having a coupling $\mu_x$ along the chain along this infinitesimal transformation.
 This probability distribution depends also on the SDRG step we are considering, which is characterized by the energy scale $\tilde{\mu}_0$ at that step. In other words, $P$ depends on $\tilde{\mu}_0$.
If we call $F$ the fraction of the post-decimation bonds with hopping parameter $\mu_x$ originated from the last renormalization step, the post-decimation distribution $P(\mu_x;\tilde{\mu}_0-d\tilde{\mu}_0 )$ is given by
\begin{equation}
\label{eq:flow_eq_0}
  P(\mu_x;\tilde{\mu}_0-d\tilde{\mu}_0 )
  =
 \frac{ P(\mu_x;\tilde{\mu}_0)+F}{1-2 d\tilde{\mu}_0 P(\tilde{\mu}_0;\tilde{\mu}_0)}\,,
\end{equation}
where the denominator quantifies the fraction of degrees of freedom that survived in the chain after decimation.
Based on the decimation procedure described in Sec.\,\ref{subsec:random_SDRG}, the fraction $F$ can be written as
\begin{eqnarray}
   && F=  d\tilde{\mu}_0 P(\tilde{\mu}_0;\tilde{\mu}_0)
    \\
    &&
  \times \int_{0}^{\tilde{\mu}_0} d\tilde{\mu}_{\rm l}d\tilde{\mu}_{\rm r}
    P(\tilde{\mu}_{\rm r};\tilde{\mu}_0)P(\tilde{\mu}_{\rm l};\tilde{\mu}_0)
    \left[
    \delta\left(\mu_x-\frac{\tilde{\mu}_{\rm l}\tilde{\mu}_{\rm r}}{\tilde{\mu}_{0}}\tanh\gamma_{23}\right)- \delta\left(\mu_x-\tilde{\mu}_{\rm r}\right)- \delta\left(\mu_x-\tilde{\mu}_{\rm l}\right)
    \right]
    \,.
    \nonumber
\end{eqnarray}
By expanding the right-hand side of (\ref{eq:flow_eq_0}) at first order in $d\tilde{\mu}_0$, we find
\begin{eqnarray}
\label{eq:flow_eq_1}
    && P(\mu_x;\tilde{\mu}_0- d\tilde{\mu}_0 )
 -P(\mu_x;\tilde{\mu}_0)\\ 
 = &&
  d\tilde{\mu}_0 P(\tilde{\mu}_0;\tilde{\mu}_0)\int_{0}^{\tilde{\mu}_0} d\tilde{\mu}_{\rm l}d\tilde{\mu}_{\rm r}
    P(\tilde{\mu}_{\rm r};\tilde{\mu}_0)P(\tilde{\mu}_{\rm l};\tilde{\mu}_0)
    \delta\left(\mu_x-\frac{\tilde{\mu}_{\rm l}\tilde{\mu}_{\rm r}}{\tilde{\mu}_{0}}\tanh\gamma_{23}\right)+\O(d\tilde{\mu}_0^2)
    \nonumber\,.
\end{eqnarray}
It is convenient to introduce the dimensionless variable $b=\frac{\mu_x}{\tilde{\mu}_0}$. Re-expressing the integration variables on the right-hand side in terms of $b$, (\ref{eq:flow_eq_1}) becomes
\begin{equation}
\label{eq:begin_floweq}
\begin{split}
    &~\tilde{\mu}_0 \frac{P(\mu_x;\tilde{\mu}_0-d\tilde{\mu}_0 )-
  P(\mu_x;\tilde{\mu}_0)}{d\tilde{\mu}_0}\\
  =&~
  P(\tilde{\mu}_0;\tilde{\mu}_0)
  \tilde{\mu}_0^2
  \int_{0}^{1} db_{\rm l}db_{\rm r}
    P(\tilde{\mu}_{\rm r};\tilde{\mu}_0)P(\tilde{\mu}_{\rm l};\tilde{\mu}_0)
    \delta\left(b-b_{\rm l}b_{\rm r}\tanh\gamma_{23}\right)\,.
\end{split}\end{equation}
Notice that, despite the rescaling of the integration variables, $P$ is still a distribution of the random variable $\mu_x$. Moreover, the dependence of $P$ on $\mu_x$ implies a dependence on $b$ in such a way that the left-hand side of (\ref{eq:begin_floweq}) can be written, in the limit $d\tilde{\mu}_0\to 0$, as
\begin{equation}
    \tilde{\mu}_0 \frac{P(\mu_x;\tilde{\mu}_0-d\tilde{\mu}_0 )-
  P(\mu_x;\tilde{\mu}_0)}{d\tilde{\mu}_0}\to
  -\tilde{\mu}_0\frac{\partial P(\mu_x;\tilde{\mu}_0)}{\partial \tilde{\mu}_0} -\tilde{\mu}_0\frac{d b}{d\tilde{\mu}_0}\frac{\partial P(\mu_x;\tilde{\mu}_0)}{\partial b}\,.
\end{equation}
Plugging back this expression in (\ref{eq:begin_floweq}), we obtain the flow equation
\begin{equation}
\label{eq:DasguptaMa}
\begin{split}
   &~-\tilde{\mu}_0\frac{\partial P(\mu_x;\tilde{\mu}_0)}{\partial \tilde{\mu}_0} +b\frac{\partial P(\mu_x;\tilde{\mu}_0)}{\partial b}
   \\
   =&~
    P(\tilde{\mu}_0;\tilde{\mu}_0)
  \tilde{\mu}_0^2
  \int_{0}^{1} db_{\rm l}db_{\rm r}
    P(\tilde{\mu}_{\rm r};\tilde{\mu}_0)P(\tilde{\mu}_{\rm l};\tilde{\mu}_0)
    \delta\left(b-b_{\rm l}b_{\rm r}\tanh\gamma_{23}\right)
    \,,
\end{split}\end{equation}
which was originally found in \cite{MDH79prl,MDH80prb}.
To write the flow equation in the convenient form reported in the main text, we consider the logarithmic variables defined in (\ref{eq:logenergy}). The main advantage of using these new variables is that now the renormalization rule (\ref{eq:mu_prime_random}) for the hopping parameters is additive instead of multiplicative. Recalling the change of variable transformation rule of probability distribution functions, we have to substitute
\begin{equation}
P(\mu_x;\tilde{\mu}_0)\to    \bigg|\frac{d\beta}{d\mu_x}\bigg| P(\beta;\Gamma)=   \frac{P(\beta;\Gamma)}{\tilde{\mu}_0 b}\,.
\end{equation}
By plugging this transformation rule into (\ref{eq:DasguptaMa}), changing the variables in the integral, and using basic properties of the $\delta$-function, we find
\begin{eqnarray}
    \frac{\partial P(\beta;\Gamma)}{\partial \Gamma} &-&\frac{\partial P(\beta;\Gamma)}{\partial \beta}
    \\
 &=&
    P(0;\Gamma)
  \int_{0}^{\infty} d\beta_{\rm l}d\beta_{\rm r}
    P(\beta_{\rm r};\Gamma)P(\beta_{\rm l};\Gamma) e^{\beta_{\rm l}+\beta_{\rm r}+\zeta(\Gamma)-\beta}
    \delta\left(\beta-\beta_{\rm l}-\beta_{\rm r}-\zeta(\Gamma)\right)\,.
    \nonumber
\end{eqnarray}
The exponential in the integral on the right-hand side becomes one due to the delta function, and we finally obtain the flow equation (\ref{eq:RGequationP_XXX}) in the main text.

\bibliographystyle{nb}
\bibliography{refs}

\end{document}